\documentclass[12pt,preprint]{aastex}
\usepackage{natbib}
\usepackage{amsmath}
\newcommand{\noprint}[1]{}

\shorttitle{Multiplicity of Massive Stars in Cyg OB2}
\shortauthors{Caballero-Nieves et al.}
\begin{document}


\title{A High Angular Resolution Survey of Massive Stars in Cygnus~OB2: 
Results from the Hubble Space Telescope Fine Guidance Sensors}

\author{S.~M. Caballero-Nieves\altaffilmark{1,2},
E.~P. Nelan\altaffilmark{3}, D.~R. Gies\altaffilmark{2},
D.~J. Wallace\altaffilmark{4}, K. DeGioia-Eastwood\altaffilmark{5},
A. Herrero\altaffilmark{6},
W.-C. Jao\altaffilmark{2}, B.~D. Mason\altaffilmark{7},
P. Massey\altaffilmark{8}, A.~F.~J. Moffat\altaffilmark{9},
N.~R. Walborn\altaffilmark{3}}

\altaffiltext{1}{Current Address: Department of Physics and Astronomy
University of Sheffield, 
Hounsfield Road, Sheffield S3 7RH, UK;
s.caballero@shef.ac.uk}

\altaffiltext{2}{Center for High Angular Resolution Astronomy,
Department of Physics and Astronomy, 
Georgia State University, P. O. Box 4106, Atlanta, GA  30302-4106, USA;
gies@chara.gsu.edu, jao@chara.gsu.edu}

\altaffiltext{3}{Space Telescope Science Institute,
3700 San Martin Drive, Baltimore, MD 21218, USA;
nelan@stsci.edu, walborn@stsci.edu}

\altaffiltext{4}{Department of Natural Sciences, University of South
Carolina Beaufort, 1 University Boulevard, Bluffton, SC 29909, USA;
debra.j.wallace@nasa.gov}

\altaffiltext{5}{Department of Physics and Astronomy, Northern Arizona
University, P.O. Box 6010, Flagstaff, AZ 86011-6010, USA;
kathy.eastwood@nau.edu}
 
\altaffiltext{6}{Instituto de Astrof\'{i}sica de Canarias, 
C/ Via Lactea s/n, E-38280 La Laguna, Spain and
Departamento de Astrof\'{i}sica, Universidad de La Laguna, 
Avda. Astrof\'{i}sico Francisco S\'{a}nchez, 
2, E-38205 La Laguna, Spain; ahd@iac.es}

\altaffiltext{7}{U. S. Naval Observatory, 3450 Massachusetts Avenue,
NW, Washington, DC 20392-5420, USA; brian.mason@usno.navy.mil}

\altaffiltext{8}{Lowell Observatory, 1400 West Mars Hill Road,
Flagstaff, AZ 86001, USA; massey@lowell.edu}

\altaffiltext{9}{D\'{e}partement de physique, Universit\'{e} de
Montr\'{e}al, CP 6128, Succ. Centre-Ville, Montr\'{e}al, QC H3C 3J7, Canada;
moffat@astro.umontreal.ca}

\slugcomment{Draft 09/13/2013}


\begin{abstract}
We present results of a high angular resolution survey of massive OB
stars in the Cygnus OB2 association that we conducted with the Fine
Guidance Sensor 1R (FGS1r) on the Hubble Space Telescope. FGS1r is
able to resolve binary systems with a magnitude difference $\Delta V <
4$ down to separations as small as $0\farcs01$.  The sample includes
58 of the brighter members of Cyg~OB2, one of the closest examples of
an environment containing a large number of very young and massive
stars.  We resolved binary companions for 12 targets and confirmed the
triple nature of one other target, and we offer evidence of marginally
resolved companions for two additional stars.  We confirm the binary
nature of 11 of these systems from complementary adaptive optics
imaging observations.  The overall binary frequency in our study is
22\% to 26\% corresponding to orbital periods ranging from
20 - 20,000 years. When combined with the known short-period
spectroscopic binaries, the results supports the hypothesis that the
binary fraction among massive stars is $> 60$\%.
One of the new discoveries is a companion to the hypergiant star
MT~304 = Cyg OB2-12, and future measurements of orbital motion should
provide mass estimates for this very luminous star.
\end{abstract}



\section{Introduction}                              

Massive stars ($\gtrsim 10 M_{\odot}$) play a fundamental role in the
evolution of the universe, from influencing galactic dynamics and
structure to triggering star formation through their spectacularly
violent deaths. It has been well established that massive stars have a
higher binary frequency than lower mass stars \citep{mas09, rag10},
and the binary fraction may be as high as 100\% for massive stars in
clusters and associations \citep{mas98, mas09, kou07, chi12}.  Because
most massive stars are born in clusters and associations, their
multiplicity properties offer important clues about their formation
processes \citep{zin07}.  However, our knowledge about the numbers and
distributions of binary and multiple stars is incomplete because the
systems are generally so distant that we cannot detect binaries in the
separation realm between spectroscopically detected systems and
angularly resolved systems, i.e., those with small angular separations
and periods in the range of years to decades \citep{mas98}.  We need
milliarcsecond (mas) resolution to start to fill in this observational
period gap \citep{san08}.

The Orion Nebula cluster provides the closest example ($d=0.41$ kpc;
\citealt{men07}) of an environment with a modest number of massive O-stars
\citep{pre99, wei99, clo12}.  However, to explore a large sample of
very massive stars, the next closest environment is the association
Cygnus~OB2 at a distance of $d = 1.40 \pm 0.08$~kpc \citep{ryg12}.
The Cyg~OB2 association has approximately $2600\pm 400$ members
\citep{kno00} with about 100 O-stars within the central 1$^{\circ}$
\citep{com02, wri10}, making it one of the largest concentrations of
OB stars in the Galaxy. It is home to some of the most massive ($M >
100M_{\odot}$; \citealt{mas91, kno00, her01}) and intrinsically
brightest stars (MT~304; \citealt{mas91}) known in our Galaxy,
including two of the rare O3-type stars (MT~417 and MT~457;
\citealt{wal02, wal73b}, respectively).  The close binary properties
of the Cyg~OB2 stars have been studied extensively in the
spectroscopic survey of \citet{kim12b} (and references therein).  A
few wider systems have been identified through high angular resolution
speckle \citep{mas09} and imaging observations \citep{mai10}.

The Fine Guidance Sensors (FGS) aboard the {\it Hubble Space
  Telescope} (HST) provide us with the means to search for visual
binaries among the optically faint, massive stars of Cyg~OB2.  FGS
TRANS mode observations can resolve systems with separations of
$0\farcs01 - 1\arcsec$, or 14-1400 AU (or 2800 AU in the case of
MT~417 with a $2\farcs5$ scan length) and differential magnitudes less
than about 4~mag \citep{hor06, nel12}.  At these separations we are
sampling systems where the components would evolve independently of
each other but are nevertheless important in understanding massive
star formation. Due to the short length of the scans ($\sim 1\arcsec$)
the sources we find are highly probable to be true companions. We will
discuss the probability of chance alignments in our sample in a future
paper (Caballero-Nieves et al., in prep.).  \citet{nel04} used the FGS
instrument to search for binaries among 23 massive stars in the Carina
Nebula region ($d\approx 2.5$ kpc), and they discovered five new
binaries, including a companion of the very massive star HD~93129A at
a separation of 53~mas (133 AU).  Here we present the results of an
FGS survey for visual binaries around 58 stars in Cyg~OB2.  We compare
the results with those from an infrared adaptive optics survey that
will be presented in a future paper.  In section~\ref{fgs-samp} we
describe the sample selection, observations, and the FGS reduction
pipeline and process.  Section~\ref{fgs-det} describes how binary
stars are detected, and section~\ref{fgs-mod} details the fitting
routines used to determine the system parameters (differential
magnitude, angular separation, and position angle). The detection
limits for our sample are also discussed there.  Our results are
presented in section~\ref{fgs-results}, and a discussion of the
multiplicity of our sample and future work is given in
section~\ref{fgs-conc}.


\section{Sample and Observations}\label{fgs-samp}     

The stars were selected from among the brightest, most massive stars
in Cyg~OB2 as cataloged in \citet{sch58}, \citet{mas91}, and
\citet{com02}.  All the stars in our sample have published spectral
classifications and are brighter than $V = 14$~mag, well within the
detection limits of FGS.  The observations were scheduled under HST
proposal 10612 (PI: D.~Gies), a SNAP program with 70 available
targets. SNAP programs are used when needed to fill gaps in HST's
observing schedule which cannot be filled by the GO class
programs. Typically 50\% of the targets in a SNAP program are actually
observed, our program faired better with 58 stars observed, listed in
Table 1. The stars are identified according to the numbering scheme
used in the optical photometric survey of \citet{mas91} (e.g., MT~138)
and the infrared spectroscopic study of \citet{com02} (e.g.,
A~23). Stars not included in those surveys are identified by the
number assigned by \citet{sch58} (e.g., SCHULTE~5 = S~5 = Cyg~OB2-5)
or by the Wolf-Rayet number from \citet{vdh01} (e.g., WR~145).
Table~\ref{t1-fgs} also lists the celestial coordinates from 2MASS
\citep{skr06}, spectral classification and reference source, $V$
magnitude, $B-V$ color, date of observation, the name of the single
star whose observation was used as the calibrator for the binary
fitting, and the number of components detected.  The stars from
\citet{com02} do not have known $B-V$ colors, but because they are
bright infrared sources, they are assumed to be redder than the other
stars in the survey.  The final column indicates known binary systems
detected through our adaptive optics program or through the
spectroscopic observations of \citet{kim12a}.  `NIRI' in the remarks
column denotes systems that we resolved in a $K$-band survey made with
the NIRI camera and Altair adaptive optics system on the Gemini North
Telescope \citep{cab12}.  `RV constant' indicates objects observed by
\citet{kob12} that did not show radial velocity variability during the
course of their survey.  Spectroscopic binaries are denoted by `SB1'
and `SB2' for those detected by single-lined or double-lined radial
velocity variations, respectively \citep{kim12a}.  Eclipsing binaries
are denoted by `EW/KE' (W~UMa type with ellipsoidal variations and
periods $P < 1$ d), `EA' (detached systems with flat maxima), and `EB'
(semi-detached with ellipsoidal light curves) \citep{kim12a}.  At the
time of writing, 32 stars out of the 58 observed are known multiple
systems; 9 are angularly resolved systems listed in the Washington
Double Star catalog \citep{mas01}, two of which are also spectroscopic
binaries from the 25 systems detected by \citet{kim12a}.

\placetable{t1-fgs}   

All the stars were observed using FGS1r in its high angular resolution
TRANSFER mode (TRANS). The FGS is a Koesters prism based, white light,
shearing interferometer that is sensitive to the tilt of the
incident wave front from a luminous object. The FGS optical train
includes a polarizing beam splitter that illuminates two mutually
orthogonal Koesters prisms which provides simultaneous sensitivity to
the wavefront tilt along the FGS's $x$- and $y$-axes. In TRANSFER mode
the the FGS $5\arcsec \times 5\arcsec$ instantaneous field of view
(IFOV) is scanned across the object of interest along a 45 degree path
in the FGS ($x$,$y$) coordinate frame, which varies the incoming
wavefront tilt along both axes. If the source is a resolved binary,
the light from the two stars will be mutually incoherent and the
observed interference fringe will differ from that of an unresolved
point source \citep{nel12}. We observed each star using ~20 scans
with a scan length of ~1\arcsec~composed of 1 milli-arcsecond steps
(for MT~417 we employed a $2\farcs5$ scan length to capture each
component of this known wide binary). All observations were made using
the F583W filter, which has a central wavelength of 5830~\AA~and a
passband of 3400~\AA. 

The resultant interference fringes, traditionally referred to as
``$S$-curves'' due to their characteristic shape, are reconstructed
from the photon counts reported by the photo-multiplier tubes (PMTs)
at each step of the scan using the FGS data reduction pipeline.  The
reconstructed individual scans are cross correlated to optimize mutual
alignment (eliminating spacecraft drift). Once aligned, the scans are
co-added and then smoothed using a piecewise spline fit to obtain the
optimal signal-to-noise ratio of the final $S$-curve. Any scan found
to be excessively noisy (from high spacecraft jitter) was deleted from
the process. We found that it was helpful to remove the low frequency,
slowly varying background of the scans because of their increasing
departure from zero difference with increasing separation from the
central fringe (caused by spatial sensitivity variations of the
photomultiplier tubes).  This was usually accomplished by subtracting
a parabolic fit of the outer fringe pattern, or wings, determined from
the mean of many calibrator (single) star scans.  However, in a few
cases the outer fringe variation was too large to be rectified this
way, and a spline fit was made of the fringe variation at
representative points along the $S$-curve.  The final $S$-curves for
all of the stars are presented in alphabetical order in Figure Set 1
(available in full in the electronic version of the paper).  The
$S$-curves appear in the middle panels for both the $x$- and $y$-axes
while the upper and lower panels summarize searches for resolved and
blended companions, respectively (see section 3).

\placefigure{figa23}   
\placefigure{figa27}   
\placefigure{figa41}   
\placefigure{figa46}   
\placefigure{figmt5}   
\placefigure{figmt59}  
\placefigure{figmt70}  
\placefigure{figmt83}  
\placefigure{figmt138} 
\placefigure{figmt145} 
\placefigure{figmt213} 
\placefigure{figmt217} 
\placefigure{figmt227} 
\placefigure{figmt250} 
\placefigure{figmt258} 
\placefigure{figmt259} 
\placefigure{figmt299} 
\placefigure{figmt304} 
\placefigure{figmt317} 
\placefigure{figmt339} 
\placefigure{figmt376} 
\placefigure{figmt390} 
\placefigure{figmt403} 
\placefigure{figmt417} 
\placefigure{figmt429} 
\placefigure{figmt431} 
\placefigure{figmt448} 
\placefigure{figmt455} 
\placefigure{figmt457} 
\placefigure{figmt462} 
\placefigure{figmt465} 
\placefigure{figmt470} 
\placefigure{figmt473} 
\placefigure{figmt480} 
\placefigure{figmt483} 
\placefigure{figmt485} 
\placefigure{figmt507} 
\placefigure{figmt516} 
\placefigure{figmt531} 
\placefigure{figmt534} 
\placefigure{figmt555} 
\placefigure{figmt556} 
\placefigure{figmt588} 
\placefigure{figmt601} 
\placefigure{figmt605} 
\placefigure{figmt611} 
\placefigure{figmt632} 
\placefigure{figmt642} 
\placefigure{figmt692} 
\placefigure{figmt696} 
\placefigure{figmt734} 
\placefigure{figmt736} 
\placefigure{figmt745} 
\placefigure{figmt771} 
\placefigure{figmt793} 
\placefigure{figs05}   
\placefigure{figs73}   
\placefigure{figwr145} 


\section{Companion Detection}\label{fgs-det} 
Binaries display an $S$-curve that appears as the sum of two offset
and rescaled fringe patterns (see section 3.2).  A binary detection is
made by comparing the $S$-curve of a target to that of a point-source
(single star).  We describe below the three detection methods we used
to determine binarity.  The comparison star, or calibrator, is taken
from unresolved sources within our sample that meet three criteria:
(1) observed using the same filter, 
(2) close in $B-V$ color, and
(3) close in observation date.
The first criterion is met by all the stars in our sample.  Criterion
2 is necessary because the appearance of a point source $S$-curve is
color dependent due to refractive optics within the FGS and the
wavelength dependent response of the PMTs.  For example, the shape of
an $S$-curve is slightly broadened for redder targets \citep{hor06}.
We chose calibrators whose colors are within $\pm 0.2$~mag from the
target $B-V$, except for a few very red stars (section~\ref{fgs-mod}).
The appearance of the $S$-curves can change with time due to the
settling of the instruments and adjustments from servicing missions.
Our observing program was not affected by any changes made due to a
servicing mission, but our large range of observation dates (2005
December to 2008 June) spans a range where the long term changes in
FGS1r, while small, are non-negligible.  Thus, we chose as the best
calibrator the one observed closest in time to the target.  The binary
systems we are able to resolve have components with nearly equal
brightness and hence, their colors should be comparable in most cases.
Consequently, we adopted the same calibrator to model both components.

\subsection{Visual Inspection}

The $S$-curve or transfer function of a binary consists of the
normalized superposition of the individual transfer functions of two
point sources.  For widely separated systems, the scan will show two
shifted transfer functions, whose relative amplitudes are directly
related to the magnitude difference (see eq.~\ref{eq3-fgs}). For
closer systems, the $S$-curve can look obviously different from that
of a single star (such as the case of MT~429 shown in Fig.~1.25),
because the composite fringe is formed from two overlapping
fringes. Thus, a direct comparison of the smoothed, co-aligned, target
$S$-curve with that of a single star provides the first way to
identify binaries.

Another indicator of a resolved binary is the relative fringe
amplitude, calculated as the value of $sppr$, or the $S$-curve
peak-to-peak ratio.  The ratio of the extrema points of the fringe to
that of a single star calibrator fringe can be expressed as
\begin{equation}\label{eq1-fgs}
 sppr = \frac{S_{\rm obs,max} - S_{\rm obs,min}}{S_{\rm F583W,max} -
   S_{\rm F583W,min}}
\end{equation}
where, for bright stars ($V < 12$), $S_{\rm F583W,max} = 0.61$ and
$S_{\rm F583W,min} = -0.55$ along the $x$-axis and $S_{\rm F583W,max}
= 0.30$ and $S_{\rm F583W,min} = -0.65$ along the $y$-axis for typical
single star observations made with the F583W filter.  This parameter
is listed in the middle panels of Figure Set 1 using denominator
values from the mean of the selected calibrator $S$-curves.  The cases
where $sppr \leq 0.9$ indicate that the target is a resolved binary in
which destructive interference causes a decline in the overall fringe
amplitude.  However, if $0.9 \leq sppr \leq 1.0$, as would be the case
for a binary with a small angular separation or large magnitude
difference of the components, a more thorough analysis is needed.
Note that the $x$- and $y$- axes have different values; this is due to
the effect of the HST spherical aberration and the alignment of the
interferometer optical elements relative to the HST optical axis. Also
note that the $S_{\rm F853W,max}$ and min values are adjusted for the
magnitude of the target since the $S$-curve is also affected by the
PMT dark current, which becomes an increasing larger percentage of the
total counts as fainter targets are observed. This effect is
negligible for $V < 12$, but is significant for fainter objects.
Note, the initial $sppr$ value used as a quick-look tool for a
companion was computed from the scans that have not been de-jittered,
cross correlated, or smoothed as they are for the more thorough
analysis described in the next sections. The $sppr$ values quoted in
Figure Set 1 are calculated after this analysis has been
performed, but do not differ significantly from the initial $sppr$
value used.

We produced an initial list of stars that appeared single according to
the visual inspection and the $sppr$ filter as possible calibrators
for the analysis described in the following sections. From there, we
narrowed the list through an iterative process of selecting which
stars met the following criteria for point sources, and we used those
as the calibrators for modeling purposes.

\subsection{Marginally Resolved Systems and Detection Limits}\label{fgs-close}

Close systems (projected separation $\lesssim 25$~mas), where the two
$S$-curves are not well separated, can slightly reduce the fringe
amplitude and widen the fringe shape. If the fringe pattern of a
single calibrator star is $S(x)$, then the observed fringe pattern for
more than one star will be
\begin{equation}\label{eq2-fgs}
S(x)_{obs} = \sum_{i=1}^n f_i ~S(x-x_i)
\end{equation}
where each of $n$ stars has a flux fraction $f_i = \frac{F_i}{\sum F_j}$
and a relative projected offset position $x_i$.  For a binary
star with a companion flux ratio $r=\frac{F_2}{F_1}$ and a projected
separation $\Delta x$, the observed pattern simplifies to
\begin{equation}\label{eq3-fgs}
S(x)_{obs} = \frac{1}{1+r} S(x) + \frac{r}{1+r} S(x-\Delta x).
\end{equation}
An analytical representation of the difference between the binary and
calibrator $S$-curves can be estimated by making a second-order
expansion for small offset $\epsilon$,
\begin{equation}\label{eq4-fgs}
S(x+\epsilon) = S(x) + \epsilon S'(x) + \frac{1}{2} \epsilon ^{2}S''(x)
\end{equation}
where $S'$ and $S''$ are the first and second derivatives of the
$S$-curve.  In the frame of reference where $S(0)=0$ for the binary,
the primary and secondary $S$-curves will be respectively shifted by
amounts $\epsilon_1 = -(\frac{r}{1+r}) \Delta x$ and $\epsilon_2 =
+(\frac{1}{1+r}) \Delta x$, where $\Delta x$ is the projected
separation of secondary from primary.  Then the difference between the
marginally resolved binary and calibrator $S$-curves is
\begin{equation}\label{eq5-fgs}
\begin{split}
S(x)_{bin} - S(x)_{cal} &
  = \frac{1}{1+r} S(x + \epsilon_1) + \frac{r}{1+r} S(x + \epsilon_2) - S(x) \\
  & = \frac{1}{2} \frac{r}{(1+r)^2} (\Delta x)^2 S''(x).
\end{split}
\end{equation}
This second-order expression has several important features.  First,
the observed difference in the core of the $S$-curve will appear to
have the same functional shape as the second derivative of the
$S$-curve, so we can directly search for companions with overlapping
fringes by looking for a difference that has a second derivative
shape.  Second, the amplitude of the difference depends on a product
involving both the separation $\Delta x$ and the flux ratio $r$, so in
the absence of other information, neither parameter can be determined
uniquely.  Third, the amplitude of the difference depends on the
separation squared, so no information can be reliably extracted on the
direction of the companion from the primary.

Figure~\ref{f2-fgs} shows examples of such $S$-curve differences for
model binaries.  The dashed line shows the difference for a model of
equally bright stars ($r = 1$) with a separation of $0\farcs015$ made
from a mean $S$-curve from a collection of calibrator $x$-axis scans.
According to the analytical expression above, the coefficient leading
the second derivative is $0.015^2 /8 = 2.8 \times
10^{-5}$~arcsec$^{2}$, and the solid line shows the product of this
coefficient and a numerical solution of the second derivative of the
calibrator $S$-curve (smoothed by convolution with a Gaussian of FWHM
= $0\farcs005$).  The good match between the detailed model and
analytical solution verifies the second derivative character of the
difference curve.  The same coefficient is found for $r=0.5$ and
$\Delta x = 0\farcs0159$, and the dotted-dashed line shows the
difference of the binary and calibrator curves for these binary
parameters.  Again, the agreement between this model and the
analytical curve shows that two models with the same product $a =
\frac{1}{2} \frac{r}{(1+r)^2} (\Delta x)^2$ have very similar
$S$-curves.

\placefigure{f2-fgs}

Therefore, in the case of marginally resolved systems the difference
between the $S$-curves of a suspected binary and a single star should
look like that of the second derivative of the point-source transfer
function scaled by the coefficient product term $a$.  Unless the flux
ratio is determined independently, there is not a unique solution for
$r$ and $\Delta x$.  The method was applied by considering the
difference between the target and calibrator $S$-curves over the range
within $\pm 100$~mas of the center of the fringe.  The coefficient was
then estimated by a least-squares fit of equation~\ref{eq5-fgs} over
the restricted range.  The coefficient $a$ was determined in practice
with an ensemble of like-color calibrators, and the criterion for
detection was set by a mean coefficient $a$ with a positive value
greater than $4\sigma$, where $\sigma$ is the standard deviation of
the coefficient derived using different calibrators.

\subsection{Wide Binary Detection}\label{fgs-wide}

The next situation to consider is the case where the absolute
projected separation of the binary companion is comparable to or
greater than the width of the fringe ($\approx 50$~mas or 70~AU for
stars at the distance of Cyg OB2) and the secondary may be faint. The
best approach in these cases is to calculate the cross-correlation
function (CCF) of a target $S$-curve with that of a calibrator
star. This method has the advantage of using more of the $S$-curve
than just the extrema points, and it potentially helps unravel those
cases where the fringes overlap.  The cost of this approach is a
slight decrease in the working angular resolution limit (but see above
for a discussion of binaries with blended $S$-curves).  The top panel
of Figure~\ref{f3-fgs} shows an example of a model $S$-curve of a
binary star with a projected separation of $+0\farcs07$ and a flux
ratio $r=\frac{F_{2}}{F_{1}} = 0.1$ (constructed using calibrator
$x$-axis scans).  The dotted line shows the calibrator $S$-curve while
the solid line shows that for the target binary.  The fringe patterns
overlap significantly at this separation, and the main differences are
a dilution of the main fringe pattern and a change in outer fringe
structure near $x=+0\farcs08$.  The lower panel shows the CCFs of the
target with the calibrator (solid line) and of the calibrator with
itself (dotted line).  Unlike the $S$-curves, the CCFs show one main
peak for each stellar component.  In order to isolate the companion,
the calibrator CCF is shifted to the peak of the target CCF and scaled
to the peak of the target CCF.  The shifted and scaled calibrator CCF
is then subtracted from the target CCF to produce the residual CCF
shown as a dashed line in an expanded scale in the lower panel (and
offset by $-0.8$ for clarity).  Now the peak from the companion is
clearly visible at the offset position of $\Delta x = +0\farcs07$.

\placefigure{f3-fgs}  

We need a working criterion to establish whether or not a peak in the
residual CCF makes a significant detection of a companion.  Because
the dominant source of uncertainty in the shape of the $S$-curves is
the inherent scatter between observations of the calibrators, the
criterion was set by running the CCF procedure for any given target
with an ensemble of calibrator $S$-curves for stars of similar color
(usually a set of ten calibrators). Then the detection criterion was
set by requiring the peak in the mean of the residual CCFs to exceed
$4\sigma (x)$, where $\sigma(x)$ is the standard deviation of the
residual CCFs at the peak position $x$.

Figure Set 1 shows transfer function plots for all 58 stars in our
sample. The figures show the $x$- (left) and $y$-axis (right)
rectified $S$-curves in the central panel. Also shown as a dashed line
is a preliminary model fit based upon the components derived from the
CCF analysis and the mean $S$-curve of the calibrator set
selected. The top panel plots the mean of the residual
cross-correlation functions of the target with the calibrator (solid,
black lines) after the peak of the primary has been subtracted
off. The vertical dashed lines indicate the position of each component
resolved. The solid, gray lines show the standard deviation of the
residual cross-correlation functions, and a peak is considered
significant only if the mean residual CCF (black) exceeds the standard
deviation (gray) by four times. The bottom panels show the difference
curves between the star and model calibrator $S$-curves (solid
line). The shaded, gray region is the uncertainty envelope determined
from the standard deviation at each point along the $S$-curves for the
calibrators. In the case where the second derivative test resolved a
blended pair (e.g. MT~304, Fig. \ref{figmt304}), the second derivative
of the calibrator $S$-curve is overplotted and scaled by the $a$
coefficient (solid, gray line).

The detection threshold for a binary of a given separation is
determined by comparing multiple examples of a model binary based upon
the calibrator observations to the calibrator $S$-curves.  This was
done using a set of 21 calibrator observations of similar color stars
for both the model and calibrator curves (for a total of 420 test
cases), and the faintest flux ratio was set by models that met the
$4\sigma$ detection criterion.  The positive (solid line) and negative
(dashed line) branches are folded onto one separation axis in
Figure~\ref{f4-fgs} in the left and right panels for the $x$- and
$y$-axis scans, respectively.  Here the limiting flux ratio, $r$, is
shown as a magnitude difference $\Delta m = -2.5 \log r$ for trial
separations, and any binary brighter than the limit (i.e., below the
line plots) would exceed the $4\sigma$ detection criterion.  The
smallest separation detected in favorable, equal flux cases is
slightly better for positive separations, and the faintest detectable
companions have $\triangle m = 4.5$~mag at large separations.  The
oscillation in these curves seen near $x=y=0\farcs06$ is due to the
changing and relatively larger uncertainties in the calibrator
$S$-curves at such distances from their zero-crossing.  The results in
these figures compare well with the advertised limits in Figure 3.3 of
the {\it Fine Guidance Sensor Instrument Handbook} \citep{nel12}
except for the case of marginally resolved binaries that we discussed
above. 

Note that the CCF method loses its effectiveness for very close
companions. The peaks in the residual CCFs in the binary models where
$ \rho \lesssim 25$~mas are rarely found at separations of less than
$0\farcs04$ even in favorable cases.  This is due to the fact that
blending of the fringe patterns becomes so severe that the calibrator
CCF is positioned at the maximum that occurs between the actual
positions of the components, and consequently the residual CCF shows
two peaks: one for the companion and a mirror one for the primary.  At
the smallest separations where the method can be applied ($0\farcs016$
in $x$ and $0\farcs017$ in $y$), the two peaks in the residual CCF
approach equal intensity, and the method can no longer distinguish the
direction of primary to secondary.  Nevertheless, the appearance of a
double-peaked feature in the residual CCF offers evidence for the
presence of a marginally resolved companion that supplements the
second derivative test.

\placefigure{f4-fgs}  

The detection limits using the second derivative method were estimated
by running the scheme with multiple binary models from a large sample
of calibrators.  The working criteria from these models for $4\sigma$
detection are $a > 1.2 \times 10^{-5}$ and $1.6 \times 10^{-5}$
arcsec$^{2}$ for the $x$- and $y$-axes, respectively.  These limits
are shown in Figure~\ref{f4-fgs} as dotted lines that trace the
upper envelope for detection by the second derivative approach.  In
the best cases ($r=1$), the estimates suggest that binaries as close
as $0\farcs010$ can be detected with FGS TRANS mode scans.  The second
derivative and CCF methods are probably both sensitive to binary
detection in the $0\farcs020$ -- $0\farcs025$ (28 -- 100 AU) range.

\subsection{Off-scan Components}

There is also the case of very widely separated binaries, where the
light of the companion falls within the instrumental FOV
($5^{\prime\prime} \times 5^{\prime\prime}$), but the projected
separation is greater than the length of the scan. This is the case
for our observation of MT~531 where the system is resolved along the
$y$-axis, but the secondary is off the scan along the $x$-axis (see
Fig.~\ref{figmt531}). The CCF method will fail to detect the companion
because its peak lies beyond the recorded scan.  However, such
binaries will still cause the $S$-curve of the primary to appear with
an amplitude reduced by a factor $\frac{1}{(1+r)}$, and hence the
ratio of target to calibrator $S$-curve amplitude ($sppr$) provides an
additional criterion to check for very wide binaries.  In practice,
the scatter between calibrator $S$-curves of similar color indicates
that a $4\sigma$ detection may be claimed if the target $S$-curve
amplitude is less than $92\%$ of the mean calibrator $S$-curve
amplitude.  Recall that overlapping fringes due to a binary may also
cause the amplitude to decline, but in this case, the fringe will also
be widened.  Consequently, one can differentiate between the off scan
and blended cases by the appearance of the difference curve.  The
dotted line in Figure~\ref{f2-fgs} shows that the difference curve for
a binary with an off-scan companion appears like a negatively scaled
version of the $S$-curve itself, which looks very different from the
second derivative curve.  Thus, it is important to inspect the shapes
of the difference curves in order to decide if a positive detection
indicates the prescence of an off-scan companion or a very close
companion.

We were able to make a second check for wide companions through
inspection of results from our adaptive optics (AO) survey using the
Near InfraRed Imager and Spectrograph (NIRI) at the Gemini North
Observatory \citep{cab12}. The NIRI observations have a larger dynamic
range (i.e., can detect fainter companions) and have a larger field of
view, so any system observed with FGS with $\rho \gtrsim 50$~mas will
also be detected in the NIRI image.  The infrared images allow us to
identify those companions with projected separations longer than the
scan length, but still within the FOV of the FGS.  With the help of
the NIRI observations, we were able to conclude that distant
companions influence the FGS results for MT~59, MT~138 and MT~531.


\section{Model Fitting}\label{fgs-mod} 

If a binary was detected, then we fit the target $S$-curve with a
model binary formed from a calibrator $S$-curve in order to determine
the relative brightness and projected separation along an axis.  We
used two routines to calculate the best fit model. The first is
BINARY\_FIT which is part of the STScI reduction package based upon
algorithms developed by Space Telescope Astrometry Team at Lowell
Observatory \citep{fra91}. In addition, we developed the Interactive
Data Language (IDL) routine TRIPFIT for the special cases where
BINARY\_FIT was not able to converge to a satisfactory solution.  Both
BINARY\_FIT and TRIPFIT use the $x$ and $y$ projected separations and
the telescope roll angle to determine the binary separation $\rho$ and
position angle $\theta$ (measured east from north).

Each binary was compared to models based upon four calibrators.  The
calibrators were selected to have a $B-V$ color within $\pm 0.2$~mag
of the target's color.  The only exception was MT~304, with a $B-V =
3.35$, which is the reddest star in our sample by more than 1~mag. In
this case we selected calibrators from the reddest, single star in our
sample, MT~448, and the stars from \citet{com02} (A~23, A~27, A~41,
A~46), which are red objects according to their brightness in the
infrared. We adopted the fit made using the calibrator closest in
color and observation date to determine the system parameters. The
spread in results from fits made using the other calibrators was used
to determine the fitting parameter uncertainties.  The same calibrator
was used to model both the primary and secondary $S$-curves, i.e., we
assumed that any color difference between the components is
negligible.  Our calibrator selections are indicated in column 9 of
Table 1.

The program BINARY\_FIT uses a least squares approach to determine the
projected separation and magnitude difference of the system
\citep{nel11b}.  BINARY\_FIT fits scans from one axis at a time,
starting from initial estimates for the projected separation along the
axis and the differential magnitude.  If the results for differential
magnitude from the $x$- and $y$-axis solutions agreed within 0.2 mag,
then the individual separation results were adopted as fit, but we
report the magnitude difference for the axis solution with the larger
separation (usually more reliable).  On the other hand, if the
differential magnitudes differed by more than 0.2 mag, then we adopted
the value from the axis solution with the greater separation, and then
re-fit the scan for the other axis by setting the adopted magnitude
difference.

We encountered several cases where BINARY\_FIT could not be used.  (1)
The fitting code is limited by the scan length of the calibrator. If
the separation of the binary is larger than the scan length of the
calibrator, the program is not able to recreate a binary wide enough
to model properly the target.  (2) BINARY\_FIT only considers
solutions where both components are within the scan length. For
example, if the companion is recorded on one axis but lies off the
scan of the other, then the program will converge to a solution of a
very close system for the axis where the star is absent. (3)
BINARY\_FIT cannot be applied to systems with more than two
components, such as the triple MT~417.

We coded the IDL program TRIPFIT to find
the best fit model for triple systems using a Levenberg-Marquardt
least-squares method. The program is capable of making fits of triple
systems, binaries, and off-scan components. The user selects initial
estimates of the positions of the components and the differential
magnitudes. The program then fits the $S$-curves one axis at a time
and returns the best fit for the differential magnitudes and
separations. Both BINARY\_FIT and TRIPFIT return similar solutions
when modeling a system. However, for most cases we adopted the results
from the BINARY\_FIT models because it uses the information from both
axis simultaneously in determining the best fit. The cases where we
could not converge on a satisfactory solution with BINARY\_FIT, we
report the results using TRIPFIT and those systems are noted in
section \ref{fgs-results}.

There are three sources of uncertainty in the analysis of FGS data.
(1) Internal errors arise from photometric shot noise, which is
important for stars with $V > 14$, and spacecraft jitter that cannot
be removed using the guide star centroids reported by the guiding
FGSs.  (2) The $S$-curves slowly evolve over time, which is due to
small changes in the alignment of the Koesters prisms relative to the
optical axis of HST, made sensitive due to the spherical aberration of
the HST primary mirror. By choosing calibrator stars that are observed
close in time (on the order of 1 year) to the science observations
this evolutionary effect is mitigated. (3) Systematic differences
between the calibrators exist.  To investigate how the photometric
noise and jitter influence the derive parameters of a binary, we
selected four binary systems (MT~5, MT~429, MT~605, and MT~632) and
binned their scans into three or four independent subsets. Each subset
includes approximately five scans that were shifted, co-added and
smoothed using the same approach applied to the complete set of
scans. The resultant subset of co-aligned scans were then fit with the
best calibrator using BINARY\_FIT. The small spread in the values of
separation and magnitude shows that the internal error is not a
significant source of uncertainty for the binary parameters (except in
the case of MT~5; see section \ref{fgs-results}).  Consequently, the
binary parameter uncertainties are dominated by the differences
between the calibrator scans, and in Table~\ref{t2-fgs} we report
uncertainties based upon the standard deviation of the parameter fits
made with the different calibrator $S$-curves.


\section{Multiplicity Results}\label{fgs-results}  

Table~\ref{t2-fgs} lists the model fitting results for separation and
differential magnitude for the 13 resolved systems in our Cyg~OB2
sample.  The cited errors are from a comparison of results from fits
made with different calibrators as described in
section~\ref{fgs-mod}. There were two additional cases where only the
second derivative analysis indicated a possible close binary system
that was partially resolved on only one axis.  The derived
$a$-coefficient was $a=(15.2 \pm 3.4)\times 10^{-6}$ arcsec$^{2}$
along the $y$-axis for MT~227 and $a=(25.5 \pm 3.6)\times 10^{-6}$
arcsec$^{2}$ along the $x$-axis for MT~317.  These two marginally
resolved systems are objects of interest for follow-up analysis.  We
describe below the individual cases for the fully resolved systems.

\placetable{t2-fgs}   

\noindent{\sl A~41.} The CCF analysis of A~41 (see Fig.
\ref{figa41}) reveals the presence of a faint companion in the $y$-axis scan.
This companion was also found in the $K$-band NIRI results at a
separation of $\rho = 0\farcs35$, essentially the same as the $y$-axis
projected separation.  This implies that the companion's projected
separation along the $x$-axis is within the fringe of the
primary. This, along with its relative faintness ($\delta V \approx
3.3$), make detection along the $x$-axis challenging.  Nevertheless,
we fit the $S$-curves of both axes with BINARY\_FIT to obtain a
separation and position angle consistent with the NIRI results.

\noindent{\sl MT~5.} (see Fig. \ref{figmt5}) After splitting the
$S$-curve data into subsets, the internal error was found to be a
non-negligible source of uncertainty in this case. The uncertainties
listed for MT~5 in Table~\ref{t2-fgs} reflect both the spread among
the calibrators and between the subsets of MT~5 (added in quadrature).
Note that the second derivative test for the $x$-axis ($a =7.9\sigma$)
indicates that the primary itself may have a close companion making
this a triple system, but the test is consistent with a single primary
for the $y$-axis scan.

\noindent{\sl MT~59.}  TRIPFIT was used to model MT~59 (see Fig.
\ref{figmt59}) because the binary is too widely separated in the
$x$-axis for BINARY\_FIT and because the secondary is positioned
beyond the recorded scan in the $y$-axis.  In the NIRI adaptive optics
image, MT~59 has a companion at $\rho = 1\farcs20$ with $\Delta K =
2.75$. This total separation is related to the projected separations
by $\rho^{2} = \Delta x^{2} + \Delta y^{2}$, so we expect that
secondary would have a projected separation along the $y$-axis of
$\Delta y = - 0\farcs91$, which is beyond the scan limits.

\noindent{\sl MT~138.}  We determined that MT~138 (see Fig.
\ref{figmt138}) is resolved along the $y$-axis with $\Delta y =
+0\farcs066$ and $\Delta V = 2.79$~mag.  This magnitude difference is
consistent with that from a close companion in the NIRI image at $\rho
= 1\farcs5$ and $\Delta K = 3.18$~mag.  According to the FGS aperture
position angle (or telescope roll angle), the NIRI component is
expected to appear at $\Delta y = +0\farcs070$ and $\Delta x =
+1\farcs338$, which agrees with the BINARY\_FIT result along the
$y$-axis and puts the companion well off the $x$-axis scan.

\noindent{\sl MT~304.} This target (Schulte~12 = Cyg OB2-12; see
Fig. \ref{figmt304}) shows evidence of a very close companion in the
CCF and second derivative tests, with values of $a = 8.81\sigma$ and
$a = 4.68\sigma$ for the $x$- and $y$-axes respectively.  The
differential magnitude from the BINARY\_FIT analysis of the $x$-axis
scan was used as a constraint in the fit of the $y$-axis scan, to
arrive at $\Delta V = 2.31 \pm 0.21$.  For close systems, where the
projected separation is less than the size of the fringe ($\rho < 15$
mas), there is an ambiguity in the ``parity'' of the secondary star's
position, i.e., solutions with the secondary to the left or right of
the primary are indistinguishable from one another. With a separation
along the $x$-axis of $61.8 \pm 3.1$ mas, and $y$-axis solutions
yielding 9.7 mas or -14.9 mas, the position angle is $283.5$~or~$305.9
\pm 3.3$ degrees, respectively.  We were not able to detect the
counterpart in the NIRI image, because the separation ($\rho =
63.6$~mas) is below the limiting resolution for NIRI ($\approx
80$~mas), but this companion has been observed using other
interferometric techniques (R. Millan-Gabet, private communication).
MT~304 is an early B-type hypergiant of very high luminosity, but it
may not belong to the class of Luminous Blue Variables because its
flux and spectral appearance are relatively constant \citep{cla12}.
The companion we find is too faint to alter the conclusion about the
star's high luminosity.  If we assume that the projected separation
corresponds to the apastron separation in a highly elliptical orbit,
then the orbital period is $P\approx 30$~yr, for $M_1 + M_2 = 120
M_\odot$ \citep{cla12} and $d=1.4$ kpc.  Consequently, additional high
angular resolution observations over the next few decades may lead to
a mass measurement of this extraordinary star.

\noindent{\sl MT~417.} This star (Schulte 22; see Fig.
\ref{figmt417}) is the only triple system resolved in our sample, and
we developed TRIPFIT to model the scans.  Visual inspection of the
$y$-axis $S$-curve (central, right panel of Fig.~\ref{figmt417}) shows
that there is a very faint third component near $\Delta y =
+1\farcs7$. Comparison with the NIRI images led us to conclude that
the third component fringe is blended with that of the secondary in
the $x$-axis scan.  The solution for differential magnitudes for the
secondary from the two axes did not agree with each other ($\Delta m_x
= 0.85$ and $\Delta m_y = 0.45$). This is due to the fact the second
and third component are blended in the $x$-axis scan, making the
amplitude of the secondary's fringe appear smaller. This is not the
case for the $y$-axis, where the two components are well
separated. The value listed in Table~\ref{t2-fgs} is from the
$y$-axis, which has the larger projected separation, and the error is
estimated from the standard deviation between the fits of the
different calibrators.  MT 417 A and B have classifications of
O3~If$^\star$ and O6~V((f)), respectively \citep{sot11}. Our
measurements of the Ba,Bb pair ($\rho = 0\farcs196 $, $\theta =
160^{\circ}$, and $\Delta m = 2.48$~mag) agree with previous
works. The A,B pair was resolved with speckle interferometry, and on
re-examining the 2007 speckle data using the methodology of
\citet{mas09}, the Ba,Bb pair were observed with $\rho = 0\farcs209$,
$\theta = 180\fdg0$. This pair was also resolved through AstraLux and
HST ACS/HRC imaging by \citet{mai10}. His positions and magnitude
differences for the Ba,Bb pair ($\rho = 0\farcs216$, $\theta =
181\fdg48$, and $\Delta m = 2.34$~mag) agree with the speckle data and
our results, within the uncertainties due to the blending of the Ba,Bb
components along one axis and neglect of the spatial sensitivity of
the photomultiplier tubes.

\noindent{\sl MT~429.} The primary of MT 429 (see Fig.
\ref{figmt429}) is a short-period eclipsing binary, and the presence
of the newly discovered bright third companion ($\Delta V = 1.09 \pm
0.02$) has a strong influence on the interpretation of the radial
velocity and photometric variations \citep{kim12a}.

\noindent{\sl MT~516.} The pair of MT~516 (see Fig. \ref{figmt516})
was first resolved through speckle interferometry by \citet{mas09} who
designate the binary as WSI~67.  Our position and magnitude
differences agree within the uncertainties.

\noindent{\sl MT~531.} MT~531 (see Fig. \ref{figmt531}) is an
obvious binary in $y$ and the second component is off the scan in $x$.
The NIRI image shows the infrared counterpart with $\rho= 1\farcs45$
and $\Delta K =0.65$ mag.  The NIRI position predicts that the
companion has projected separation along the $x$-axis $\Delta x =
-1\farcs38$, well beyond the recorded scan.

\noindent{\sl MT~605.} The primary of MT~605 (see Fig. \ref{figmt605})
is a double-lined spectroscopic binary \citep{kim12a}, and the
unknown, relatively bright third companion may cause line blending
difficulties for radial velocity measurements.

\noindent{\sl MT~632.} This system was an obvious binary in both axes
(see Fig. \ref{figmt632}). We resolved the secondary in the NIRI
observation and the FGS results are consistent with those from the AO
measurements.

\noindent{\sl MT~696.} A companion to this star (Schulte~27; Fig.
\ref{figmt696}) was detected using the second derivative test ($a =
8.46\sigma$) and the cross-correlation method for the $y$-axis scan.
(and was detected at the $3\sigma$ level in the second derivative test
of the $x$-axis scan).  We adopted the differential magnitude from the
$y$-axis solution, because of the greater projected separation along
the $y$-axis.  There exists a 180$^{\circ}$ ambiguity in the position
of the secondary along the $x$-axis ($\Delta x = \pm 9.7$~mas).  The
ambiguity is reflected in the two solutions for the position angle
given in Table~\ref{t2-fgs}. The primary is an eclipsing, double-lined
spectroscopic binary \citep{kim12a}, and the presence of the FGS
companion will influence the interpretation of the light curve and
spectroscopy.

\noindent{\sl SCHULTE~5.} This target (see Fig. \ref{figs05}) is an
obvious binary in the scans of both axes.  The separation is too large
along the $x$-axis for a BINARY\_FIT solution, so we applied TRIPFIT
to this system. The program arrived at similar results for the
differential magnitude derived from the $x$ and $y$ scans, and we list
the average in Table~\ref{t2-fgs}. The uncertainties were determined
from the standard deviation between the differential magnitude from
both axes.  This system was first resolved by \citet{her67}. The
primary is a short period ($P = 6.6$~d) eclipsing binary consisting of
two luminous evolved stars \citep{lin09}.  \citet{ken10} suggest that
there is another star close to A with an orbital period of $P=6.7$~y
based upon the variable radio emission. \citet{mas09} did not resolve
the B component, probably due to the magnitude difference, but our
results are consistent with recent observations by \citet{mai10}.


\section{Discussion}\label{fgs-conc}     

Our high angular resolution survey of 58 massive stars in Cyg~OB2 led
to the detection of 13 resolved systems and the partial resolution of
two other stars.  The resulting binary fraction of 22\% to 26\% is
consistent with the results from \citet{nel04}, who resolved 5 out of
23 OB stars (22\%) in the Carina Nebula cluster.  We were able to find
the infrared counterpart in our NIRI observations for 11 of the
resolved systems, but not for the marginally resolved pairs MT~304 and
MT~696.  Additional observations with ground-based interferometry
confirm the companion to MT~304.  Two of the most massive stars in the
association (MT~304 = Cyg~OB2-12 and MT~417 = Cyg~OB2-22A) have
resolved companions while the third very massive star (MT~457 =
Cyg~OB2-7) appears single.  The companion of the hypergiant MT~304
(Cyg~OB2-12) may have an orbital period of a few decades, and
continued high angular resolution observations should reveal the
companion's orbital motion.  This is potentially a very important
target for mass determination (like HD~93129A; \citealt{nel04}).

The high angular resolution capabilities of the FGS allow us to start
filling in the observational gap in the period distribution of massive
binaries \citep{mas98}.  The angular separations of the resolved
binaries correspond to binary orbital periods in the range of $20 < P
< 20,000$~yrs. The longest period spectroscopic system in Cyg OB2
observed by \citet{kob12} is just over 6 yr, only a factor of 3
smaller than our lower limit. This is probably the smallest gap to
date between spectroscopic and high angular resolution methods.
Though most of our resolved companions have long periods, their
projected separations are less than 10,000 AU.  This corresponds to an
orbital velocity that is larger than the velocity dispersion of the
association ($\sigma = 8.03$ km s$^{-1}$; \citealt{kim08e}), so these
systems are probably orbitally bound companions.

The companions we detected are relatively bright, and hence, it is
important to account for their flux in analyzing the spectra of the
primary stars.  Fainter companions are undoubtedly present in this
period range, and our complementary adaptive optics study with NIRI
\citep{cab14} will help to determine the mass ratio distribution of
lower mass and fainter companions (at least among the long period
binaries).  The systems resolved in our sample are at the limits of
what is angularly resolvable with a single aperture telescope
today. Preliminary results from the larger AO sample allows us to
perform a statistical study of chance alignments, which suggests that
the systems resolved with the FGS have $< 1$\% probability of being a
chance alignment. In a future work we will combine the results of the
spectroscopic and high angular resolution surveys of Cyg~OB2 and
provide an unprecedented census of the binary properties of massive
stars over a large range in orbital period.


\acknowledgments

We thank the staff of the Space Telescope Science Institute (STScI)
for their support in obtaining these observations.  We are also
grateful to Dr.\ John P.\ Subasavage who helped implement the FGS
reduction pipeline at GSU.  Support for {\it HST} proposal numbers
GO-9646, 9840, and 10612 was provided by NASA through a grant from the
Space Telescope Science Institute, which is operated by the
Association of Universities for Research in Astronomy, Incorporated,
under NASA contract NAS5-26555.  This work was also supported by the
National Science Foundation under grants AST-0606861 and AST-1009080.
A.~F.~J. Moffat is grateful for financial aid from the Natural
Sciences and Engineering Research Council of Canada
(NSERC). Institutional support has been provided from the GSU College
of Arts and Sciences and from the Research Program Enhancement fund of
the Board of Regents of the University System of Georgia, administered
through the GSU Office of the Vice President for Research and Economic
Development.  We are grateful for all this support.

{\it Facilities:} \facility{HST (FGS)} 



\bibliographystyle{apj}
\bibliography{apj-jour,paper}


\clearpage


\begin{deluxetable}{lcccccccccl}
\tablewidth{0pc}
\tabletypesize{\scriptsize}
\rotate
\tablenum{1}
\tablecaption{Cygnus OB2 Target List\label{t1-fgs}}
\tablewidth{0pt}
\tablehead{
\colhead{Star} &
\colhead{R.A.} &
\colhead{Dec.} &
\colhead{Spectral} &
\colhead{Class.} &
\colhead{$V$} &
\colhead{$B-V$} &
\colhead{Obs. Date} &
\colhead{Calibrator} &
\colhead{No.} &
\colhead{Remarks} \\
\colhead{Name} &
\colhead{(J2000)} &
\colhead{(J2000)} &
\colhead{Classification} &
\colhead{Ref.} &
\colhead{(mag)} &
\colhead{(mag)} &
\colhead{(BY)} &
\colhead{Name} &
\colhead{Comp.} &
\colhead{} \\
\colhead{(1)} &
\colhead{(2)} &
\colhead{(3)} &
\colhead{(4)} &
\colhead{(5)} &
\colhead{(6)} & 
\colhead{(7)} &
\colhead{(8)} &
\colhead{(9)} &
\colhead{(10)} &
\colhead{(11)} }

\startdata

A 23      & 20:30:39.71 & +41:08:49.0 & B0.7 Ib       & 1 & 11.25 & \nodata& 2007.2729 & \nodata & 1 & \nodata           \\
A 27      & 20:34:44.72 & +40:51:46.6 & B0 Ia         & 2 & 11.26 & \nodata& 2007.4675 & \nodata & 1 & \nodata           \\
A 41      & 20:31:08.38 & +42:02:42.3 & O9.7 II       & 1 & 11.70 & \nodata& 2006.0666 & \nodata & 2 & NIRI              \\
A 46      & 20:31:00.20 & +40:49:49.7 & O7 V((f))     & 1 & 11.40 & \nodata& 2008.4988 & \nodata & 1 & \nodata           \\
MT 5      & 20:30:39.82 & +41:36:50.7 & O6 V          & 3 & 12.93 & 1.64   & 2006.5055 & WR 145  & 2 & NIRI              \\
MT 59     & 20:31:10.55 & +41:31:53.5 & O8 V          & 4 & 11.18 & 1.47   & 2006.4855 & MT 601  & 2 & NIRI, SB1         \\
MT 70     & 20:31:18.33 & +41:21:21.7 & O9 II         & 4 & 12.99 & 2.10   & 2006.4928 & \nodata & 1 & SB1               \\
MT 83     & 20:31:22.04 & +41:31:28.4 & B1 I          & 3 & 10.64 & 1.18   & 2006.0001 & \nodata & 1 & SB1               \\
MT 138    & 20:31:45.40 & +41:18:26.8 & O8 I          & 3 & 12.26 & 1.99   & 2006.4854 & MT 390  & 2 & NIRI, SB1         \\
MT 145    & 20:31:49.66 & +41:28:26.5 & O9 III        & 4 & 11.52 & 1.11   & 2005.9990 & \nodata & 1 & SB1               \\
MT 213    & 20:32:13.13 & +41:27:24.6 & B0 V          & 3 & 11.95 & 1.13   & 2007.2387 & \nodata & 1 & RV constant       \\
MT 217    & 20:32:13.83 & +41:27:12.0 & O7 IIIf       & 3 & 10.23 & 1.19   & 2006.3752 & \nodata & 1 & RV constant       \\
MT 227    & 20:32:16.56 & +41:25:35.7 & O9 V          & 3 & 11.47 & 1.24   & 2006.4883 & \nodata & 1 & \nodata           \\
MT 250    & 20:32:26.08 & +41:29:39.4 & B2 III        & 3 & 12.88 & 1.06   & 2007.2942 & \nodata & 1 & RV constant       \\
MT 258    & 20:32:27.66 & +41:26:22.1 & O8 V          & 4 & 11.10 & 1.20   & 2006.4986 & \nodata & 1 & SB1               \\
MT 259    & 20:32:27.74 & +41:28:52.3 & B0 Ib         & 3 & 11.42 & 1.00   & 2006.3778 & \nodata & 1 & SB1               \\
MT 299    & 20:32:38.58 & +41:25:13.8 & O7 V          & 3 & 10.84 & 1.19   & 2006.4773 & \nodata & 1 & SB1?              \\
MT 304    & 20:32:40.96 & +41:14:29.2 & B3-4 Ia$^{+}$ & 5 & 11.10 & 3.35   & 2006.2346 & MT 448  & 2 & \nodata           \\
MT 317    & 20:32:45.46 & +41:25:37.4 & O8 V          & 3 & 10.68 & 1.25   & 2006.4959 & \nodata & 1 & RV constant       \\
MT 339    & 20:32:50.02 & +41:23:44.7 & O8 V          & 3 & 11.60 & 1.35   & 2006.5004 & \nodata & 1 & RV constant       \\
MT 376    & 20:32:59.19 & +41:24:25.5 & O8 V          & 3 & 11.91 & 1.35   & 2006.4963 & \nodata & 1 & RV constant       \\
MT 390    & 20:33:02.92 & +41:17:43.1 & O8 V          & 3 & 12.95 & 1.98   & 2006.4844 & \nodata & 1 & RV constant       \\
MT 403    & 20:33:05.27 & +41:43:36.8 & B1 V          & 3 & 12.94 & 1.49   & 2007.4728 & \nodata & 1 & \nodata           \\
MT 417    & 20:33:08.80 & +41:13:18.2 & O3~If$^\star$  & 6 & 11.55 & 2.04   & 2006.3776 & MT 771  & 3 & NIRI, SB1         \\
MT 429    & 20:33:10.51 & +41:22:22.5 & B0 V          & 4 & 12.98 & 1.56   & 2007.2895 & MT 793  & 2 & NIRI, SB1/EA      \\
MT 431    & 20:33:10.75 & +41:15:08.2 & O5:           & 4 & 10.96 & 1.81   & 2006.3065 & \nodata & 1 & SB2               \\
MT 448    & 20:33:13.26 & +41:13:28.7 & O6 V          & 3 & 13.61 & 2.15   & 2006.4894 & \nodata & 1 & SB1               \\
MT 455    & 20:33:13.69 & +41:13:05.8 & O8 V          & 3 & 12.92 & 1.81   & 2006.4879 & \nodata & 1 & \nodata           \\
MT 457    & 20:33:14.11 & +41:20:21.8 & O3~If$^\star$ & 3 & 10.55 & 1.45   & 2006.3397 & \nodata & 1 & RV constant       \\
MT 462    & 20:33:14.76 & +41:18:41.6 & O7 III-II     & 3 & 10.33 & 1.44   & 2006.5006 & \nodata & 1 & RV constant       \\
MT 465    & 20:33:15.08 & +41:18:50.5 & O5.5 I        & 4 &\phn9.06 & 1.30 & 2006.4567 & \nodata & 1 & SB2               \\
MT 470    & 20:33:15.71 & +41:20:17.2 & O9 V          & 3 & 12.50 & 1.46   & 2006.4930 & \nodata & 1 & RV constant       \\
MT 473    & 20:33:16.34 & +41:19:01.8 & O8.5 V        & 3 & 12.02 & 1.45   & 2006.4936 & \nodata & 1 & RV constant       \\
MT 480    & 20:33:17.48 & +41:17:09.3 & O7 V          & 3 & 11.88 & 1.59   & 2006.4903 & \nodata & 1 & RV constant       \\
MT 483    & 20:33:17.99 & +41:18:31.1 & O5 III        & 3 & 10.19 & 1.24   & 2006.3069 & \nodata & 1 & SB1?              \\
MT 485    & 20:33:18.03 & +41:21:36.6 & O8 V          & 3 & 12.06 & 1.51   & 2006.4881 & \nodata & 1 & SB1?              \\
MT 507    & 20:33:21.02 & +41:17:40.1 & O9 V          & 3 & 12.70 & 1.54   & 2006.4934 & \nodata & 1 & RV constant       \\
MT 516    & 20:33:23.46 & +41:09:13.0 & O5.5 V        & 3 & 11.84 & 2.20   & 2006.3067 & MT 448  & 2 & NIRI, RV constant \\
MT 531    & 20:33:25.56 & +41:33:27.0 & O8.5 V        & 3 & 11.58 & 1.57   & 2006.4961 & MT 480  & 2 & NIRI, RV constant \\
MT 534    & 20:33:26.75 & +41:10:59.5 & O8.5 V        & 3 & 13.00 & 1.87   & 2006.4892 & \nodata & 1 & RV constant       \\
MT 555    & 20:33:30.31 & +41:35:57.9 & O8 V          & 3 & 12.51 & 1.90   & 2007.2411 & \nodata & 1 & SB1               \\
MT 556    & 20:33:30.79 & +41:15:22.7 & B1 I          & 3 & 11.01 & 1.77   & 2006.4937 & \nodata & 1 & SB1               \\
MT 588    & 20:33:37.00 & +41:16:11.3 & B0 V          & 3 & 12.40 & 1.66   & 2007.2413 & \nodata & 1 & RV constant       \\
MT 601    & 20:33:39.11 & +41:19:25.9 & B0 Iab        & 3 & 11.07 & 1.47   & 2006.2571 & \nodata & 1 & SB1               \\
MT 605    & 20:33:39.80 & +41:22:52.4 & B1 V          & 4 & 11.78 & 1.19   & 2007.2469 & MT 217  & 2 & NIRI, SB2         \\
MT 611    & 20:33:40.87 & +41:30:19.0 & O7 V          & 3 & 12.77 & 1.55   & 2006.0664 & \nodata & 1 & RV constant       \\
MT 632    & 20:33:46.10 & +41:33:01.1 & O9 I          & 3 &\phn9.88 & 1.59 & 2006.4770 & MT 480  & 2 & NIRI              \\
MT 642    & 20:33:47.84 & +41:20:41.5 & B1 III        & 3 & 11.78 & 1.55   & 2007.2442 & \nodata & 1 & SB1               \\
MT 692    & 20:33:59.25 & +41:05:38.1 & B0 V          & 3 & 13.61 & 1.69   & 2007.4677 & \nodata & 1 & SB2?              \\
MT 696    & 20:33:59.53 & +41:17:35.5 & O9.5 V        & 4 & 12.32 & 1.65   & 2007.2415 & MT~588  & 2 & SB2/EW/KE         \\
MT 734    & 20:34:08.50 & +41:36:59.2 & O5 I          & 4 & 10.03 & 1.49   & 2006.4180 & \nodata & 1 & SB1               \\
MT 736    & 20:34:09.52 & +41:34:13.7 & O9 V          & 3 & 12.79 & 1.46   & 2007.2532 & \nodata & 1 & RV constant       \\
MT 745    & 20:34:13.51 & +41:35:02.7 & O7 V          & 3 & 11.91 & 1.50   & 2005.9724 & \nodata & 1 & SB1?              \\
MT 771    & 20:34:29.60 & +41:31:45.5 & O7 V          & 4 & 12.06 & 2.05   & 2007.2507 & \nodata & 1 & SB2               \\
MT 793    & 20:34:43.58 & +41:29:04.6 & B2 IIIe       & 3 & 12.29 & 1.54   & 2007.2505 & \nodata & 1 & \nodata           \\
SCHULTE 5 & 20:32:22.43 & +41:18:19.1 & O7 Ianfp      & 4 &\phn9.12 & 1.67 & 2006.5002 & MT 588  & 2 & NIRI, SB2/EB      \\
SCHULTE 73& 20:34:21.93 & +41:17:01.6 & O8 III        & 4 & 12.40 & 1.73   & 2007.2497 & \nodata & 1 & SB2               \\
WR 145    & 20:32:06.29 & +40:48:29.6 & WN7o/CE       & 7 & 11.83 & 1.63   & 2006.4932 & \nodata & 1 & SB1
\enddata                                              
\tablerefs{ Column 5:
1. \citet{han03};
2. \citet{neg08};
3. \citet{kim07};
4. \citet{kob12};
5. \citet{cla12};
6. \citet{sot11};
7. \citet{mun09}.}
\end{deluxetable}


\begin{deluxetable}{lccrcc}
\tablewidth{0pc}
\tabletypesize{\scriptsize}
\tablenum{2}
\tablecaption{Multiplicity Parameters for Resolved Systems\label{t2-fgs}}
\tablewidth{0pt}
\tablehead{
\colhead{Star} &
\colhead{$\Delta x$} &
\colhead{$\Delta y$} &
\colhead{$\rho$} &
\colhead{$\theta$} &
\colhead{$\Delta m_{\rm F583W}$} \\
\colhead{Name} &
\colhead{(mas)} &
\colhead{(mas)} &
\colhead{(mas)} &
\colhead{($^{\circ}$)} &
\colhead{(mag)} }
\startdata
A~41        &        54.4 $\pm$ 15.2&         352.6 $\pm$ 1.4 &     356.8 $\pm$ 3.1 &           276.1 $\pm$  6.5 & 3.27 $\pm$ 0.86 \\
MT 5        &      $-129.6 \pm$ 1.4 &         294.0 $\pm$ 1.1 &     321.3 $\pm$ 0.8 &            91.7 $\pm$  0.3 & 2.79 $\pm$ 0.09 \\
MT 59       &      $-751.3 \pm$ 0.8 &                 \nodata & $>$ 751.3 $\pm$ 0.8 &                    \nodata & 2.58 $\pm$ 0.04 \\
MT 138      &               \nodata &          66.0 $\pm$ 3.1 & $>$  66.0 $\pm$ 3.1 &                    \nodata & 2.79 $\pm$ 0.24 \\
MT 304      &        61.8 $\pm$ 3.1 &($-14.9$ or 9.7) $\pm$4.6&      63.6 $\pm$ 3.5 & (305.9 or 283.5) $\pm$ 3.3 & 2.31 $\pm$ 0.21 \\
MT 417 A,Ba &      $-231.3 \pm$ 1.3 &        1505.0 $\pm$ 3.9 &    1522.7 $\pm$ 3.7 &           146.9 $\pm$  0.5 & 0.45 $\pm$ 0.01 \\
MT 417 A,Bb &      $-214.1 \pm$ 5.1 &        1700.0 $\pm$ 2.5 &    1715.7 $\pm$ 2.9 &           147.9 $\pm$  0.4 & 2.66 $\pm$ 0.32 \\
MT 429      &      $ -36.2 \pm$ 0.7 &         $-95.8 \pm$ 0.3 &     101.9 $\pm$ 0.3 &            23.5 $\pm$  0.4 & 1.09 $\pm$ 0.02 \\
MT 516      &       374.5 $\pm$ 0.04&        $-618.0 \pm$ 0.1 &     722.6 $\pm$ 0.09&           325.7 $\pm$  0.1 & 0.28 $\pm$ 0.02 \\
MT 531      &               \nodata &         470.8 $\pm$ 0.08& $>$ 470.8 $\pm$ 0.08&                    \nodata & 0.50 $\pm$ 0.04 \\
MT 605      &        99.0 $\pm$ 0.15&          61.2 $\pm$ 0.5 &     116.4 $\pm$ 0.4 &           255.0 $\pm$  0.2 & 0.69 $\pm$ 0.02 \\
MT 632      &       165.0 $\pm$ 0.3 &        $-144.1 \pm$ 0.3 &     219.1 $\pm$ 0.2 &           247.2 $\pm$  0.09& 2.00 $\pm$ 0.16 \\
MT~696      & ($\pm$ 9.7) $\pm$ 1.5 &          20.6 $\pm$ 2.7 &      22.8 $\pm$ 3.0 & (175.1 or 225.6) $\pm$ 1.9 & 0.94 $\pm$ 0.40 \\
SCHULTE 5   &      $-801.1 \pm$ 2.2 &         439.6 $\pm$ 2.1 &     913.9 $\pm$ 2.3 &            56.4 $\pm$  0.11& 2.93 $\pm$ 0.29 \\
\enddata
\end{deluxetable}


\clearpage



\input{epsf}

\clearpage
\setcounter{figure}{0}
\renewcommand{\thefigure}{\arabic{figure}.1}
\begin{figure}
\begin{center}
{\includegraphics[angle=90, width=17.5cm]{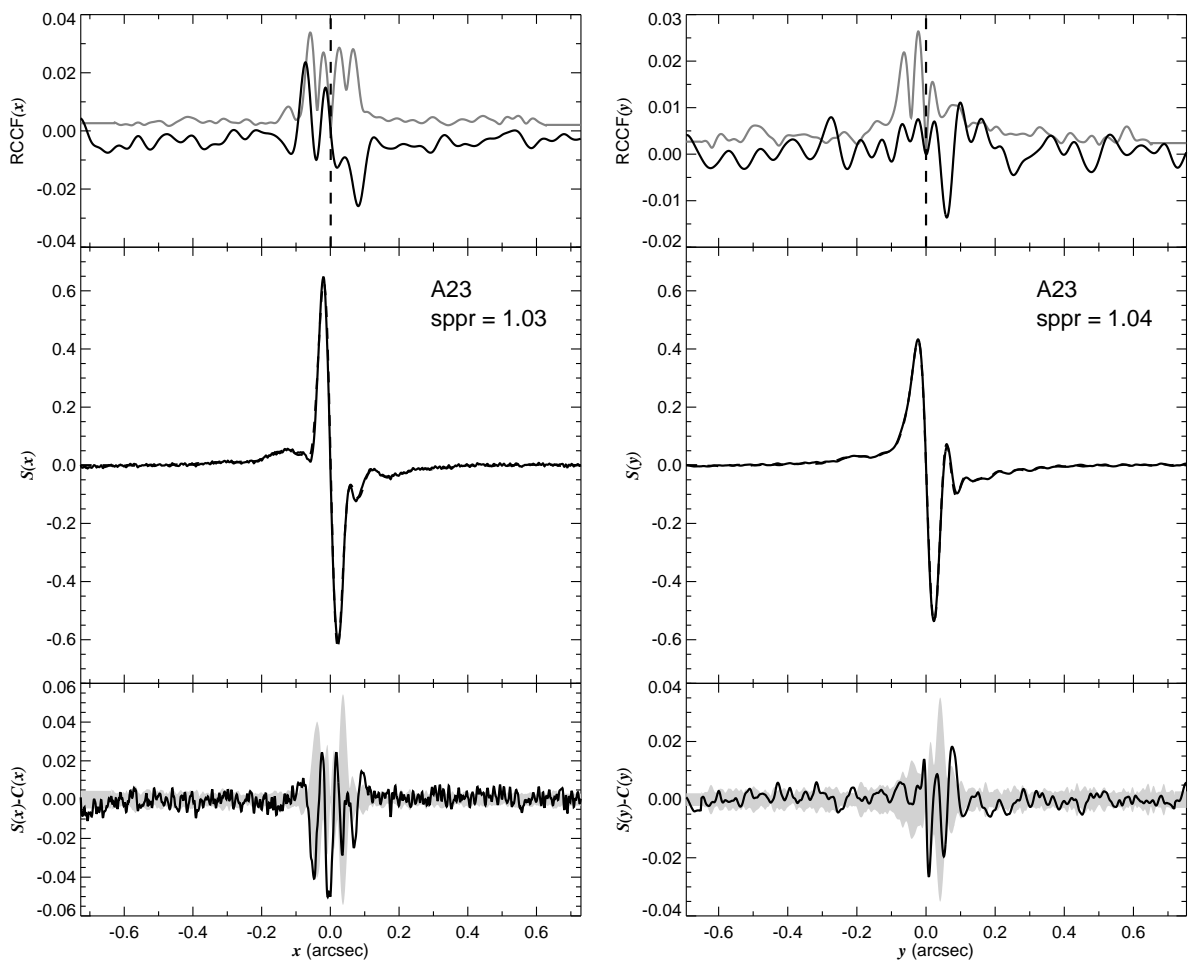}}
\end{center}
\caption{The final $S$-curves for the $x$ and $y$ orthogonal scans for
A~23.  The top panels show the mean residual cross-correlation
functions (black line: after removal of the primary component) plus
the CCF standard deviations (gray line).  The positions of components
are indicated by vertical dashed lines.  The middle panels show the
observed (solid line) and modeled (dashed line) $S$-curves.  The lower
panels display the observed minus model residuals (black line) and
standard deviation (gray region) plus a fit of the second derivative
(gray line) if a blended companion is present.\label{figa23}\label{figset1}}
\end{figure}

\clearpage
\setcounter{figure}{0}
\renewcommand{\thefigure}{\arabic{figure}.2}
\begin{figure}
\begin{center}
{\includegraphics[angle=90, width=17.5cm]{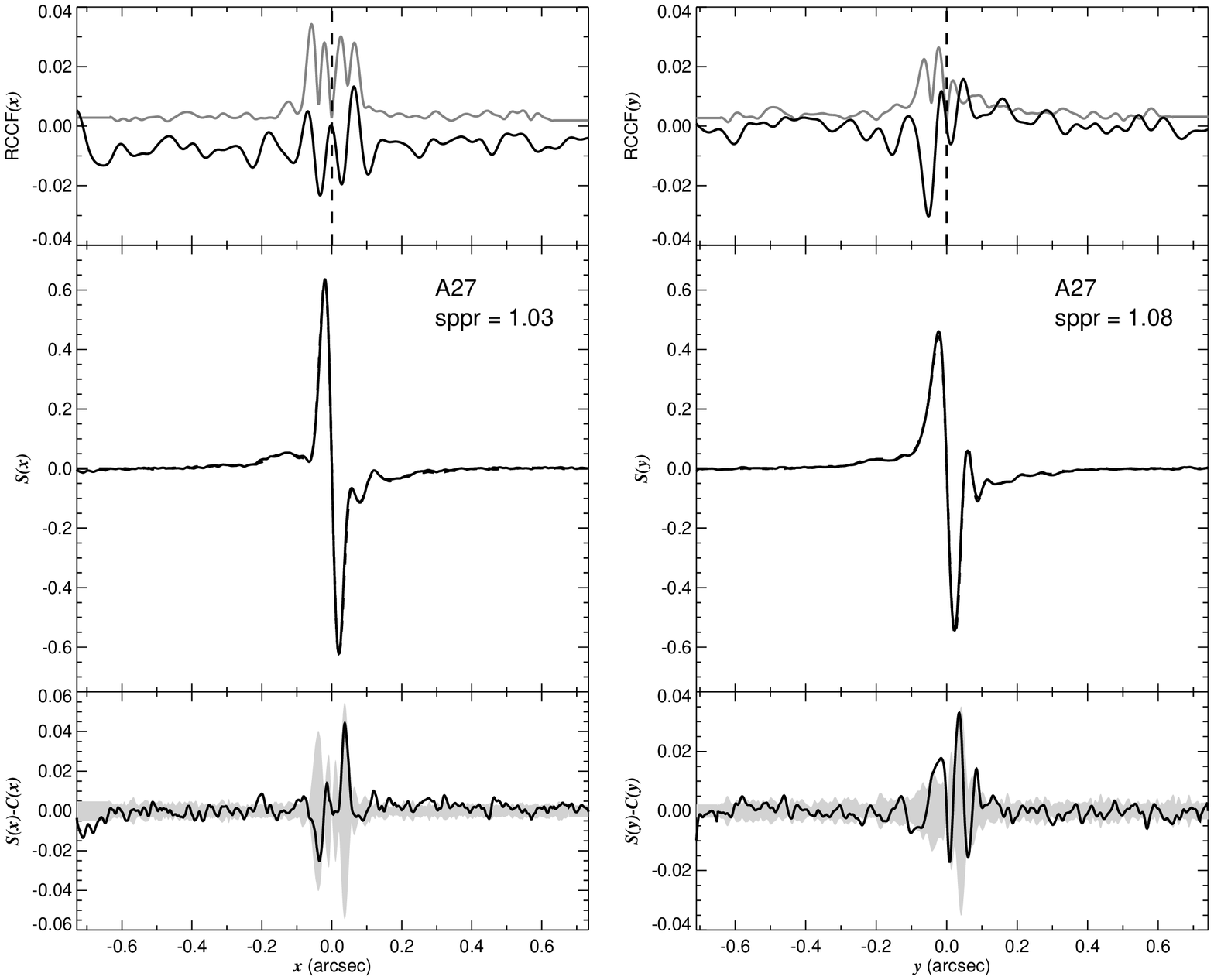}}
\end{center}
\caption{The final $S$-curves for the $x$ and $y$ orthogonal scans for A 27 
in the same format as Fig.\ 1.1.\label{figa27}}
\end{figure}

\clearpage
\setcounter{figure}{0}
\renewcommand{\thefigure}{\arabic{figure}.3}
\begin{figure}
\begin{center}
{\includegraphics[angle=90, width=17.5cm]{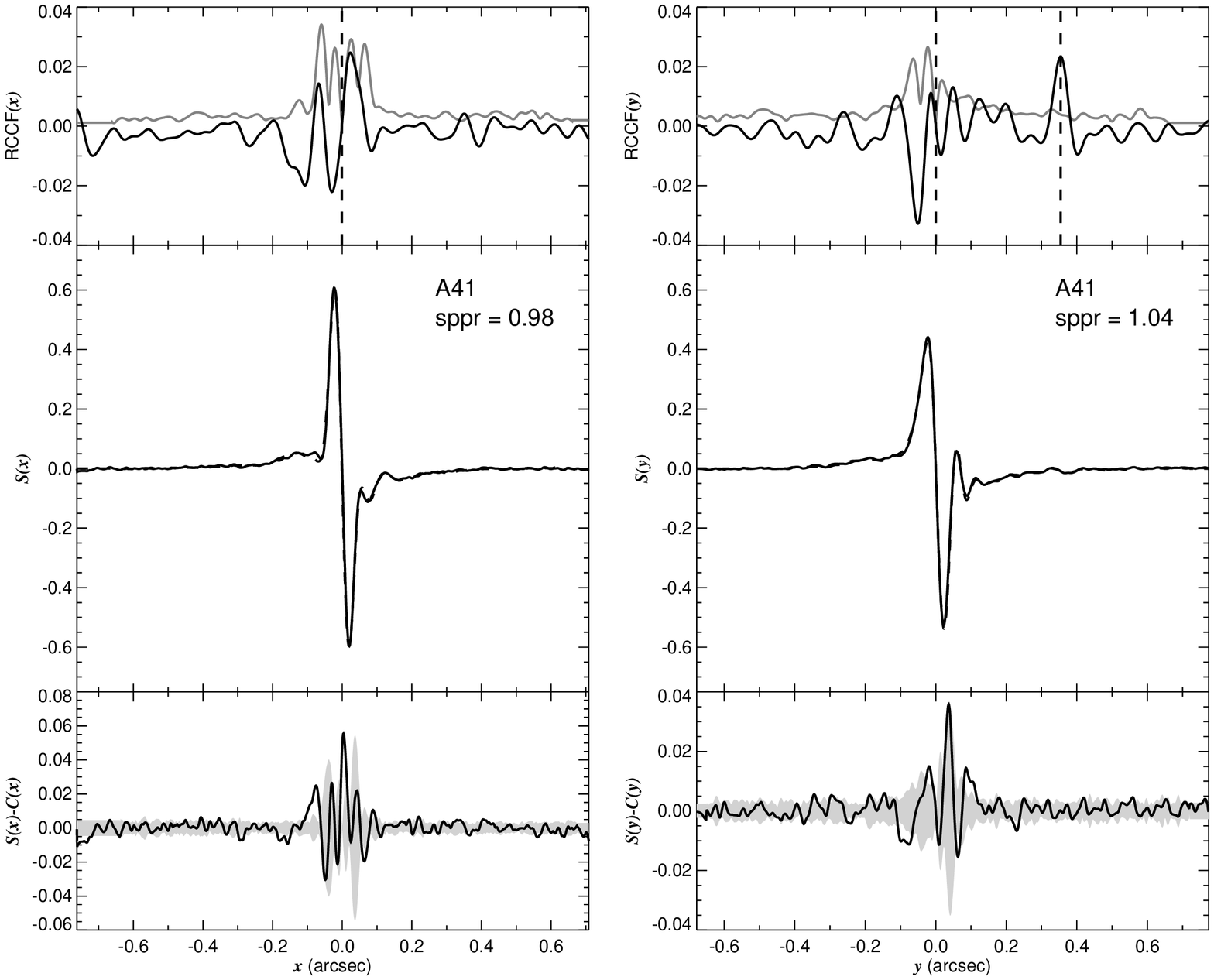}}
\end{center}
\caption{The final $S$-curves for the $x$ and $y$ orthogonal scans for A 41 
in the same format as Fig.\ 1.1.\label{figa41}}
\end{figure}

\clearpage
\setcounter{figure}{0}
\renewcommand{\thefigure}{\arabic{figure}.4}
\begin{figure}
\begin{center}
{\includegraphics[angle=90, width=17.5cm]{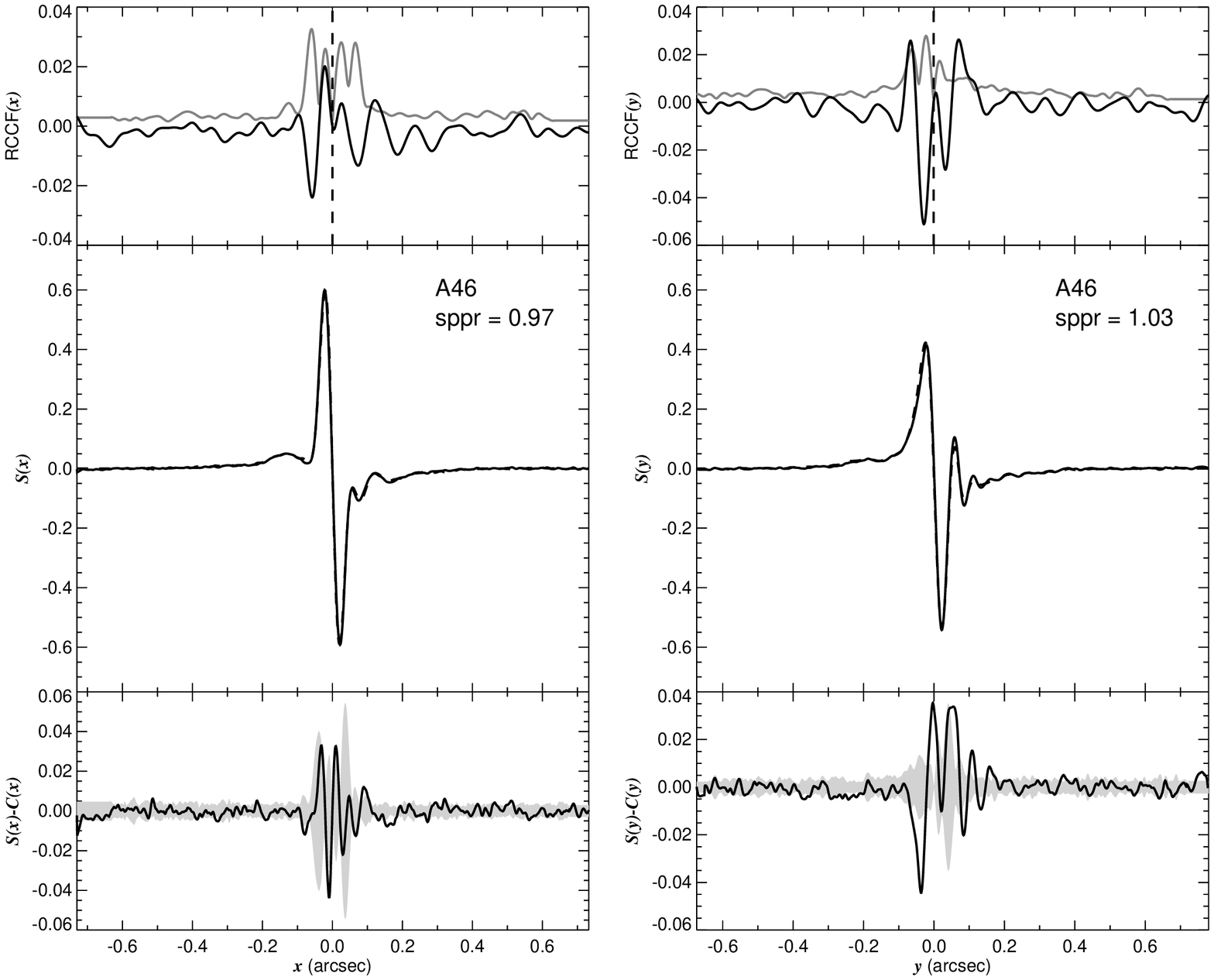}}
\end{center}
\caption{The final $S$-curves for the $x$ and $y$ orthogonal scans for A 46 
in the same format as Fig.\ 1.1.\label{figa46}}
\end{figure}

\clearpage
\setcounter{figure}{0}
\renewcommand{\thefigure}{\arabic{figure}.5}
\begin{figure}
\begin{center}
{\includegraphics[angle=90, width=17.5cm]{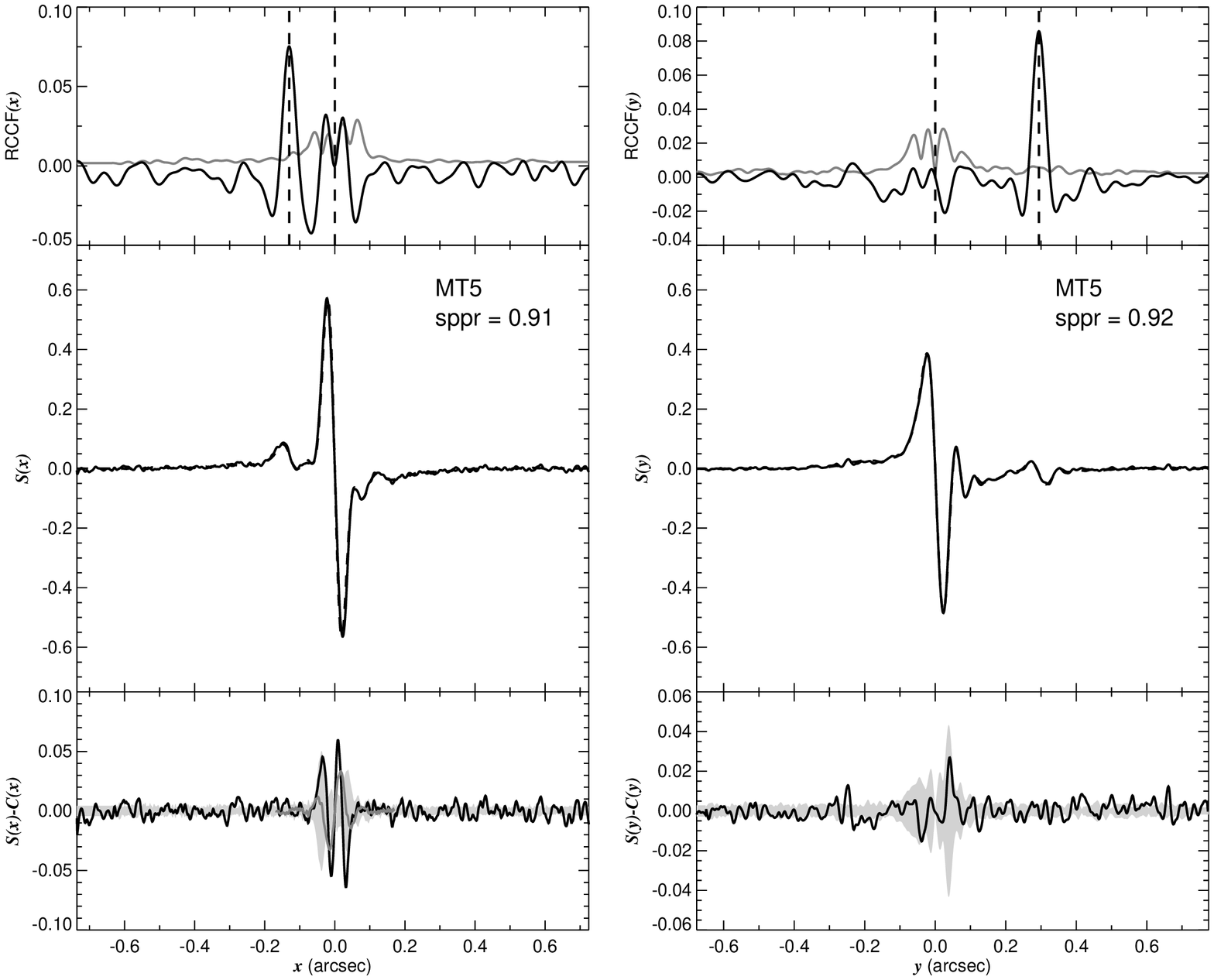}}
\end{center}
\caption{The final $S$-curves for the $x$ and $y$ orthogonal scans for MT 5 
in the same format as Fig.\ 1.1.\label{figmt5}}
\end{figure}

\clearpage
\setcounter{figure}{0}
\renewcommand{\thefigure}{\arabic{figure}.6}
\begin{figure}
\begin{center}
{\includegraphics[angle=90, width=17.5cm]{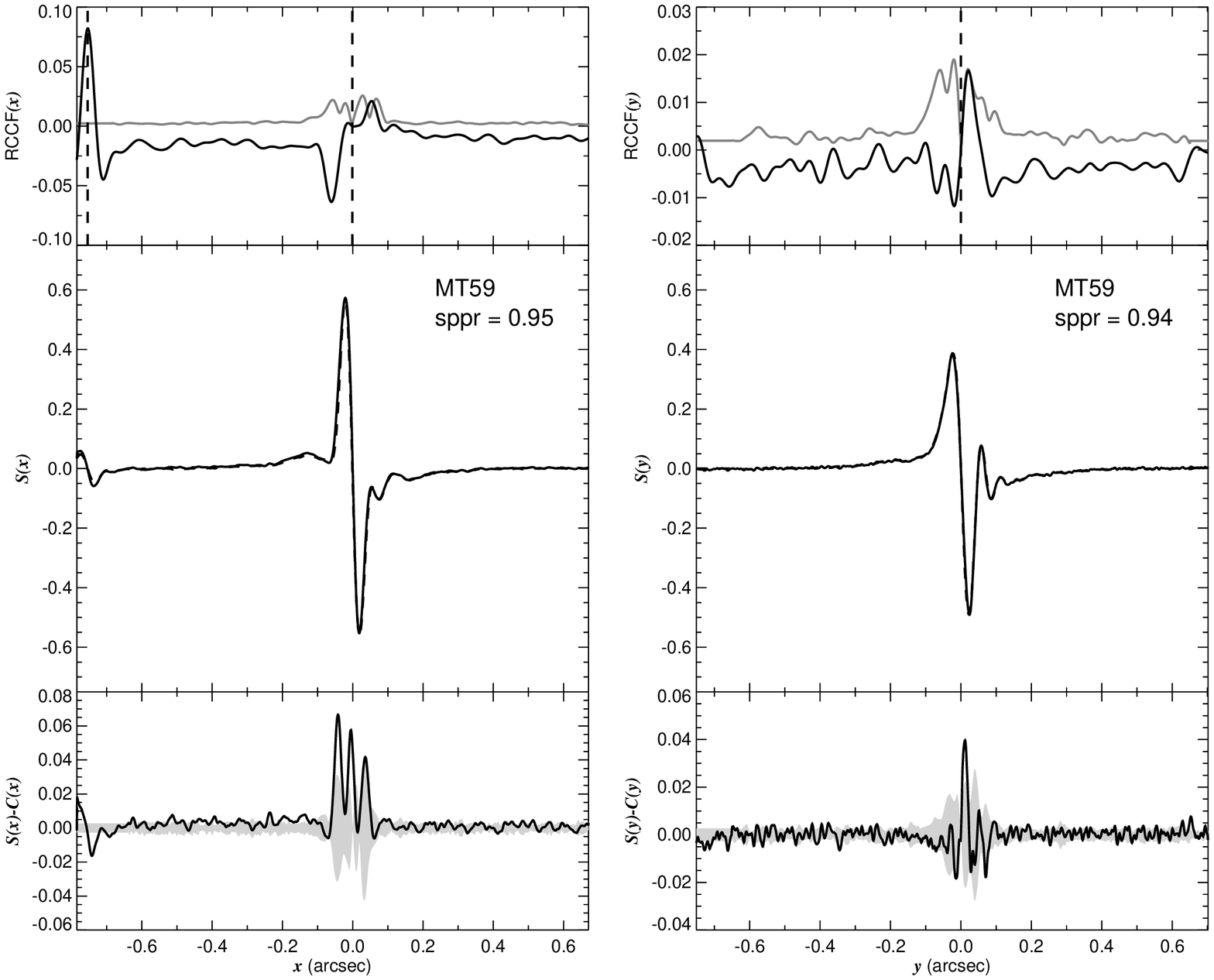}}
\end{center}
\caption{The final $S$-curves for the $x$ and $y$ orthogonal scans for MT 59 
in the same format as Fig.\ 1.1.\label{figmt59}}
\end{figure}

\clearpage
\setcounter{figure}{0}
\renewcommand{\thefigure}{\arabic{figure}.7}
\begin{figure}
\begin{center}
{\includegraphics[angle=90, width=17.5cm]{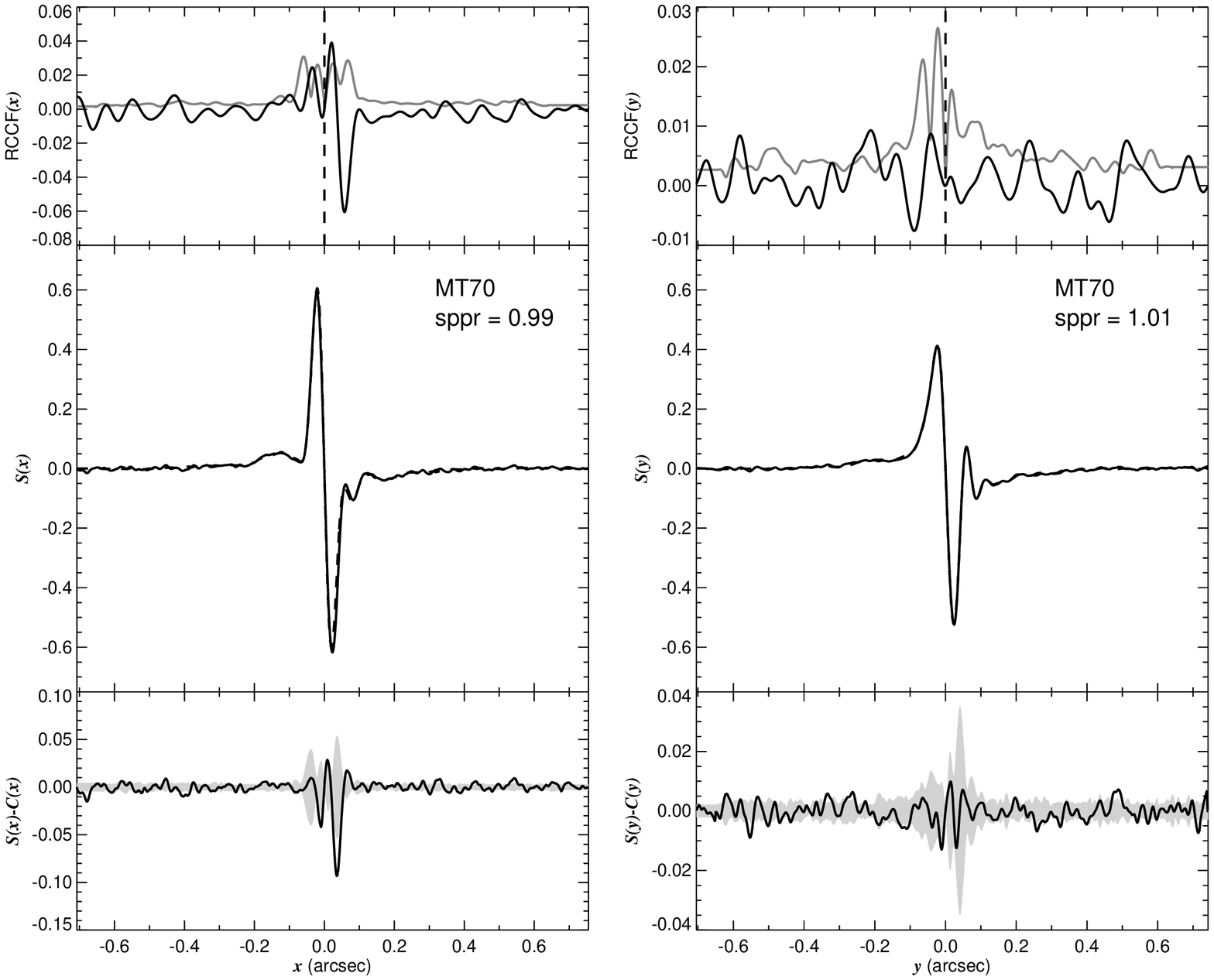}}
\end{center}
\caption{The final $S$-curves for the $x$ and $y$ orthogonal scans for MT 70 
in the same format as Fig.\ 1.1.\label{figmt70}}
\end{figure}

\clearpage
\setcounter{figure}{0}
\renewcommand{\thefigure}{\arabic{figure}.8}
\begin{figure}
\begin{center}
{\includegraphics[angle=90, width=17.5cm]{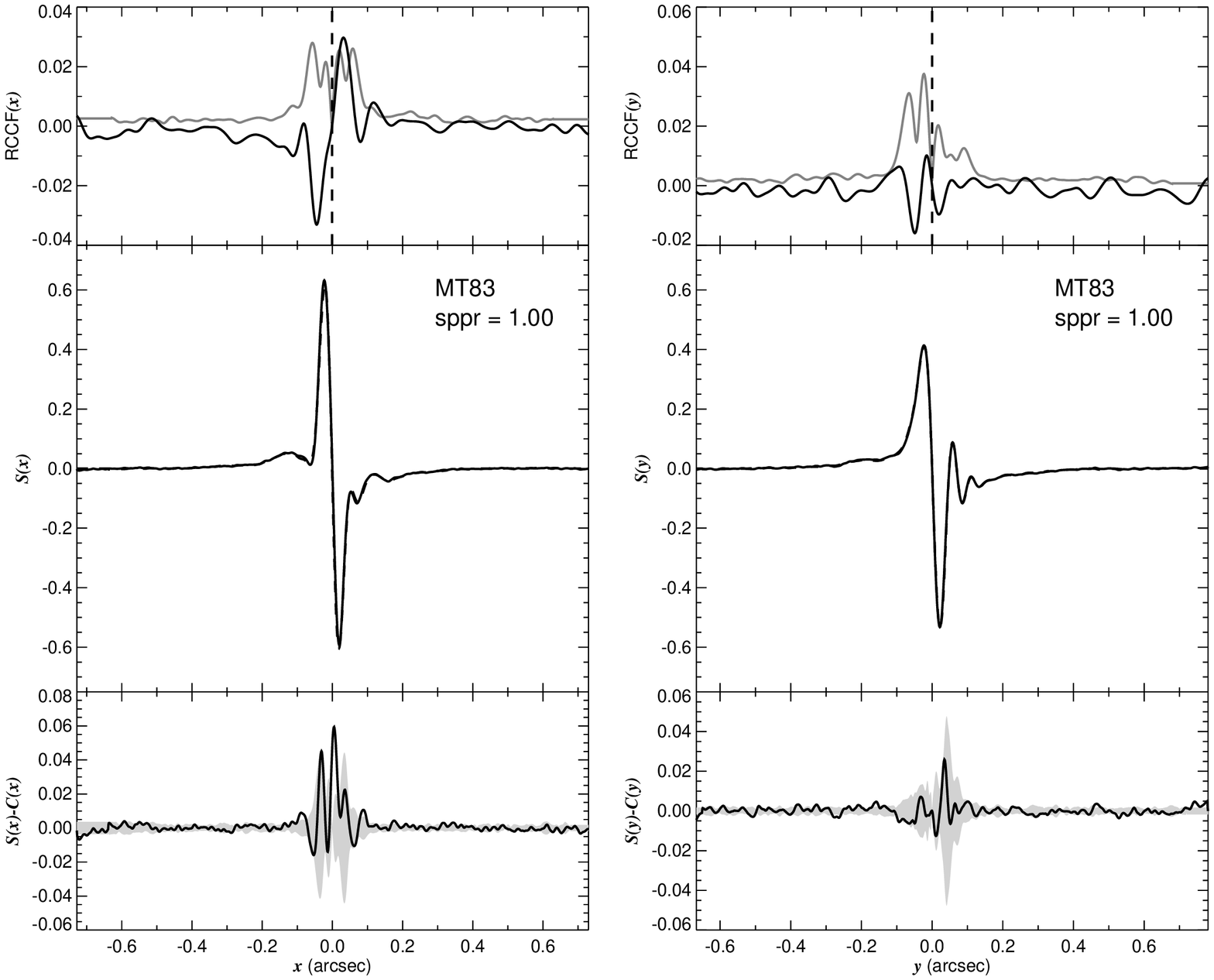}}
\end{center}
\caption{The final $S$-curves for the $x$ and $y$ orthogonal scans for MT 83 
in the same format as Fig.\ 1.1.\label{figmt83}}
\end{figure}

\clearpage
\setcounter{figure}{0}
\renewcommand{\thefigure}{\arabic{figure}.9}
\begin{figure}
\begin{center}
{\includegraphics[angle=90, width=17.5cm]{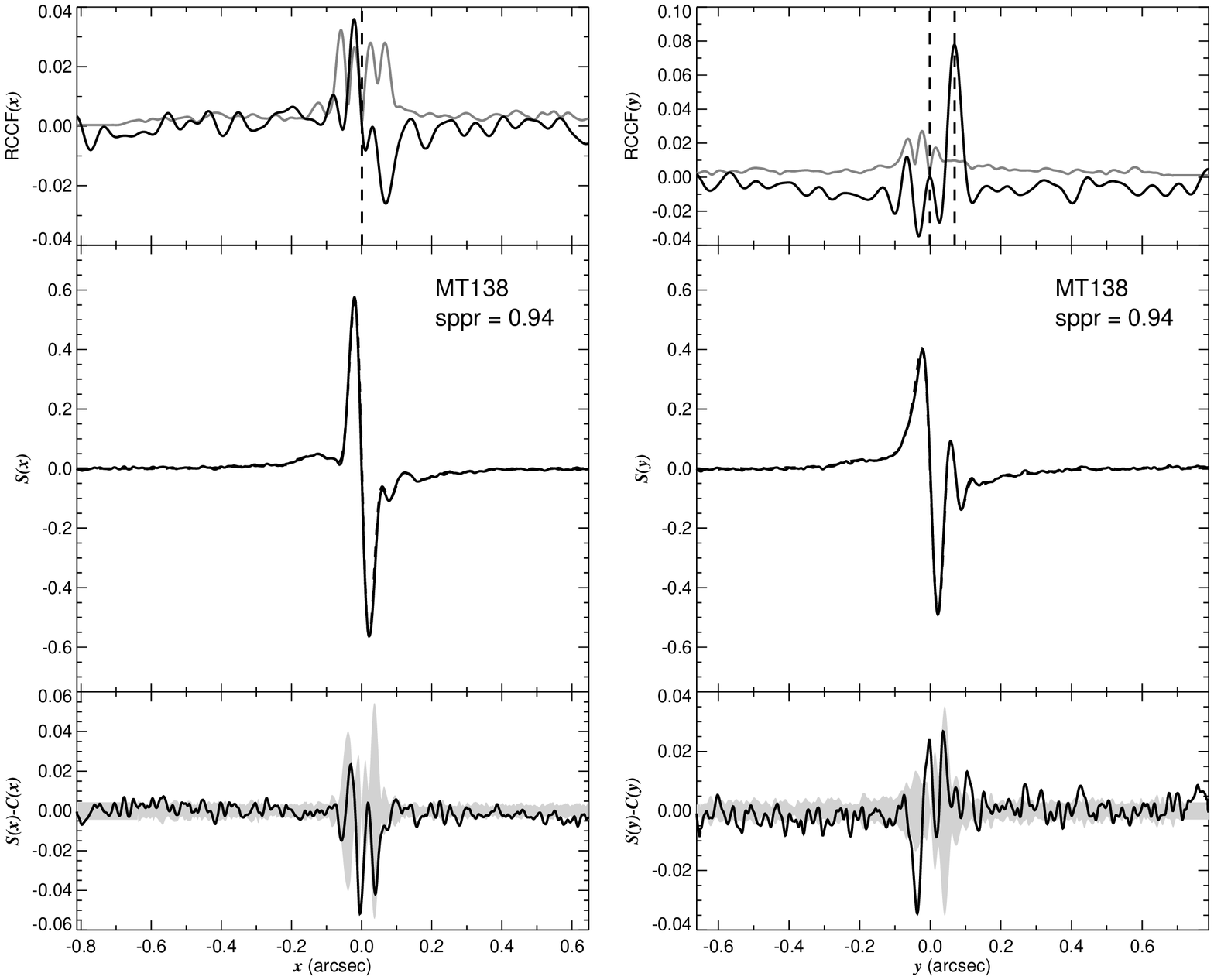}}
\end{center}
\caption{The final $S$-curves for the $x$ and $y$ orthogonal scans for MT 138
in the same format as Fig.\ 1.1.\label{figmt138}}
\end{figure}

\clearpage
\setcounter{figure}{0}
\renewcommand{\thefigure}{\arabic{figure}.10}
\begin{figure}
\begin{center}
{\includegraphics[angle=90, width=17.5cm]{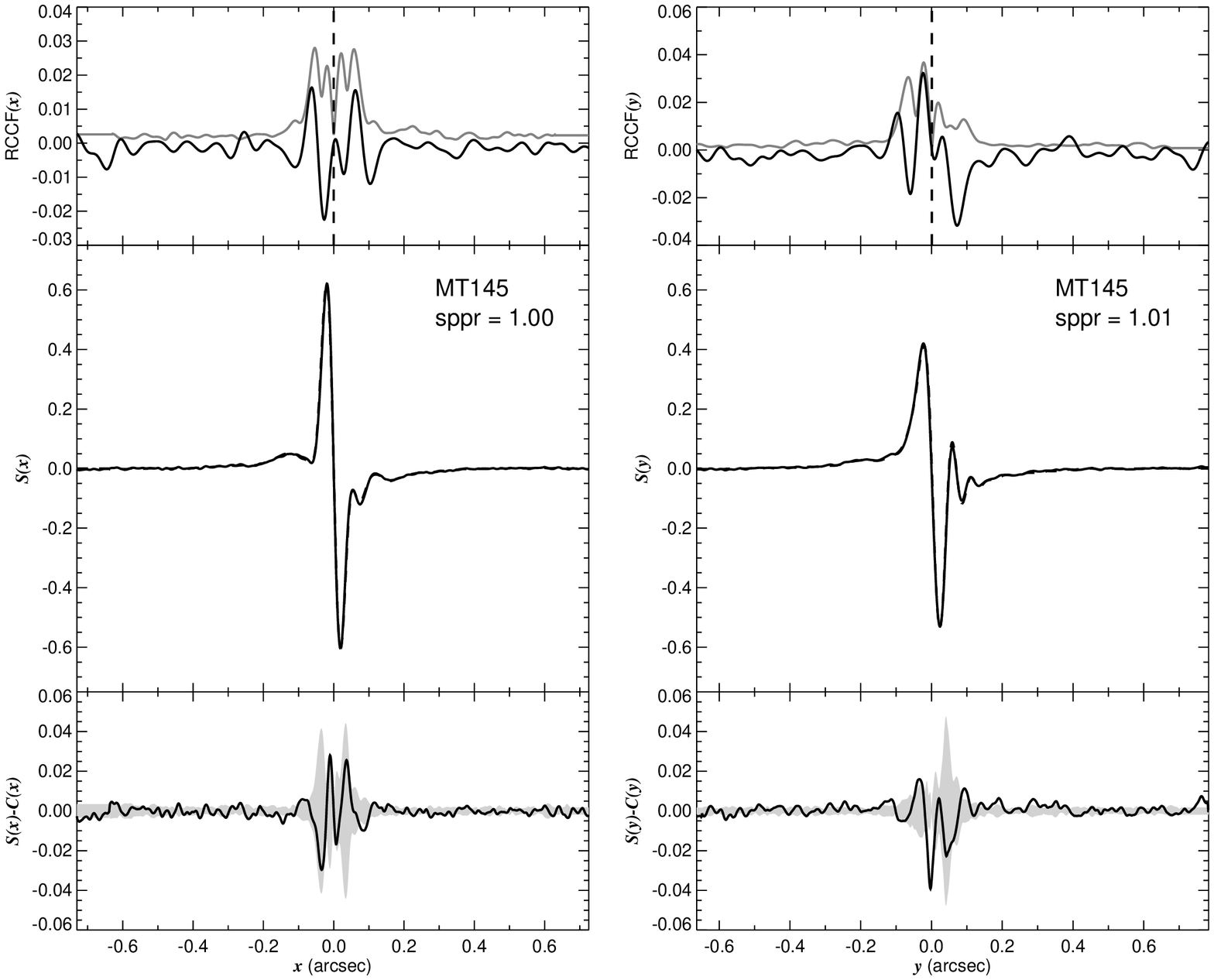}}
\end{center}
\caption{The final $S$-curves for the $x$ and $y$ orthogonal scans for MT 145 
in the same format as Fig.\ 1.1.\label{figmt415}}
\end{figure}

\clearpage
\setcounter{figure}{0}
\renewcommand{\thefigure}{\arabic{figure}.11}
\begin{figure}
\begin{center}
{\includegraphics[angle=90, width=17.5cm]{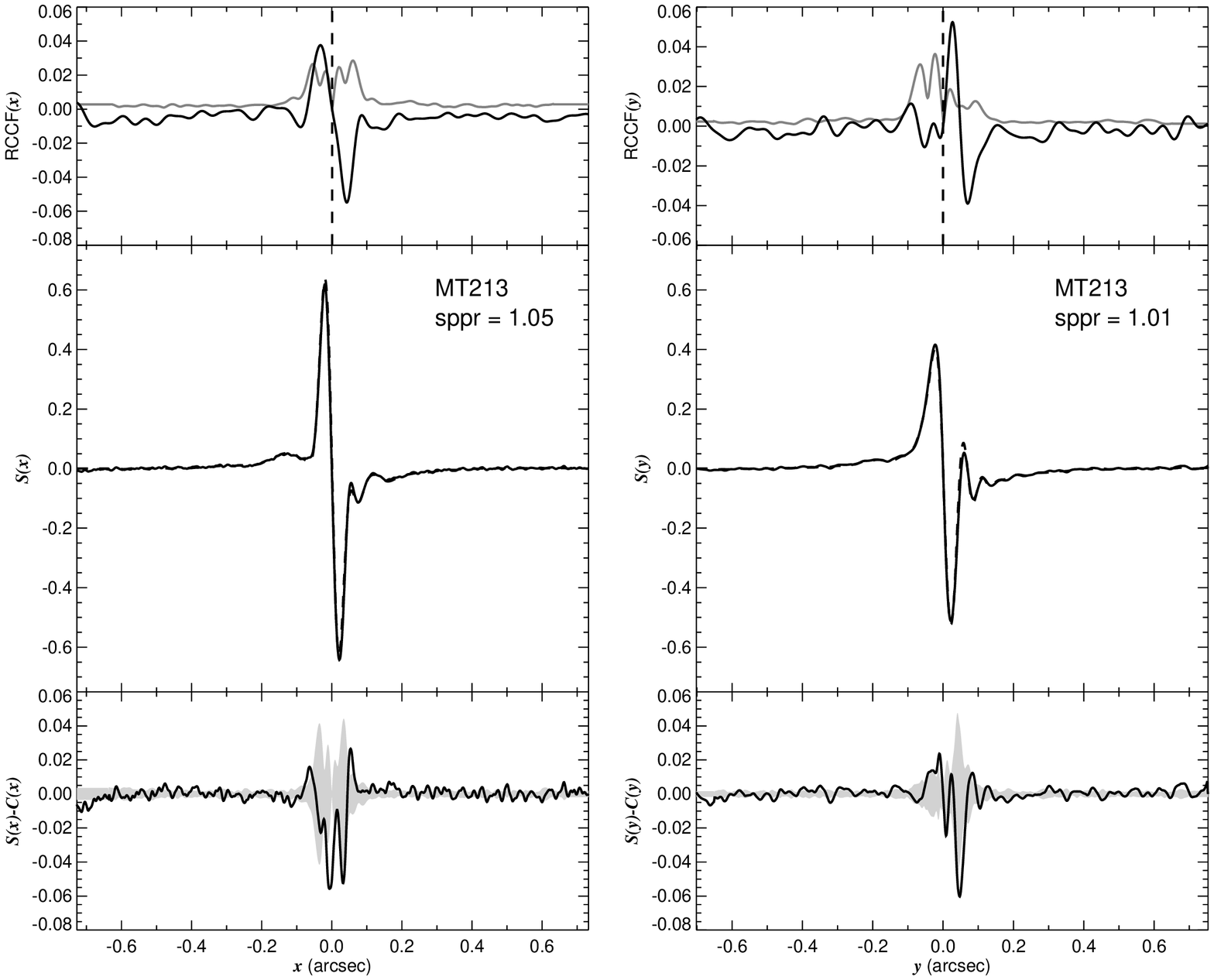}}
\end{center}
\caption{The final $S$-curves for the $x$ and $y$ orthogonal scans for MT 213 
in the same format as Fig.\ 1.1.\label{figmt213}}
\end{figure}

\clearpage
\setcounter{figure}{0}
\renewcommand{\thefigure}{\arabic{figure}.12}
\begin{figure}
\begin{center}
{\includegraphics[angle=90, width=17.5cm]{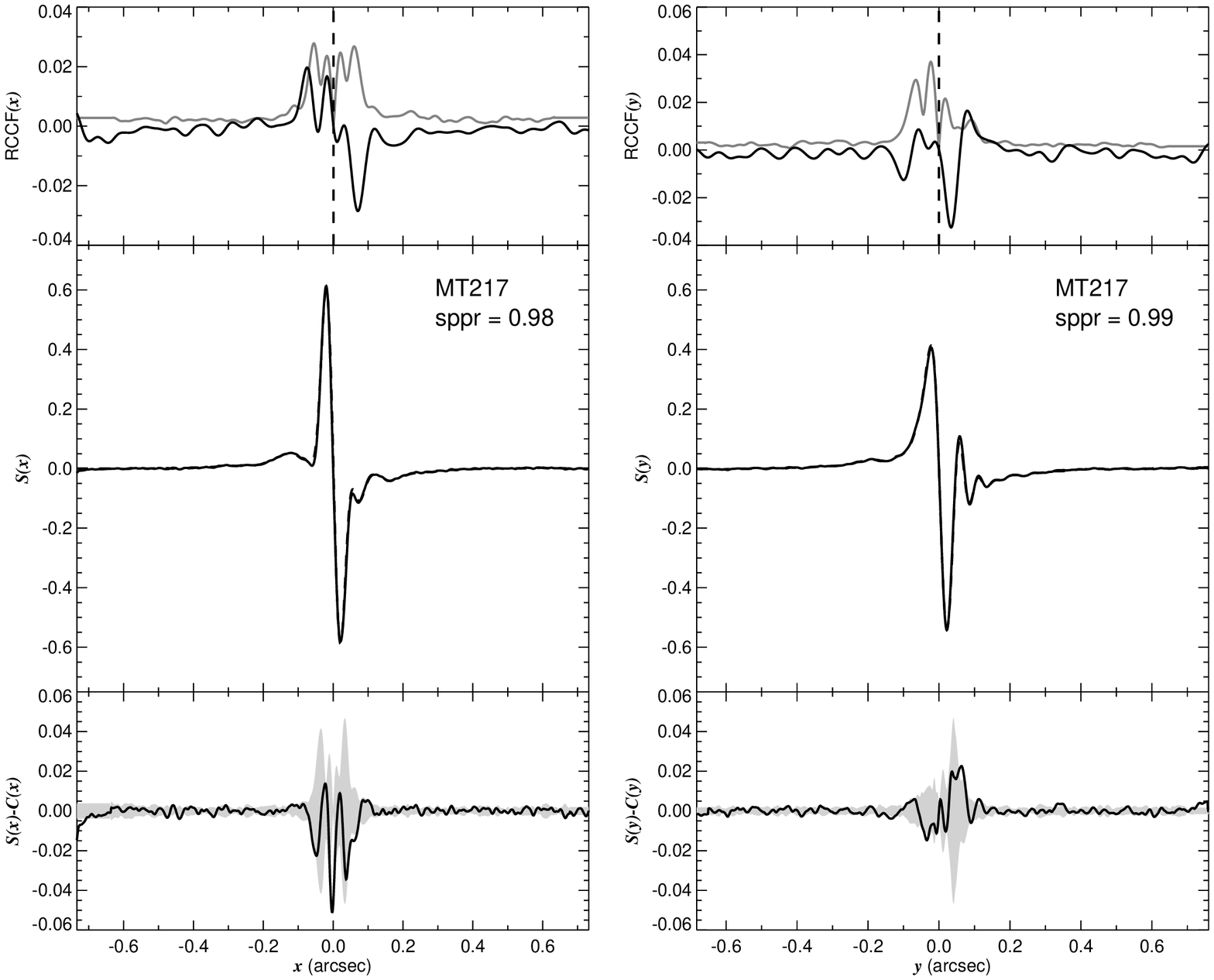}}
\end{center}
\caption{The final $S$-curves for the $x$ and $y$ orthogonal scans for MT 217
in the same format as Fig.\ 1.1.\label{figmt217}}
\end{figure}

\clearpage
\setcounter{figure}{0}
\renewcommand{\thefigure}{\arabic{figure}.13}
\begin{figure}
\begin{center}
{\includegraphics[angle=90, width=17.5cm]{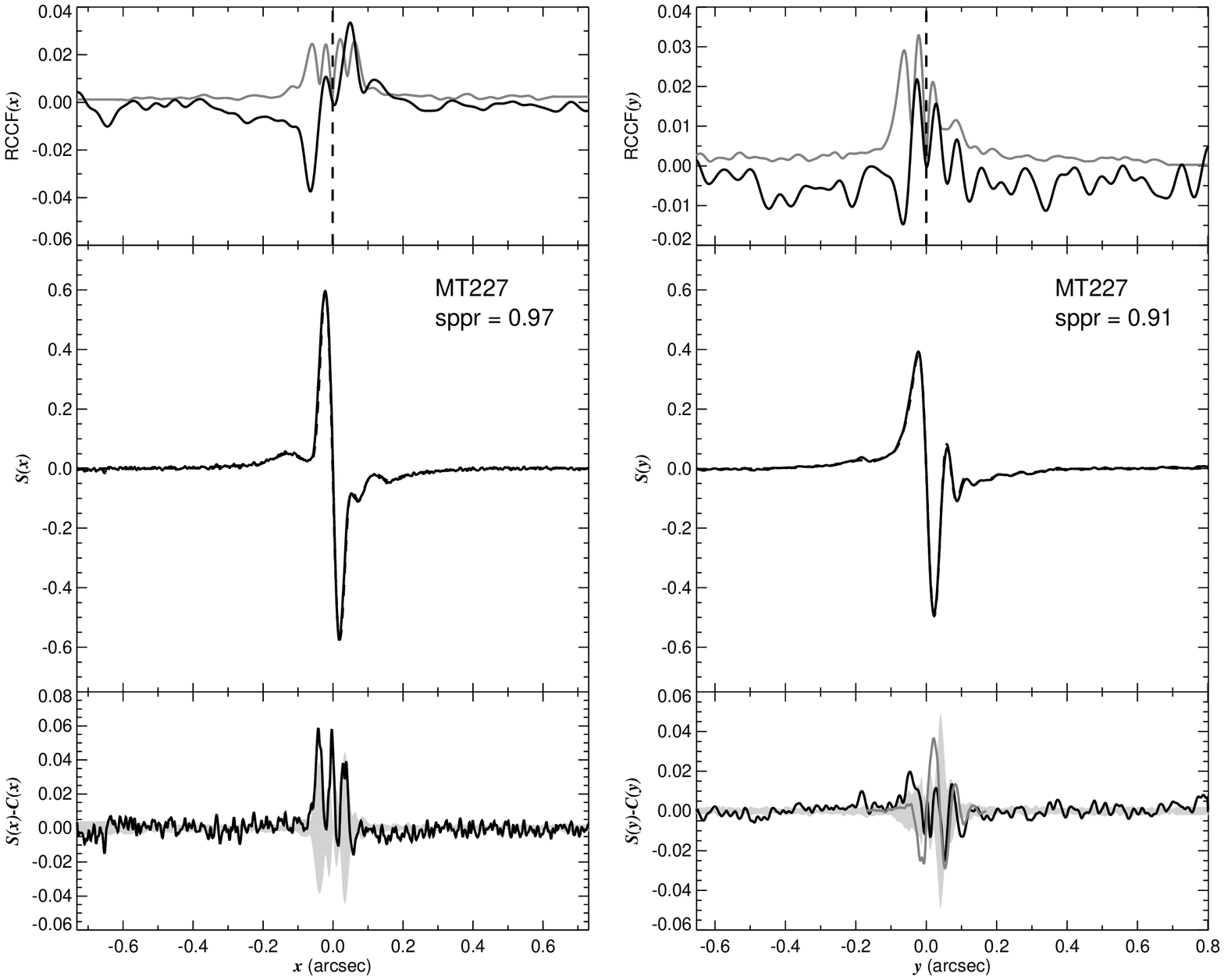}}
\end{center}
\caption{The final $S$-curves for the $x$ and $y$ orthogonal scans for MT 227
in the same format as Fig.\ 1.1.\label{figmt227}}
\end{figure}

\clearpage
\setcounter{figure}{0}
\renewcommand{\thefigure}{\arabic{figure}.14}
\begin{figure}
\begin{center}
{\includegraphics[angle=90, width=17.5cm]{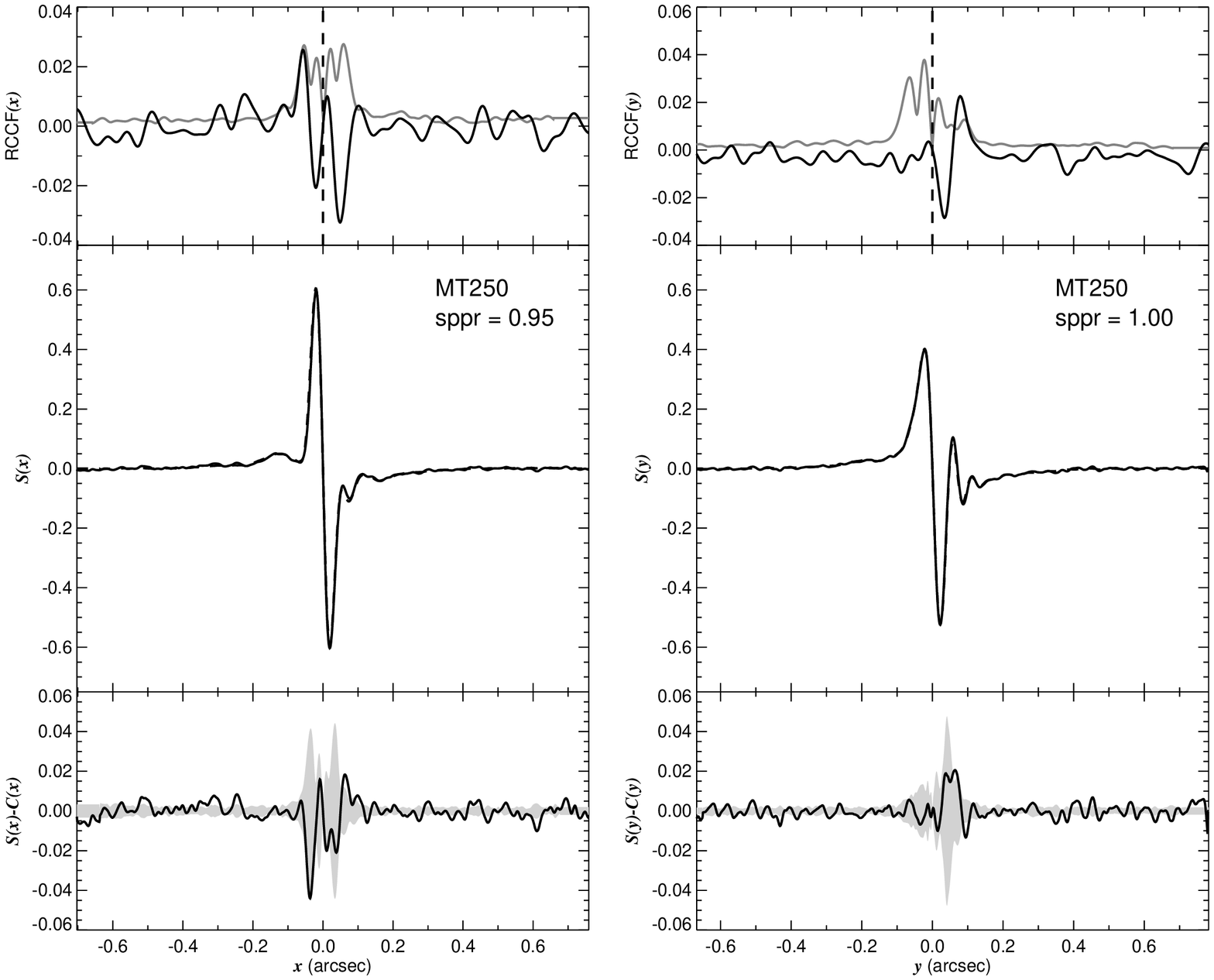}}
\end{center}
\caption{The final $S$-curves for the $x$ and $y$ orthogonal scans for MT 250 
in the same format as Fig.\ 1.1.\label{figmt250}}
\end{figure}

\clearpage
\setcounter{figure}{0}
\renewcommand{\thefigure}{\arabic{figure}.15}
\begin{figure}
\begin{center}
{\includegraphics[angle=90, width=17.5cm]{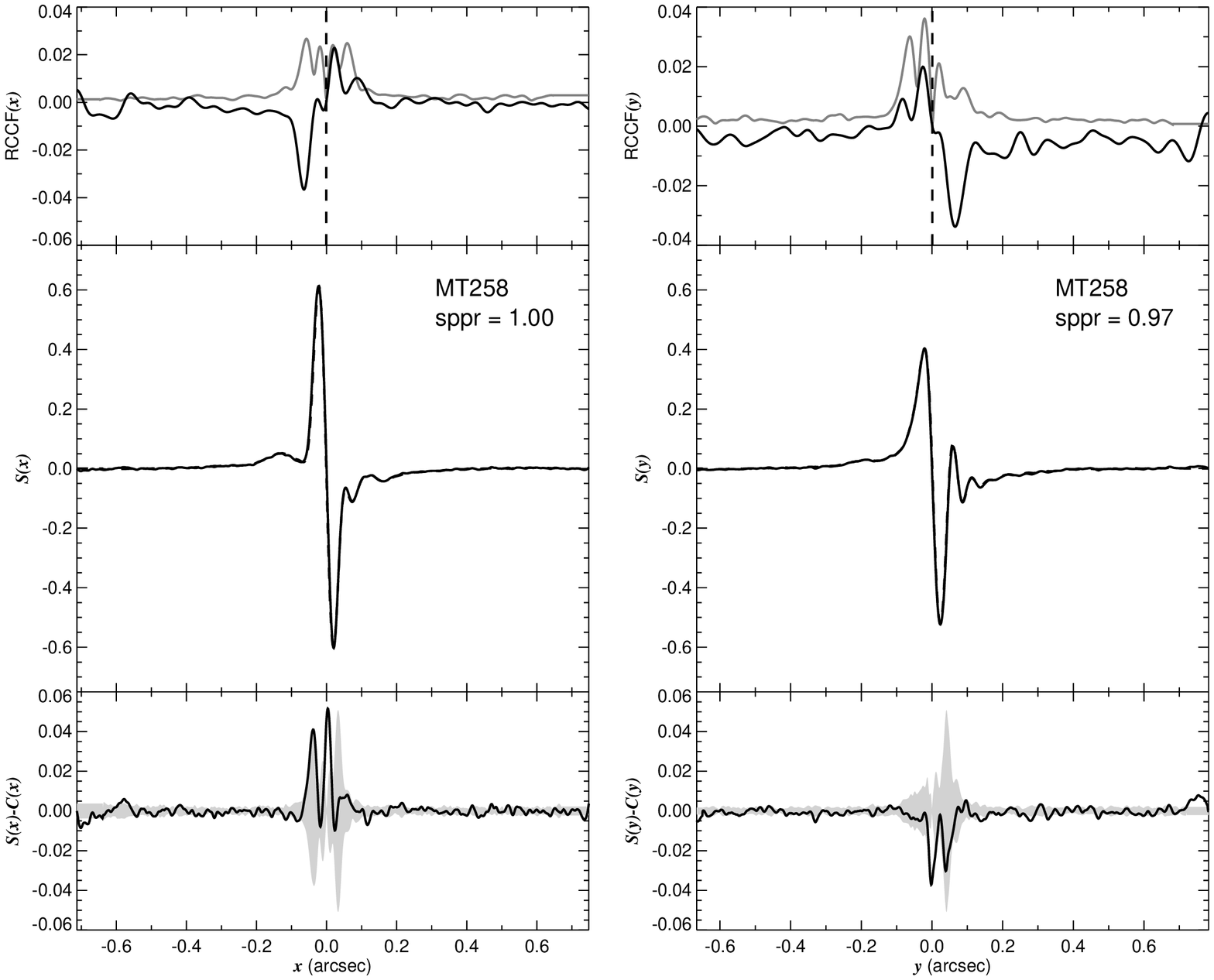}}
\end{center}
\caption{The final $S$-curves for the $x$ and $y$ orthogonal scans for MT 258 
in the same format as Fig.\ 1.1.\label{figmt258}}
\end{figure}

\clearpage
\setcounter{figure}{0}
\renewcommand{\thefigure}{\arabic{figure}.16}
\begin{figure}
\begin{center}
{\includegraphics[angle=90, width=17.5cm]{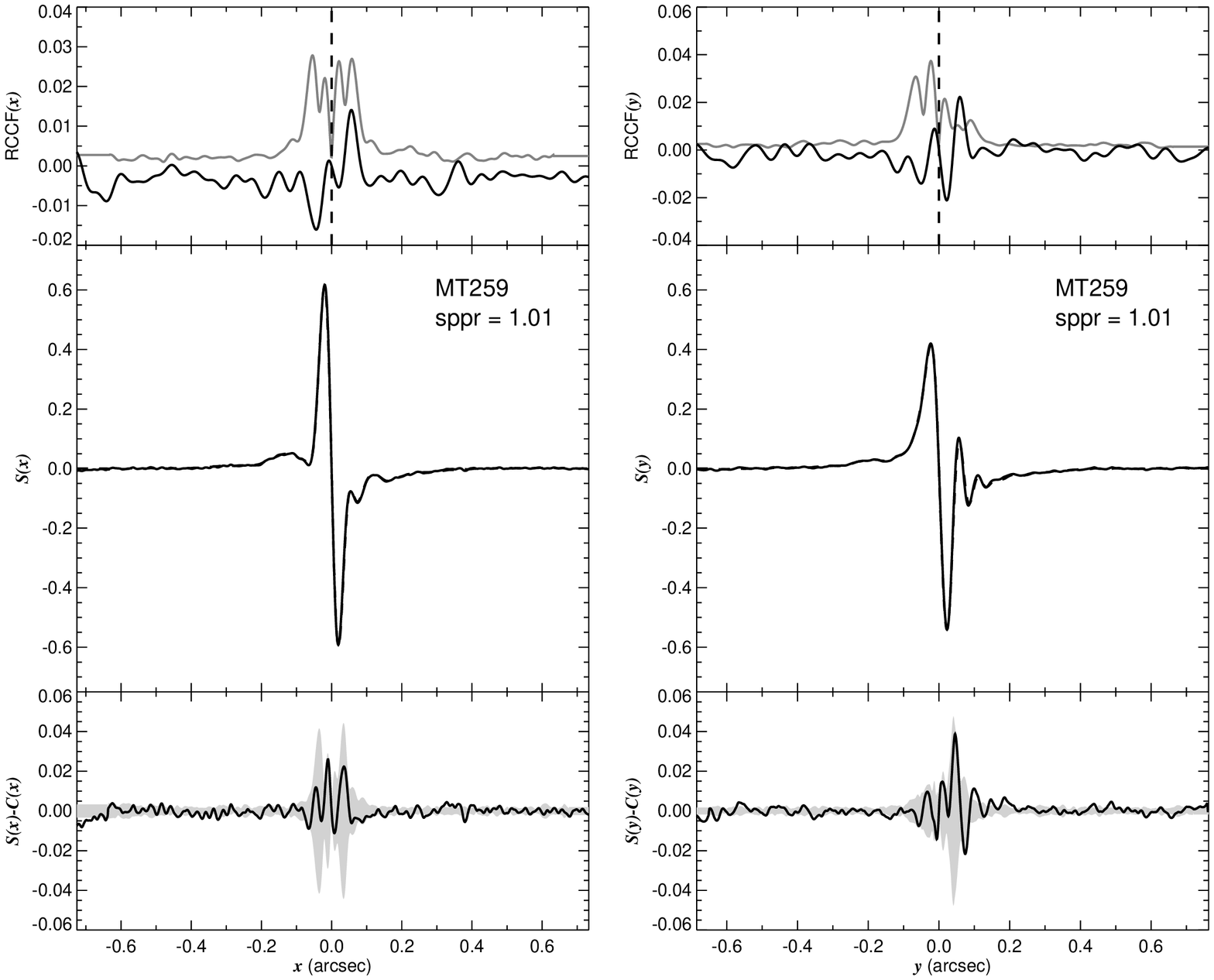}}
\end{center}
\caption{The final $S$-curves for the $x$ and $y$ orthogonal scans for MT 259 
in the same format as Fig.\ 1.1.\label{figmt259}}
\end{figure}

\clearpage
\setcounter{figure}{0}
\renewcommand{\thefigure}{\arabic{figure}.17}
\begin{figure}
\begin{center}
{\includegraphics[angle=90, width=17.5cm]{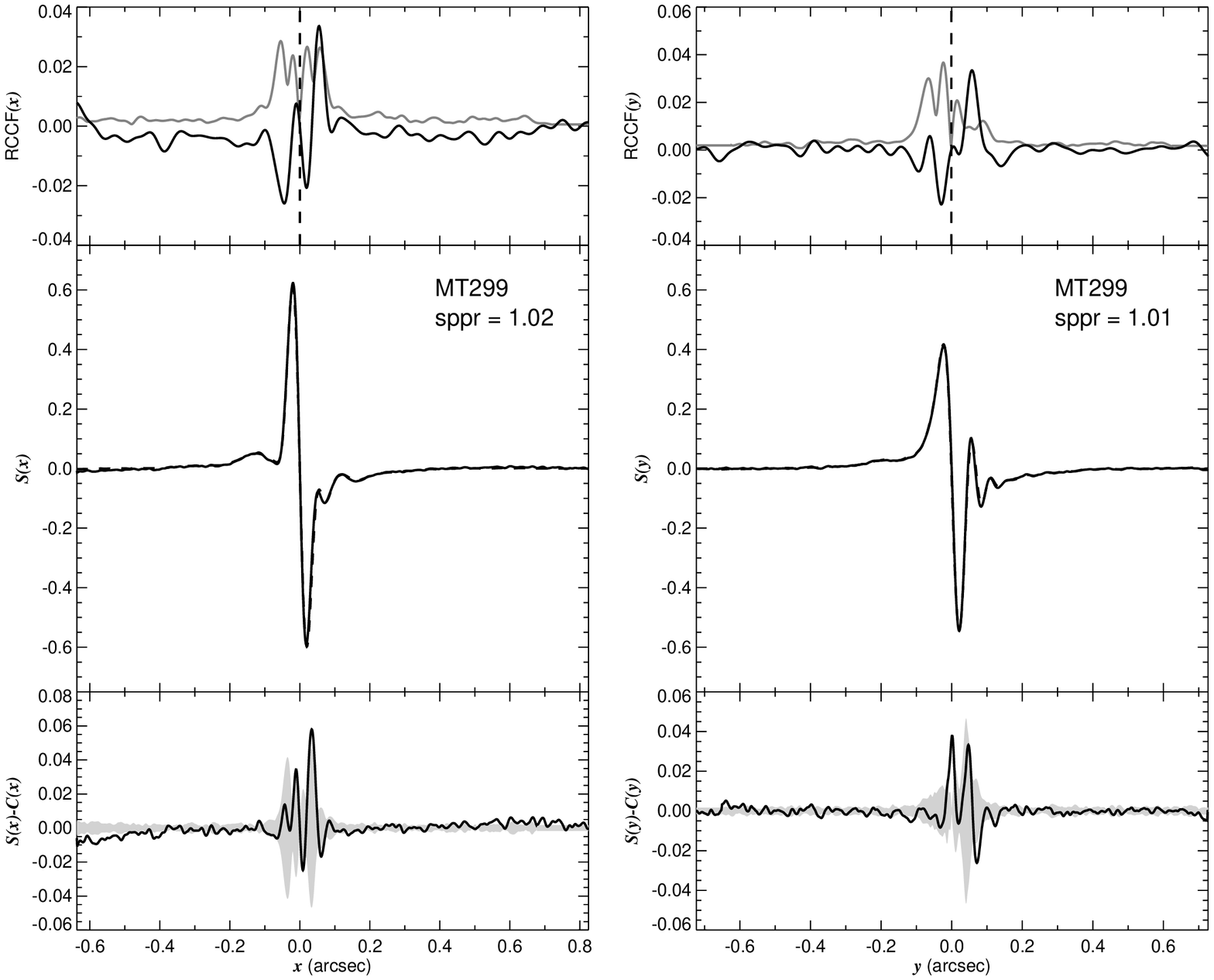}}
\end{center}
\caption{The final $S$-curves for the $x$ and $y$ orthogonal scans for MT 299
in the same format as Fig.\ 1.1.\label{figmt299}}
\end{figure}

\clearpage
\setcounter{figure}{0}
\renewcommand{\thefigure}{\arabic{figure}.18}
\begin{figure}
\begin{center}
{\includegraphics[angle=90, width=17.5cm]{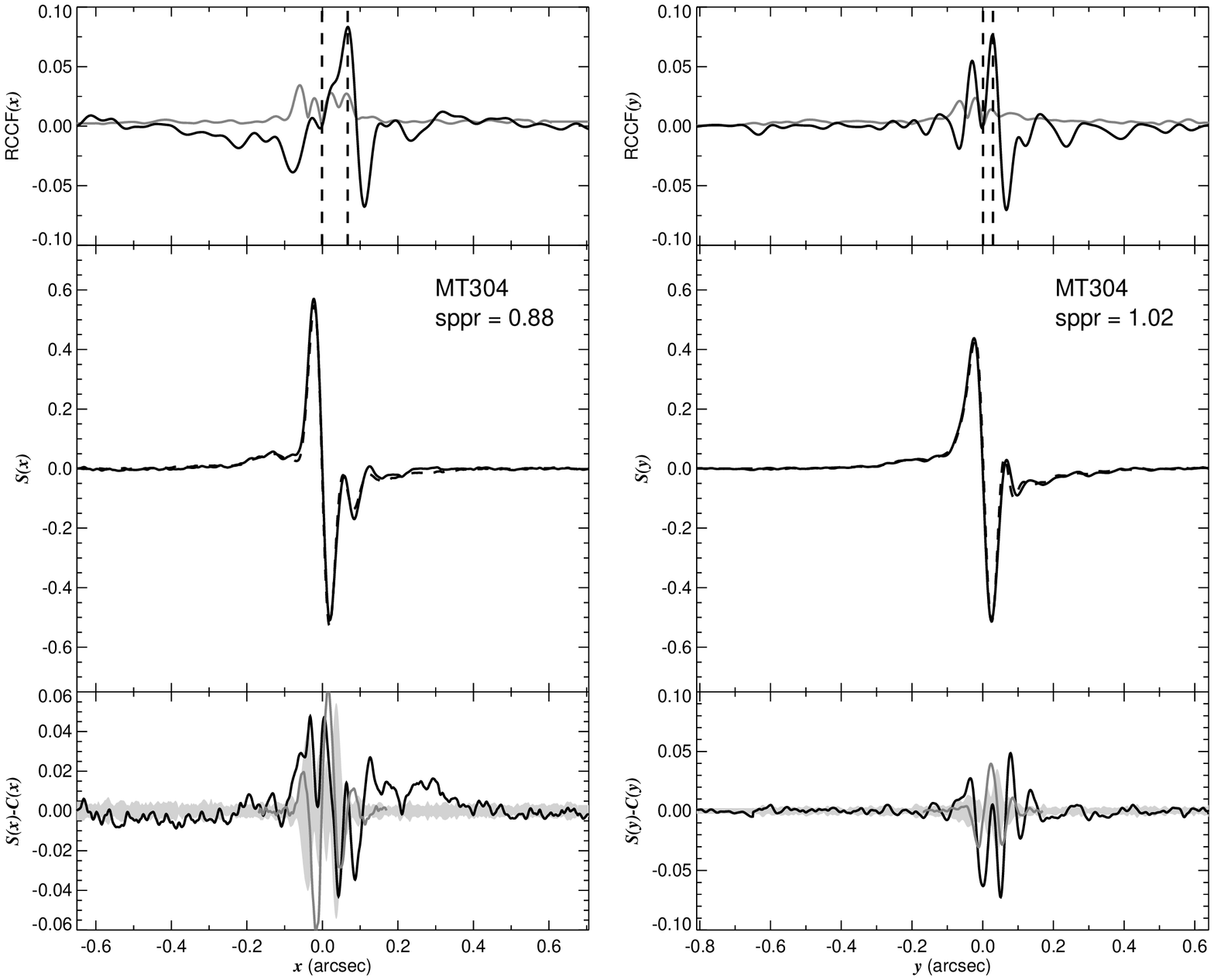}}
\end{center}
\caption{The final $S$-curves for the $x$ and $y$ orthogonal scans for MT 304 
in the same format as Fig.\ 1.1.\label{figmt304}}
\end{figure}

\clearpage
\setcounter{figure}{0}
\renewcommand{\thefigure}{\arabic{figure}.19}
\begin{figure}
\begin{center}
{\includegraphics[angle=90, width=17.5cm]{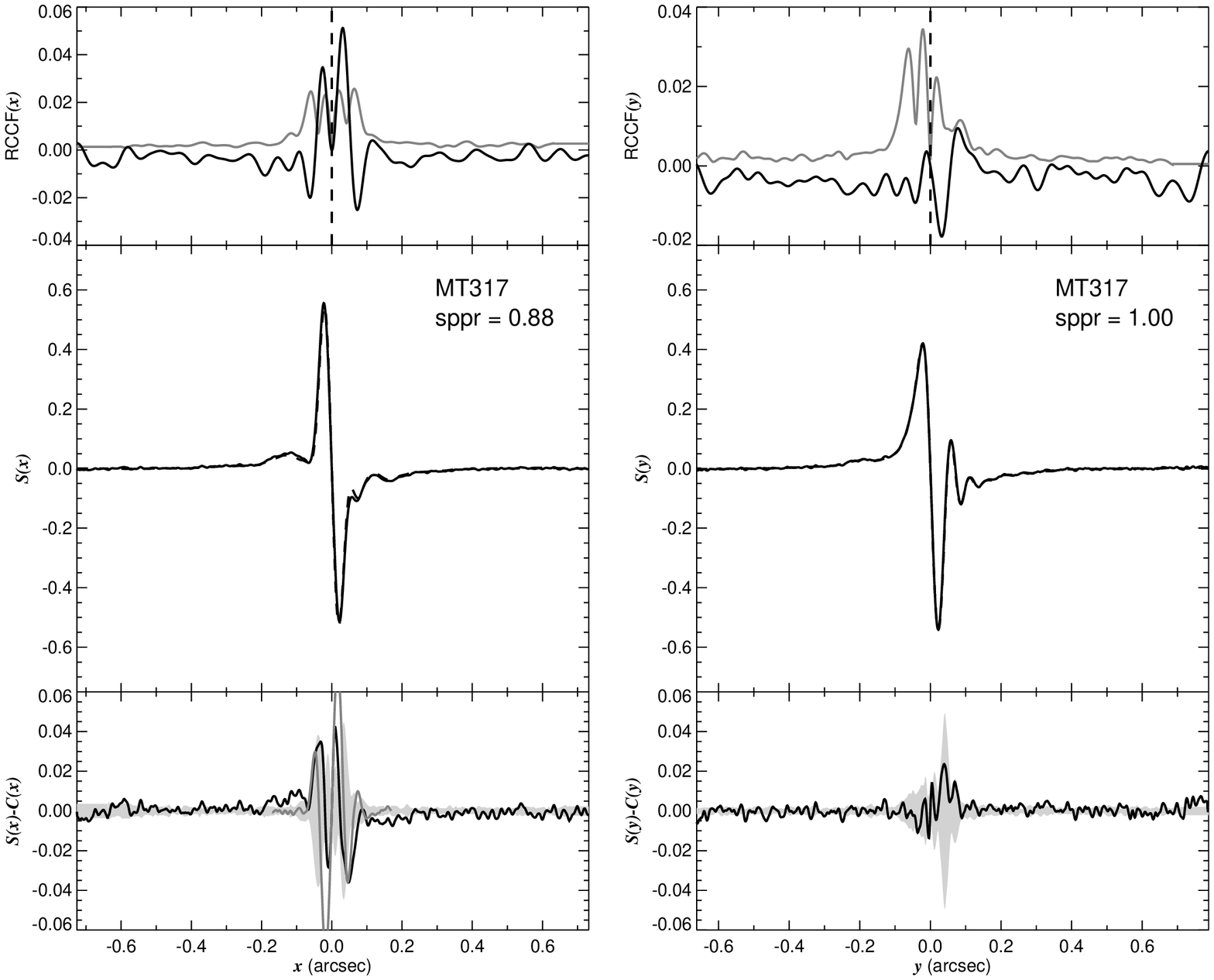}}
\end{center}
\caption{The final $S$-curves for the $x$ and $y$ orthogonal scans for MT 317
in the same format as Fig.\ 1.1.\label{figmt317}}
\end{figure}

\clearpage
\setcounter{figure}{0}
\renewcommand{\thefigure}{\arabic{figure}.20}
\begin{figure}
\begin{center}
{\includegraphics[angle=90, width=17.5cm]{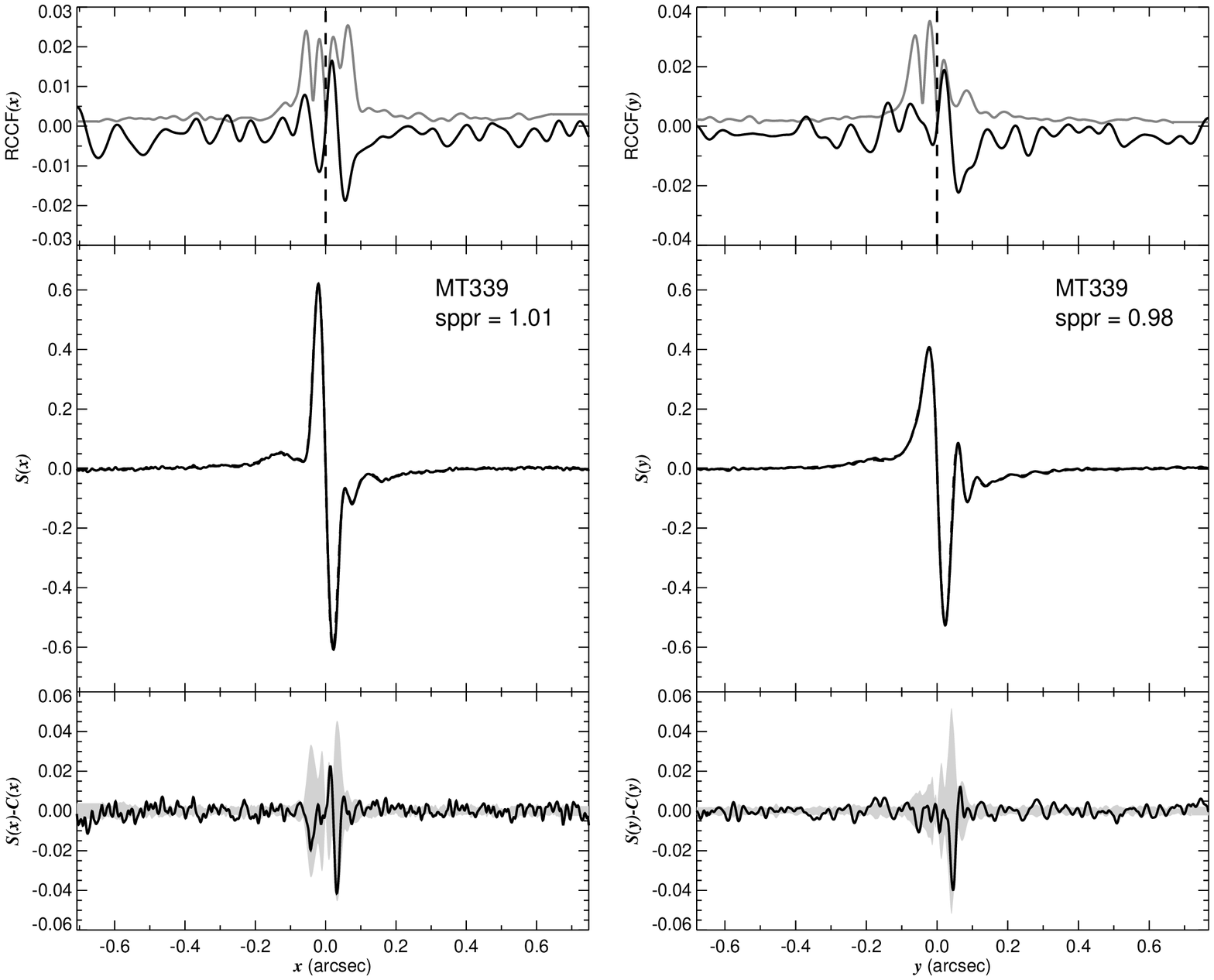}}
\end{center}
\caption{The final $S$-curves for the $x$ and $y$ orthogonal scans for MT 339 
in the same format as Fig.\ 1.1.\label{figmt339}}
\end{figure}

\clearpage
\setcounter{figure}{0}
\renewcommand{\thefigure}{\arabic{figure}.21}
\begin{figure}
\begin{center}
{\includegraphics[angle=90, width=17.5cm]{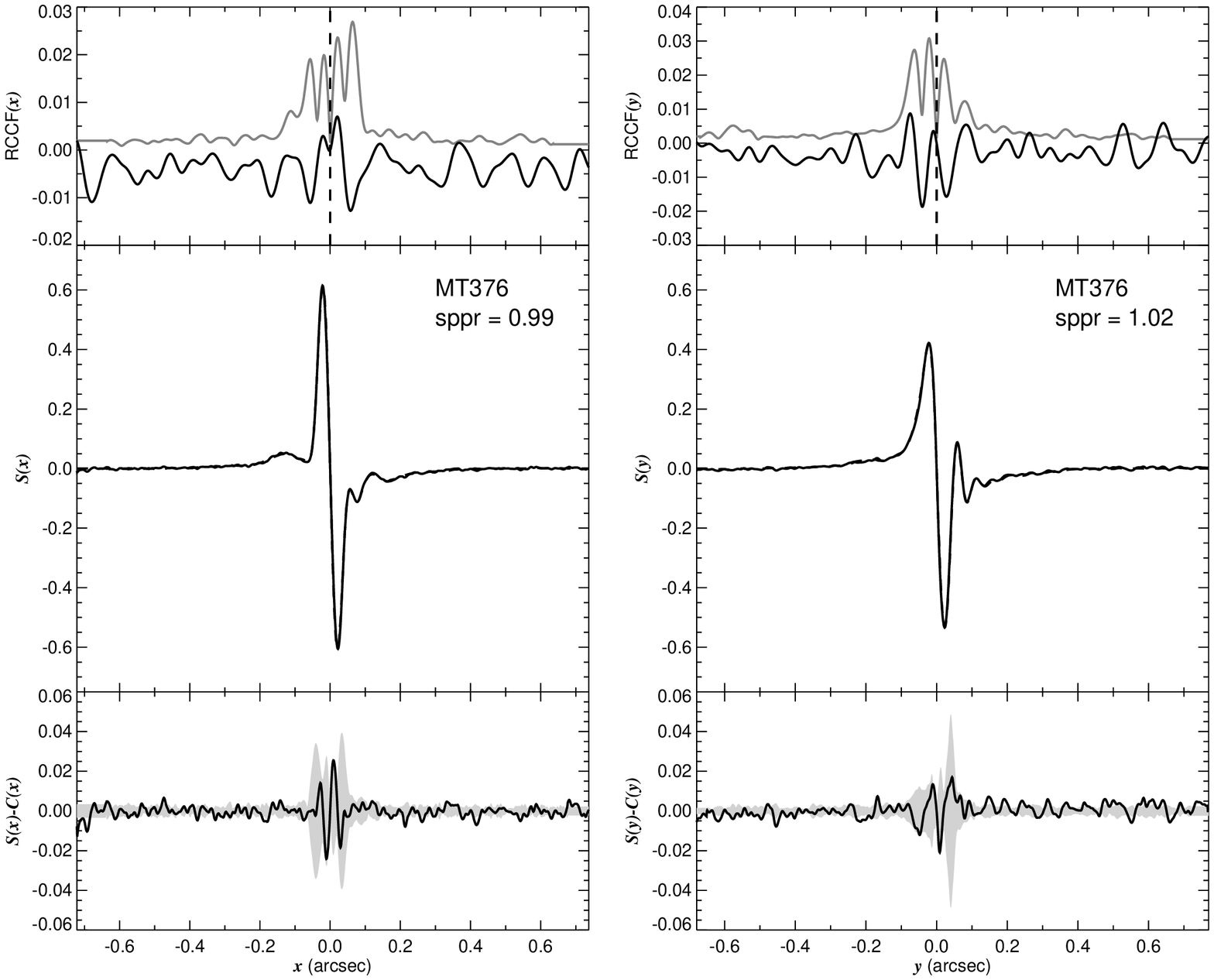}}
\end{center}
\caption{The final $S$-curves for the $x$ and $y$ orthogonal scans for MT 376 
in the same format as Fig.\ 1.1.\label{figmt376}}
\end{figure}

\clearpage
\setcounter{figure}{0}
\renewcommand{\thefigure}{\arabic{figure}.22}
\begin{figure}
\begin{center}
{\includegraphics[angle=90, width=17.5cm]{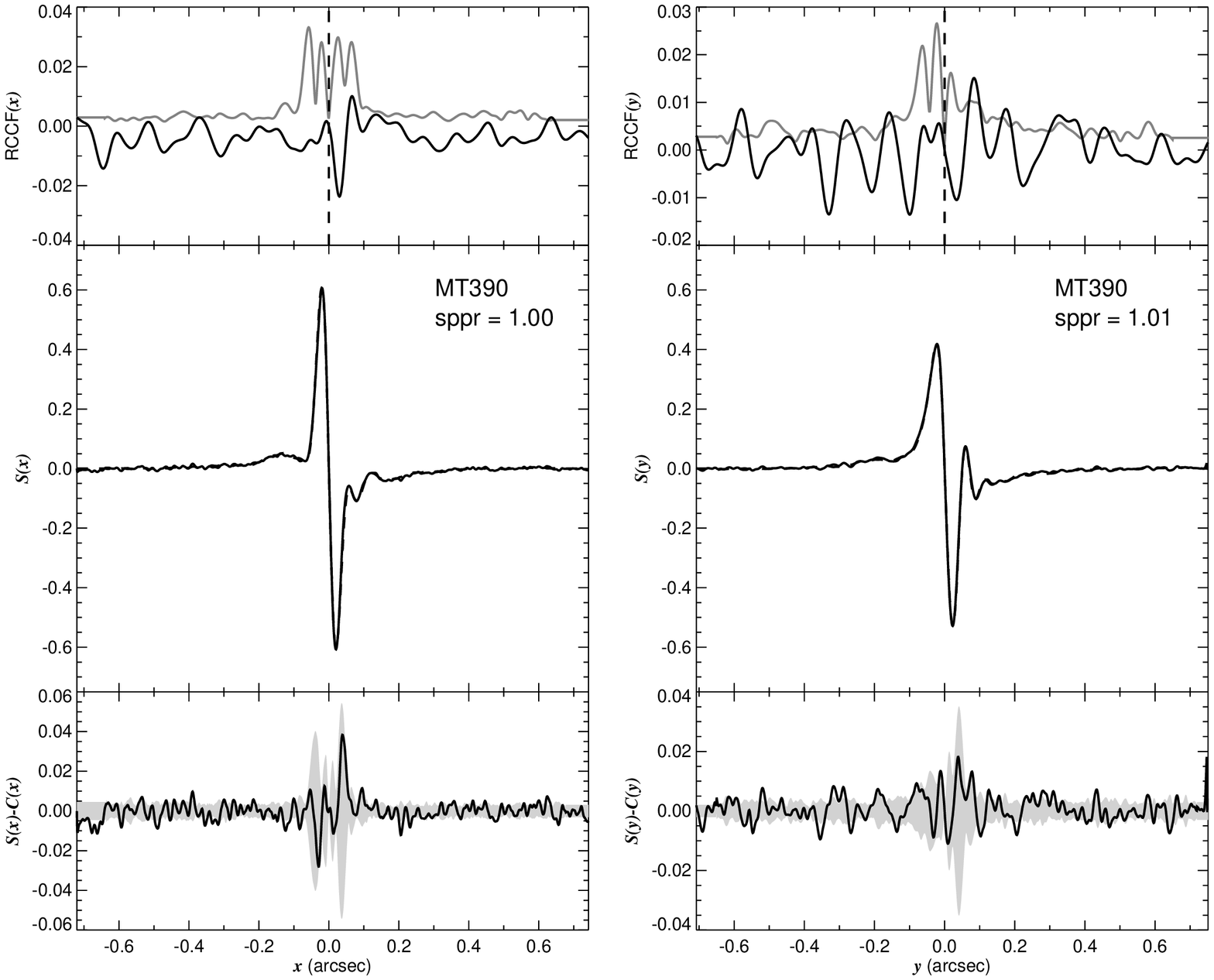}}
\end{center}
\caption{The final $S$-curves for the $x$ and $y$ orthogonal scans for MT 390 
in the same format as Fig.\ 1.1.\label{figmt390}}
\end{figure}

\clearpage
\setcounter{figure}{0}
\renewcommand{\thefigure}{\arabic{figure}.23}
\begin{figure}
\begin{center}
{\includegraphics[angle=90, width=17.5cm]{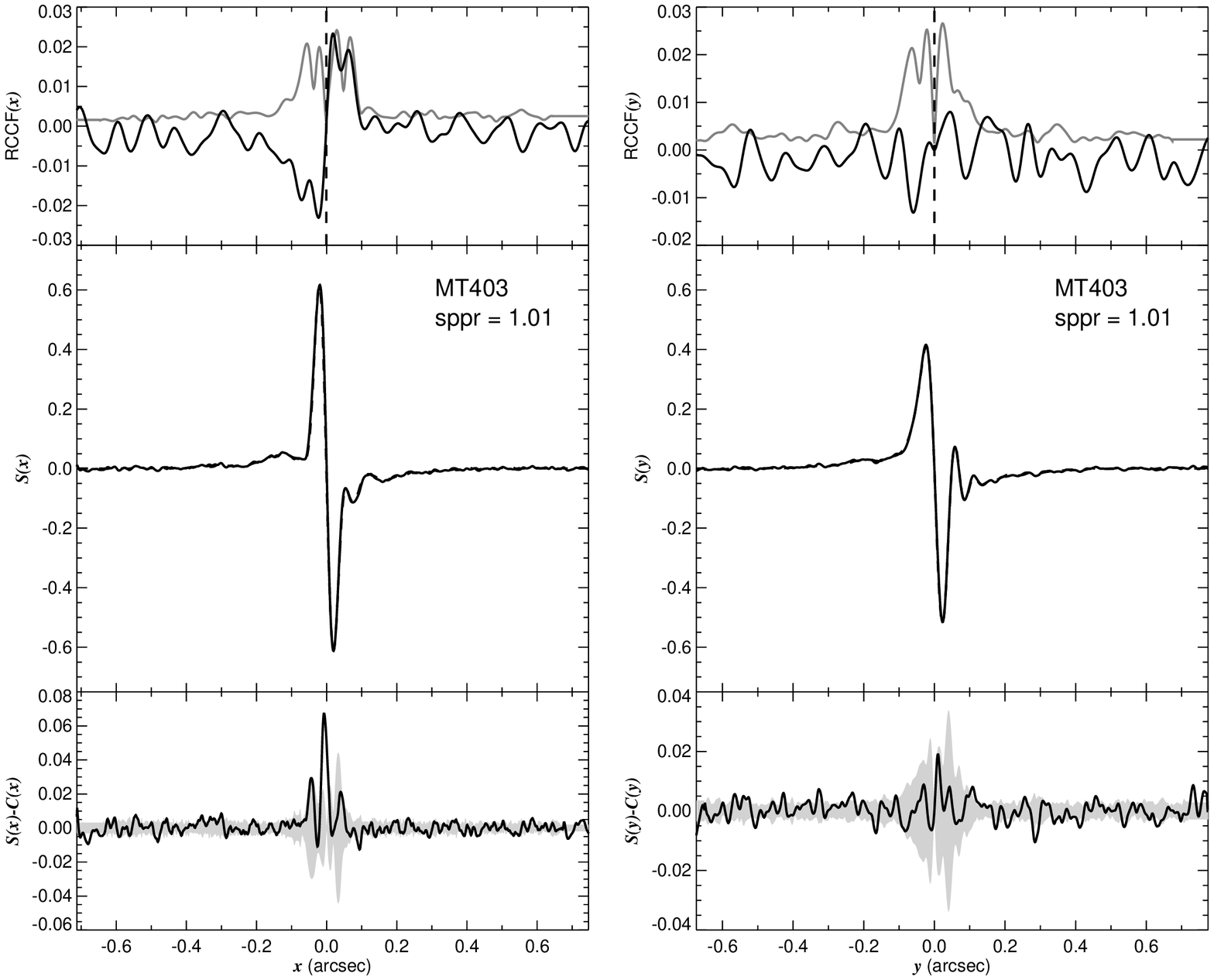}}
\end{center}
\caption{The final $S$-curves for the $x$ and $y$ orthogonal scans for MT 403
in the same format as Fig.\ 1.1.\label{figmt403}}
\end{figure}

\clearpage
\setcounter{figure}{0}
\renewcommand{\thefigure}{\arabic{figure}.24}
\begin{figure}
\begin{center}
{\includegraphics[angle=90, width=17.5cm]{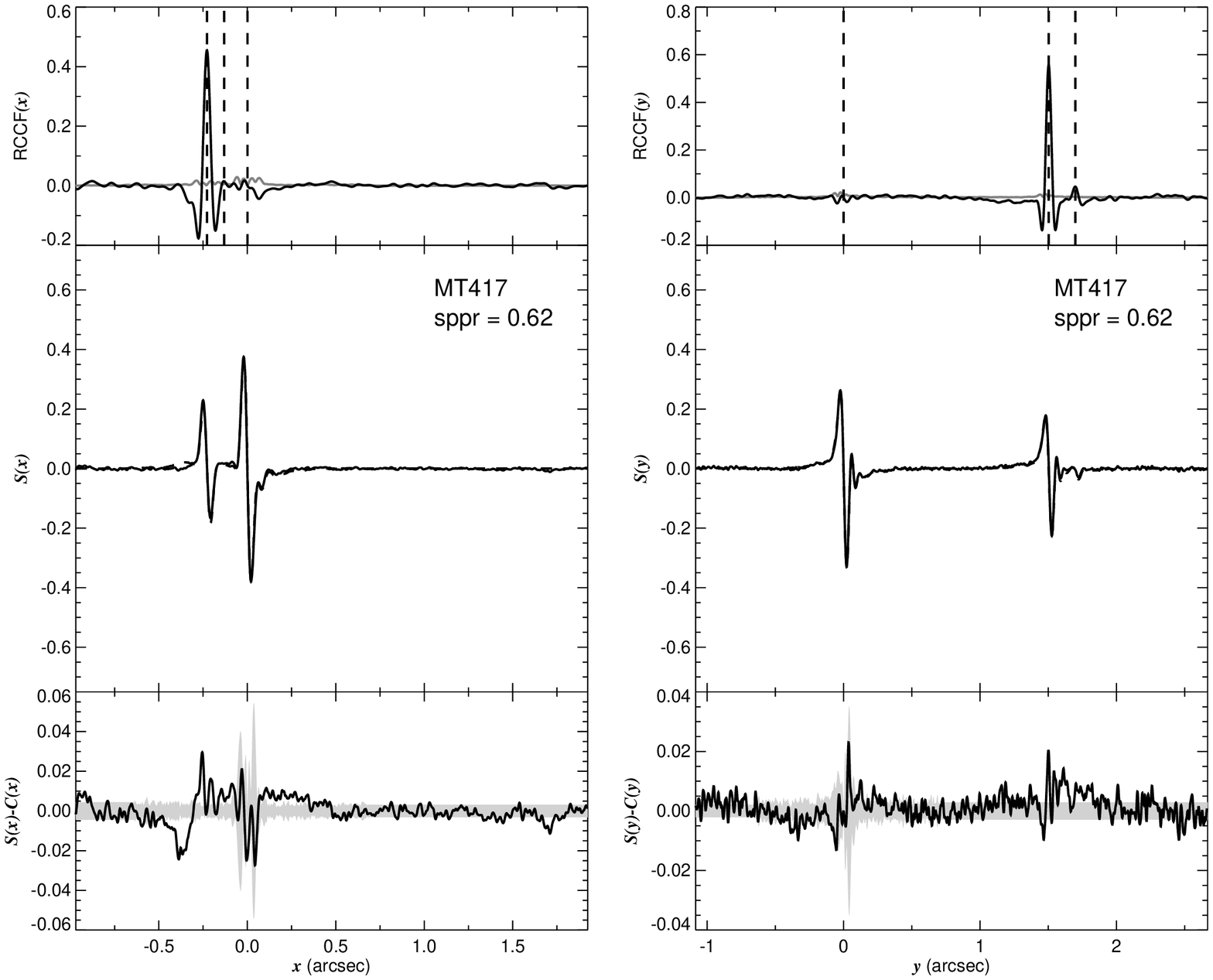}}
\end{center}
\caption{The final $S$-curves for the $x$ and $y$ orthogonal scans for MT 417 
in the same format as Fig.\ 1.1.\label{figmt417}}
\end{figure}

\clearpage
\setcounter{figure}{0}
\renewcommand{\thefigure}{\arabic{figure}.25}
\begin{figure}
\begin{center}
{\includegraphics[angle=90, width=17.5cm]{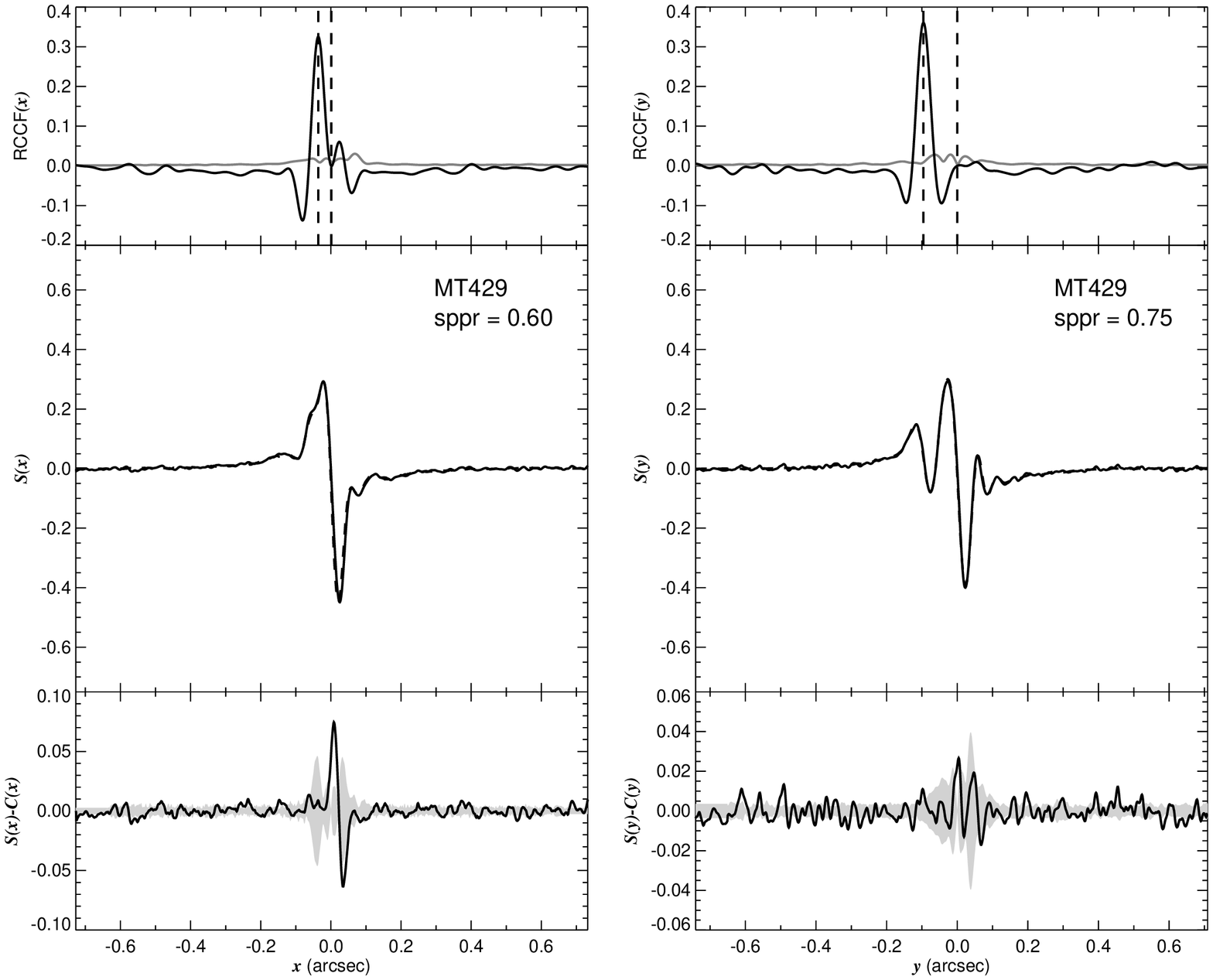}}
\end{center}
\caption{The final $S$-curves for the $x$ and $y$ orthogonal scans for MT 429 
in the same format as Fig.\ 1.1.\label{figmt429}}
\end{figure}

\clearpage
\setcounter{figure}{0}
\renewcommand{\thefigure}{\arabic{figure}.26}
\begin{figure}
\begin{center}
{\includegraphics[angle=90, width=17.5cm]{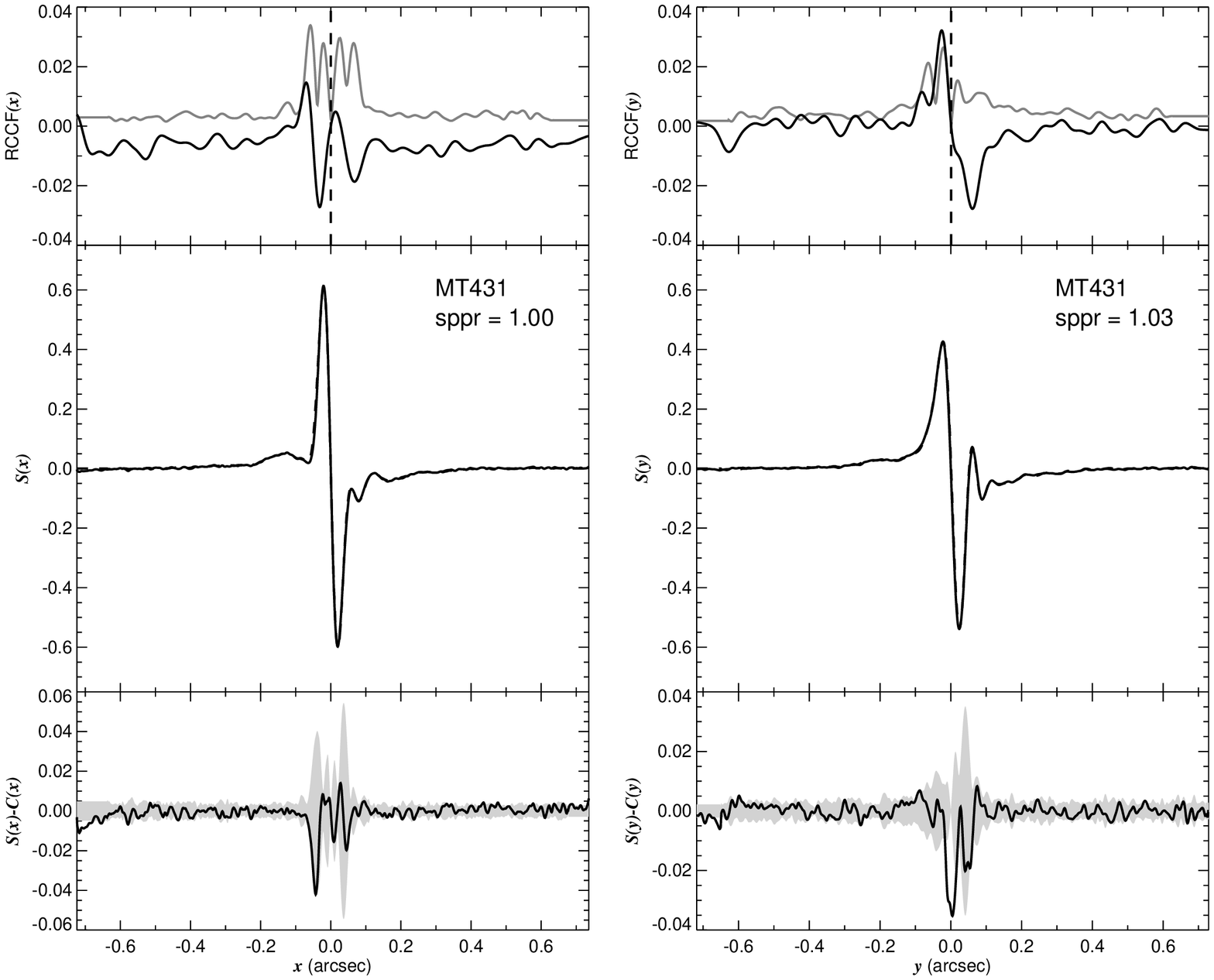}}
\end{center}
\caption{The final $S$-curves for the $x$ and $y$ orthogonal scans for MT 431
in the same format as Fig.\ 1.1.\label{figmt431}}
\end{figure}

\clearpage
\setcounter{figure}{0}
\renewcommand{\thefigure}{\arabic{figure}.27}
\begin{figure}
\begin{center}
{\includegraphics[angle=90, width=17.5cm]{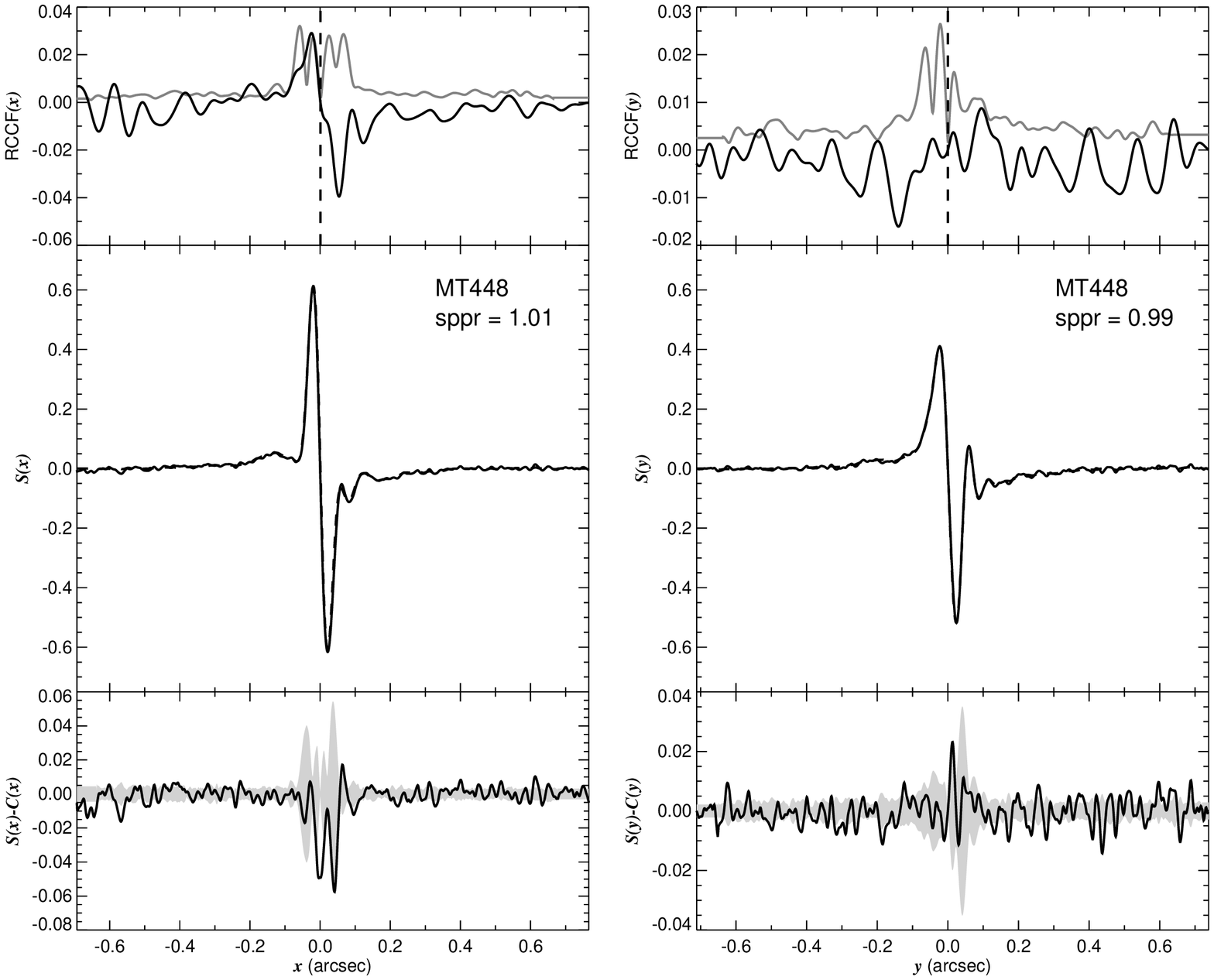}}
\end{center}
\caption{The final $S$-curves for the $x$ and $y$ orthogonal scans for MT448
in the same format as Fig.\ 1.1.\label{figmt448}}
\end{figure}

\clearpage
\setcounter{figure}{0}
\renewcommand{\thefigure}{\arabic{figure}.28}
\begin{figure}
\begin{center}
{\includegraphics[angle=90, width=17.5cm]{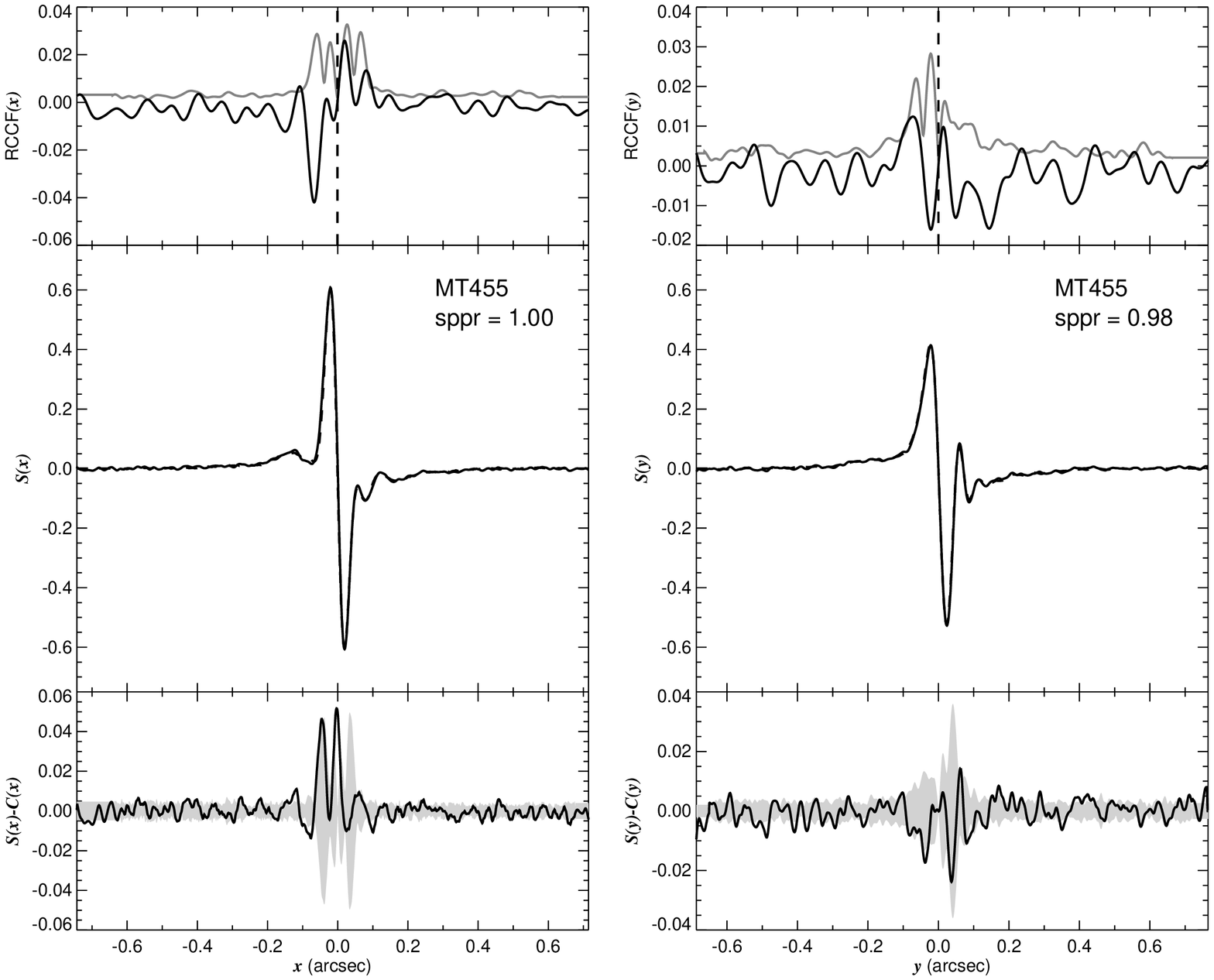}}
\end{center}
\caption{The final $S$-curves for the $x$ and $y$ orthogonal scans for MT 455 
in the same format as Fig.\ 1.1.\label{figmt455}}
\end{figure}

\clearpage
\setcounter{figure}{0}
\renewcommand{\thefigure}{\arabic{figure}.29}
\begin{figure}
\begin{center}
{\includegraphics[angle=90, width=17.5cm]{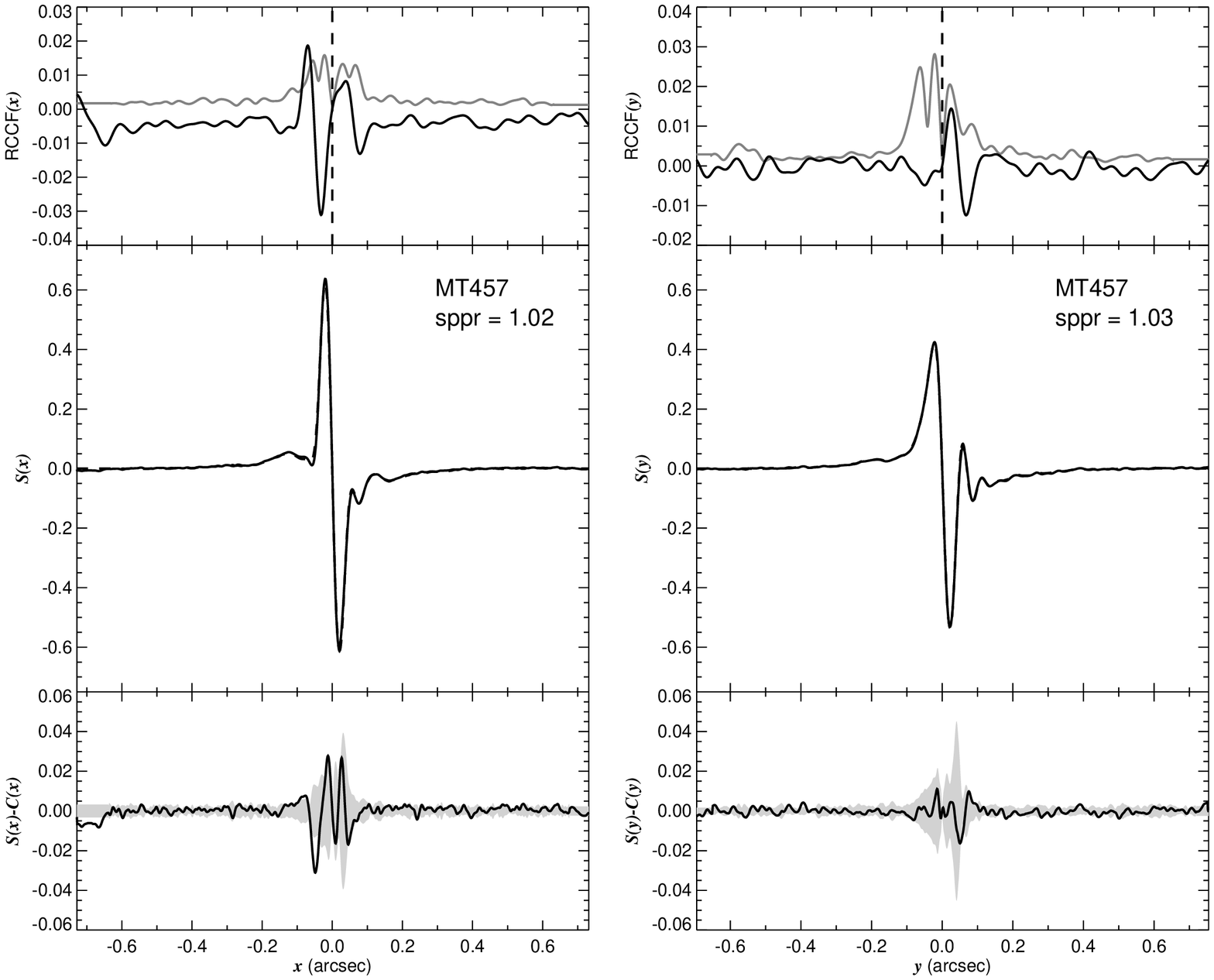}}
\end{center}
\caption{The final $S$-curves for the $x$ and $y$ orthogonal scans for MT 457 
in the same format as Fig.\ 1.1.\label{figmt457}}
\end{figure}

\clearpage
\setcounter{figure}{0}
\renewcommand{\thefigure}{\arabic{figure}.30}
\begin{figure}
\begin{center}
{\includegraphics[angle=90, width=17.5cm]{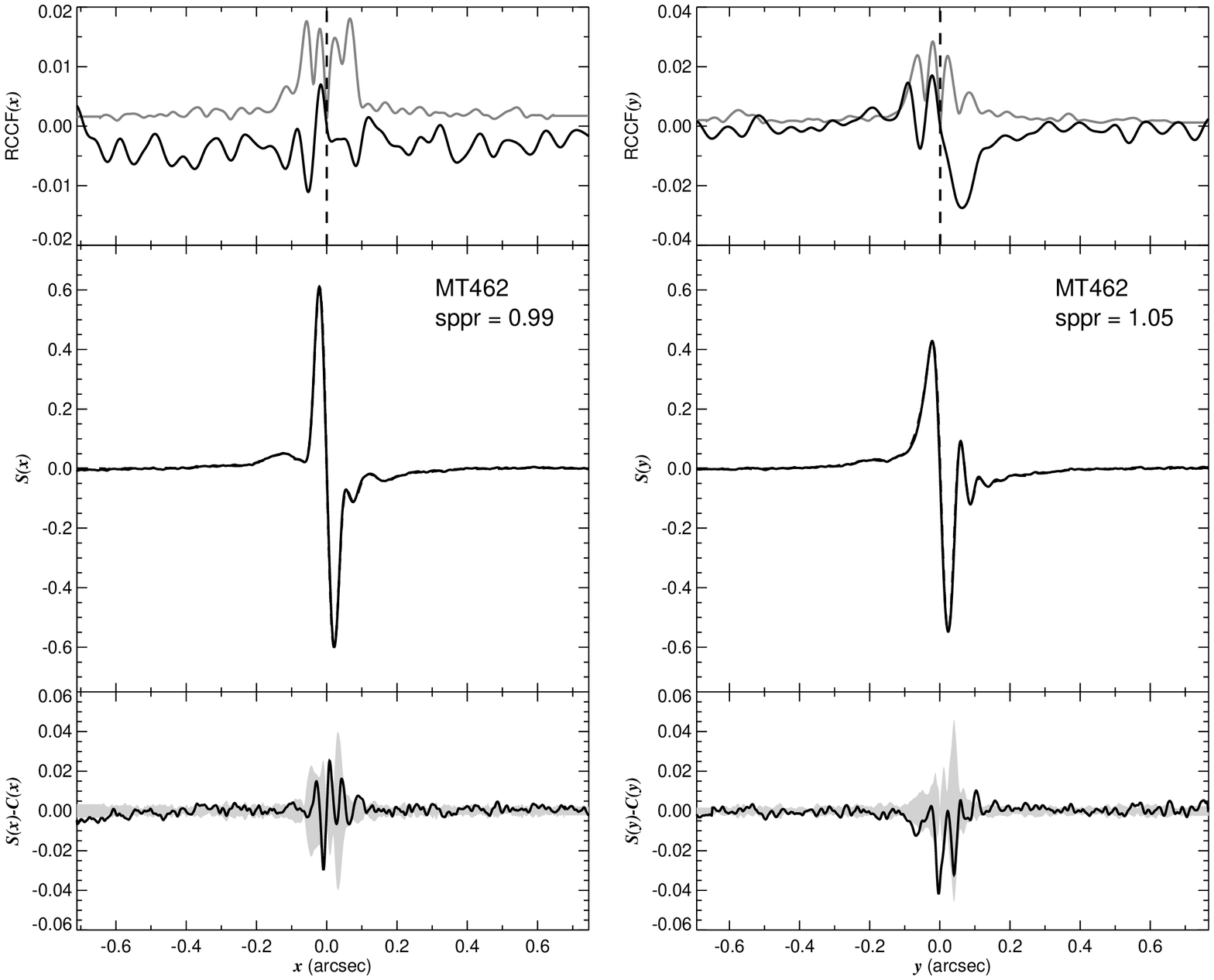}}
\end{center}
\caption{The final $S$-curves for the $x$ and $y$ orthogonal scans for MT 462
in the same format as Fig.\ 1.1.\label{figmt462}}
\end{figure}

\clearpage
\setcounter{figure}{0}
\renewcommand{\thefigure}{\arabic{figure}.31}
\begin{figure}
\begin{center}
{\includegraphics[angle=90, width=17.5cm]{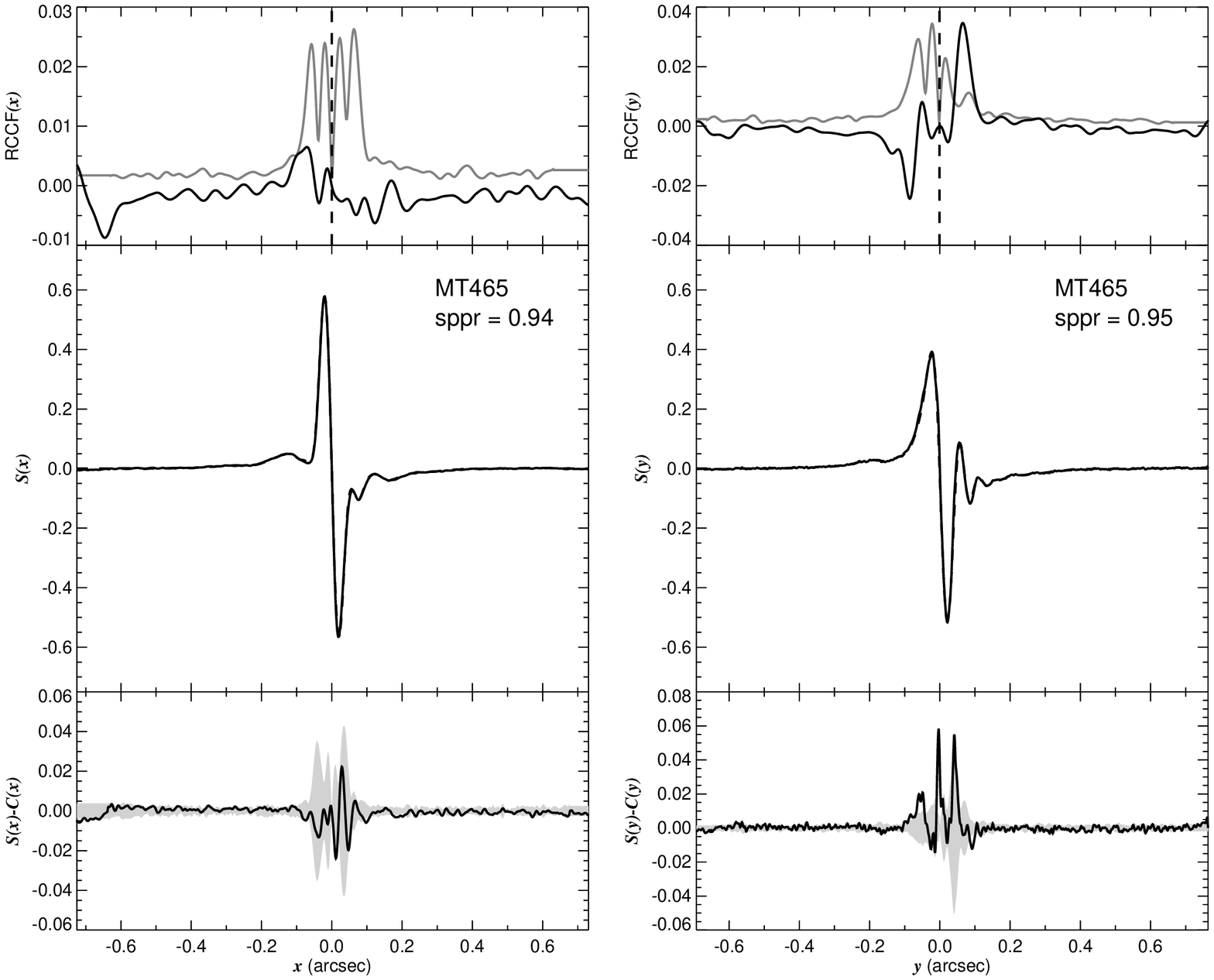}}
\end{center}
\caption{The final $S$-curves for the $x$ and $y$ orthogonal scans for MT 465
in the same format as Fig.\ 1.1.\label{figmt465}}
\end{figure}

\clearpage
\setcounter{figure}{0}
\renewcommand{\thefigure}{\arabic{figure}.32}
\begin{figure}
\begin{center}
{\includegraphics[angle=90, width=17.5cm]{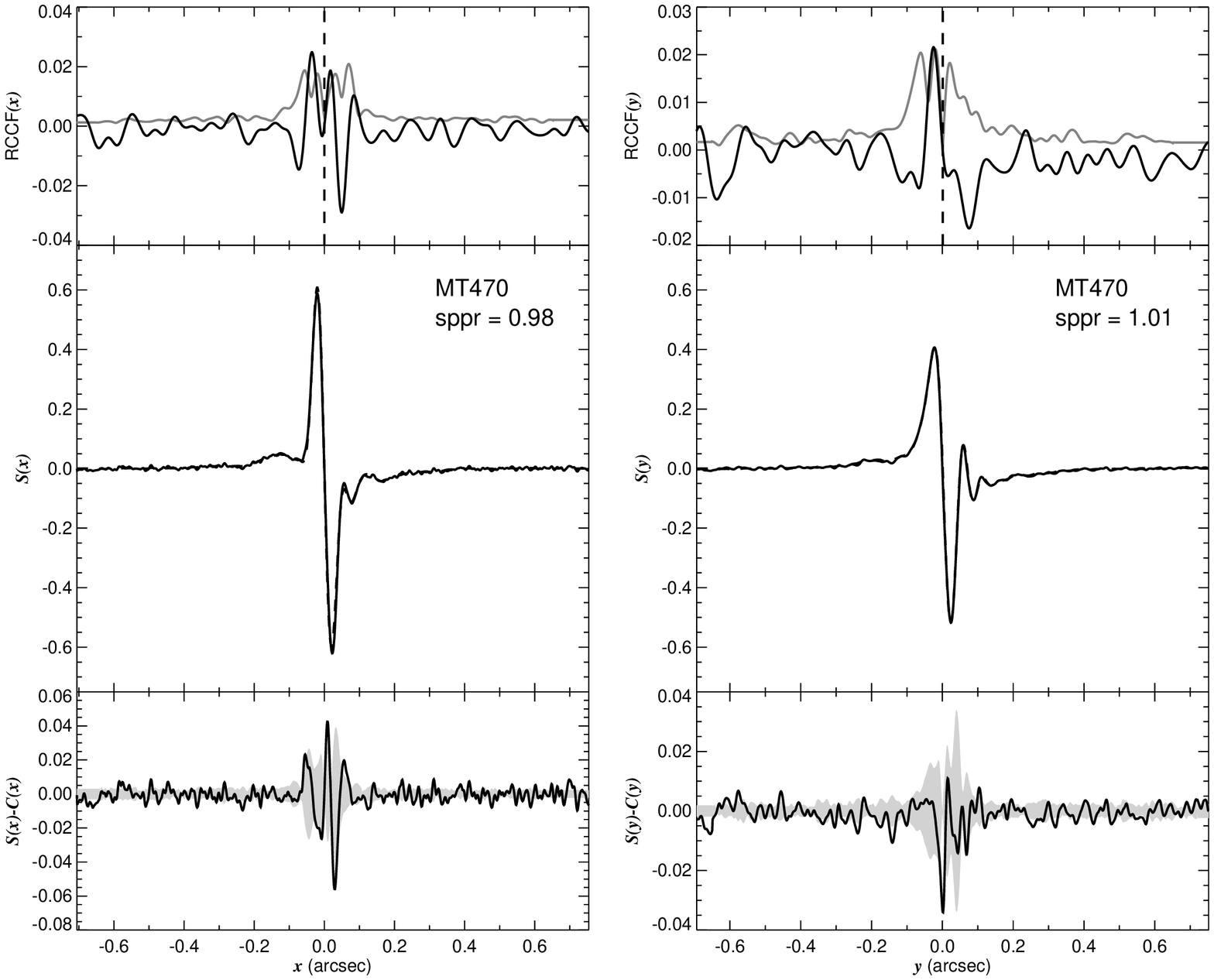}}
\end{center}
\caption{The final $S$-curves for the $x$ and $y$ orthogonal scans for MT 470 
in the same format as Fig.\ 1.1.\label{figmt470}}
\end{figure}

\clearpage
\setcounter{figure}{0}
\renewcommand{\thefigure}{\arabic{figure}.33}
\begin{figure}
\begin{center}
{\includegraphics[angle=90, width=17.5cm]{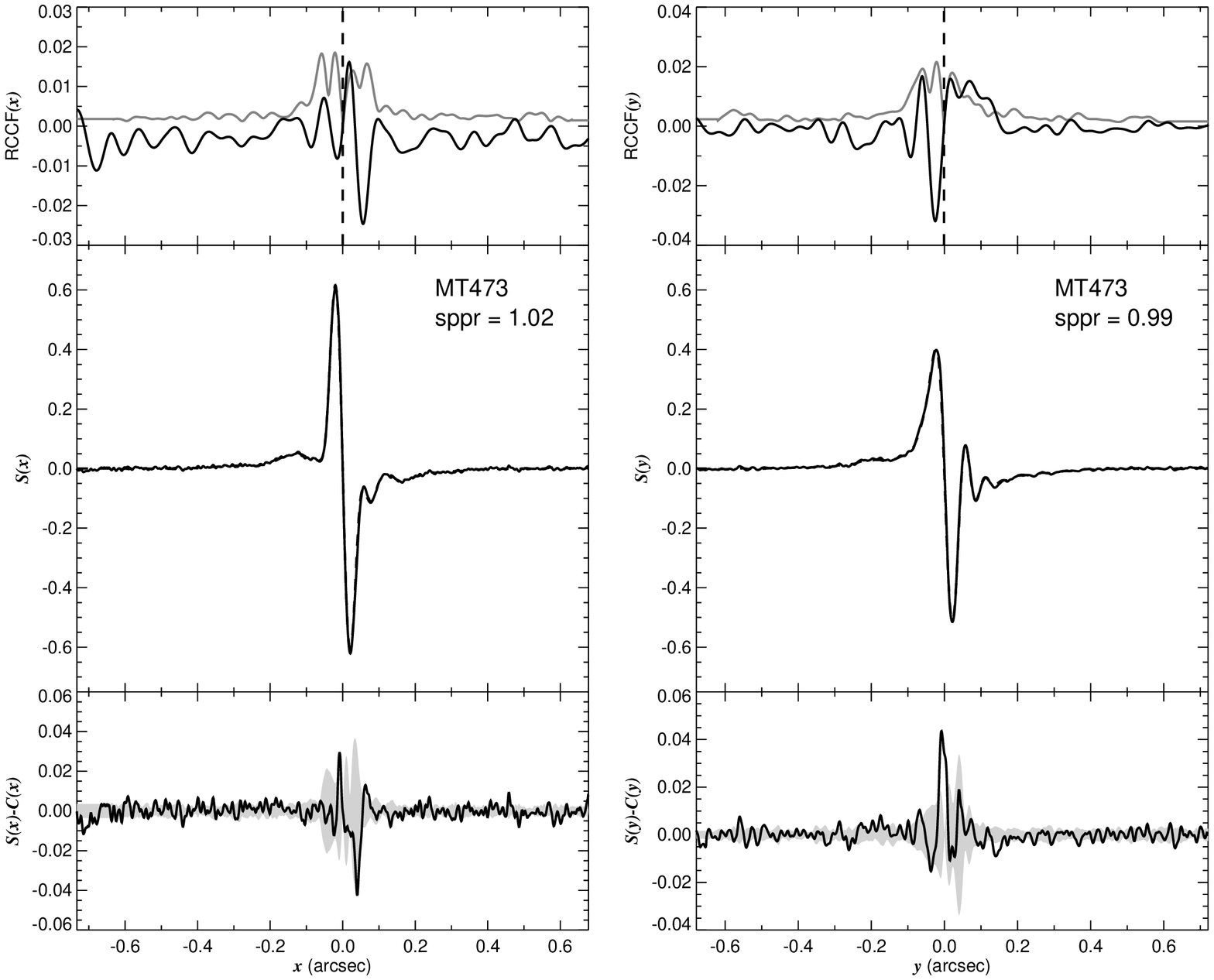}}
\end{center}
\caption{The final $S$-curves for the $x$ and $y$ orthogonal scans for MT 473
in the same format as Fig.\ 1.1.\label{figmt473}}
\end{figure}

\clearpage
\setcounter{figure}{0}
\renewcommand{\thefigure}{\arabic{figure}.34}
\begin{figure}
\begin{center}
{\includegraphics[angle=90, width=17.5cm]{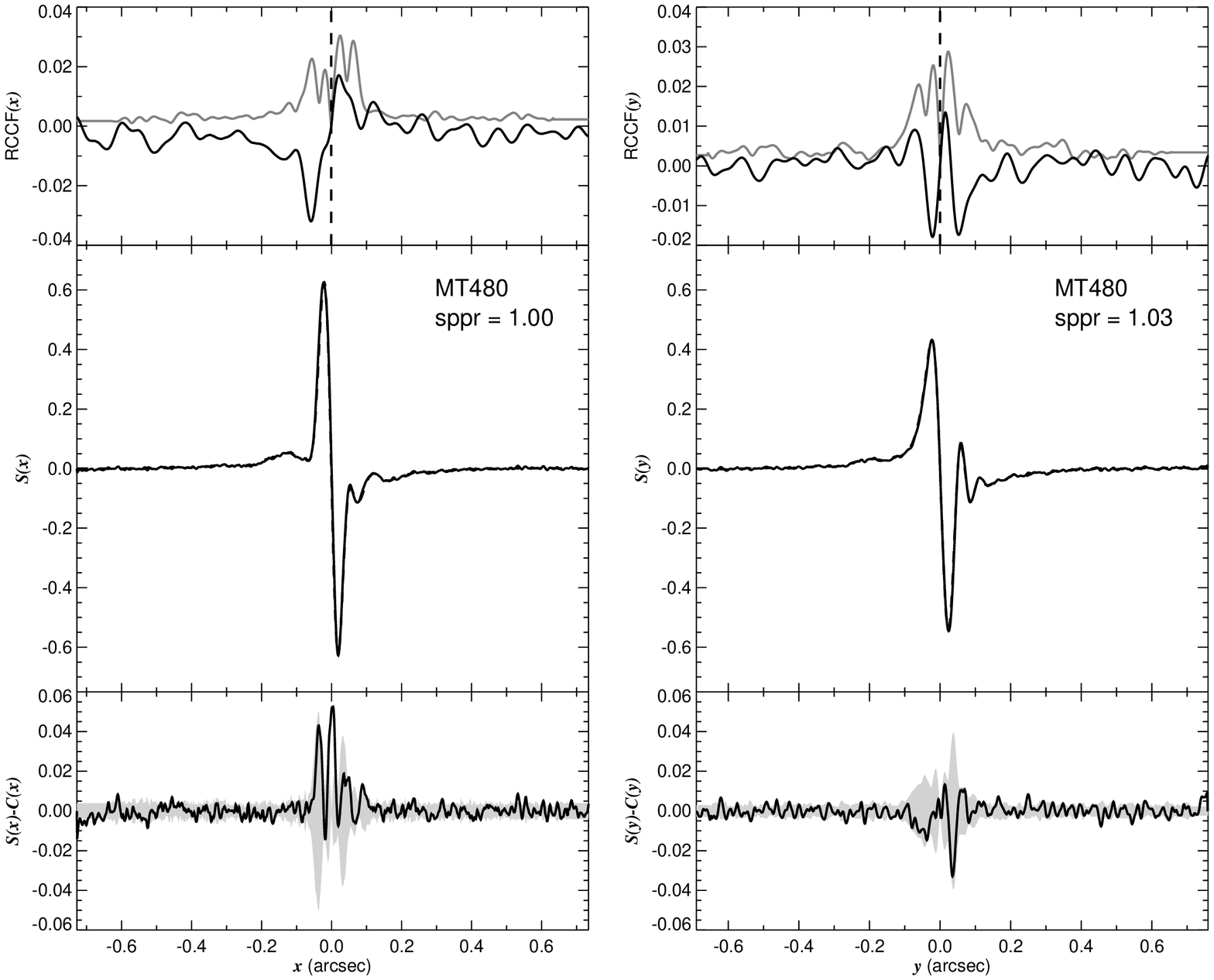}}
\end{center}
\caption{The final $S$-curves for the $x$ and $y$ orthogonal scans for MT 480
in the same format as Fig.\ 1.1.\label{figmt480}}
\end{figure}

\clearpage
\setcounter{figure}{0}
\renewcommand{\thefigure}{\arabic{figure}.35}
\begin{figure}
\begin{center}
{\includegraphics[angle=90, width=17.5cm]{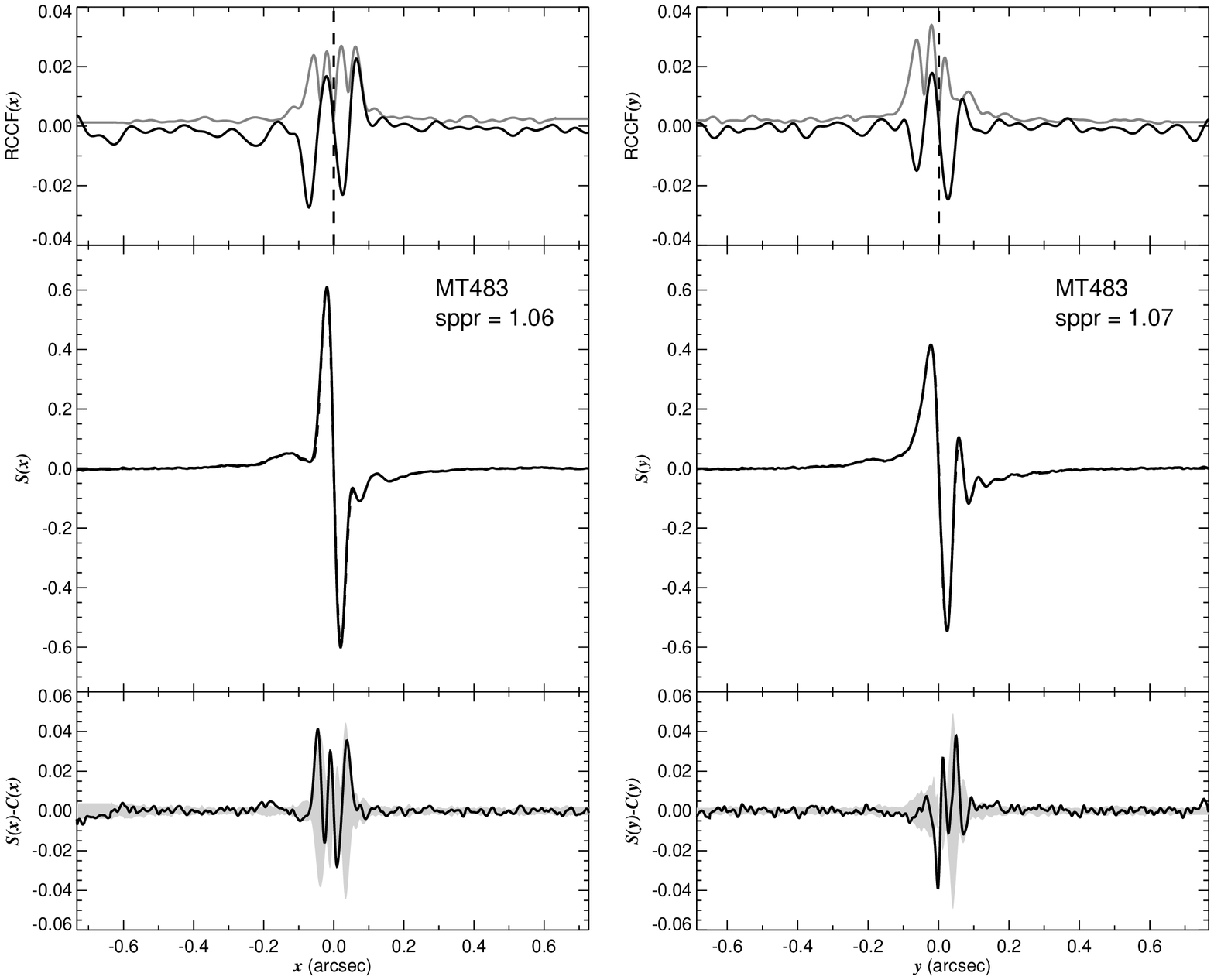}}
\end{center}
\caption{The final $S$-curves for the $x$ and $y$ orthogonal scans for MT 483
in the same format as Fig.\ 1.1.\label{figmt483}}
\end{figure}

\clearpage
\setcounter{figure}{0}
\renewcommand{\thefigure}{\arabic{figure}.36}
\begin{figure}
\begin{center}
{\includegraphics[angle=90, width=17.5cm]{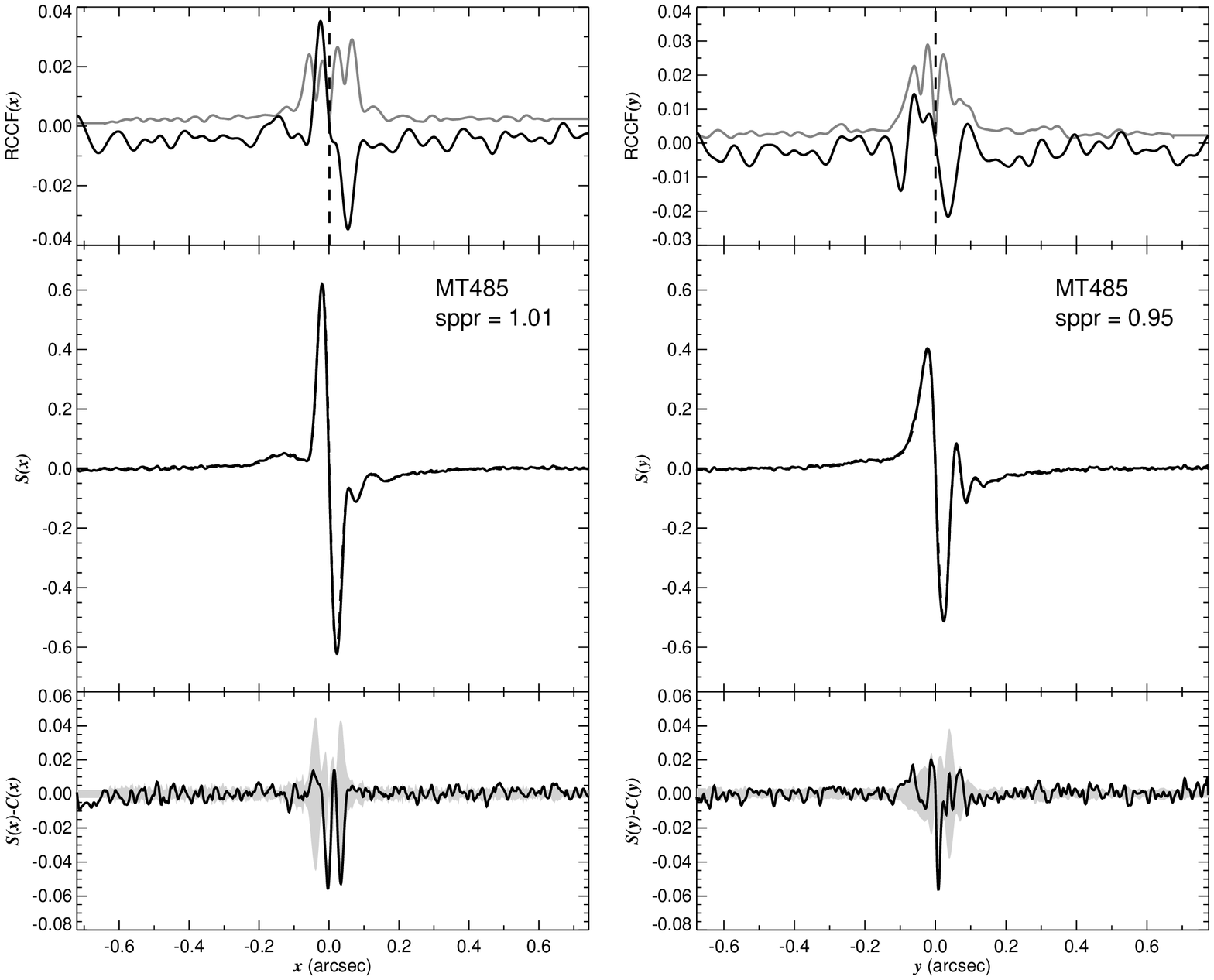}}
\end{center}
\caption{The final $S$-curves for the $x$ and $y$ orthogonal scans for MT 485
in the same format as Fig.\ 1.1.\label{figmt485}}
\end{figure}

\clearpage
\setcounter{figure}{0}
\renewcommand{\thefigure}{\arabic{figure}.37}
\begin{figure}
\begin{center}
{\includegraphics[angle=90, width=17.5cm]{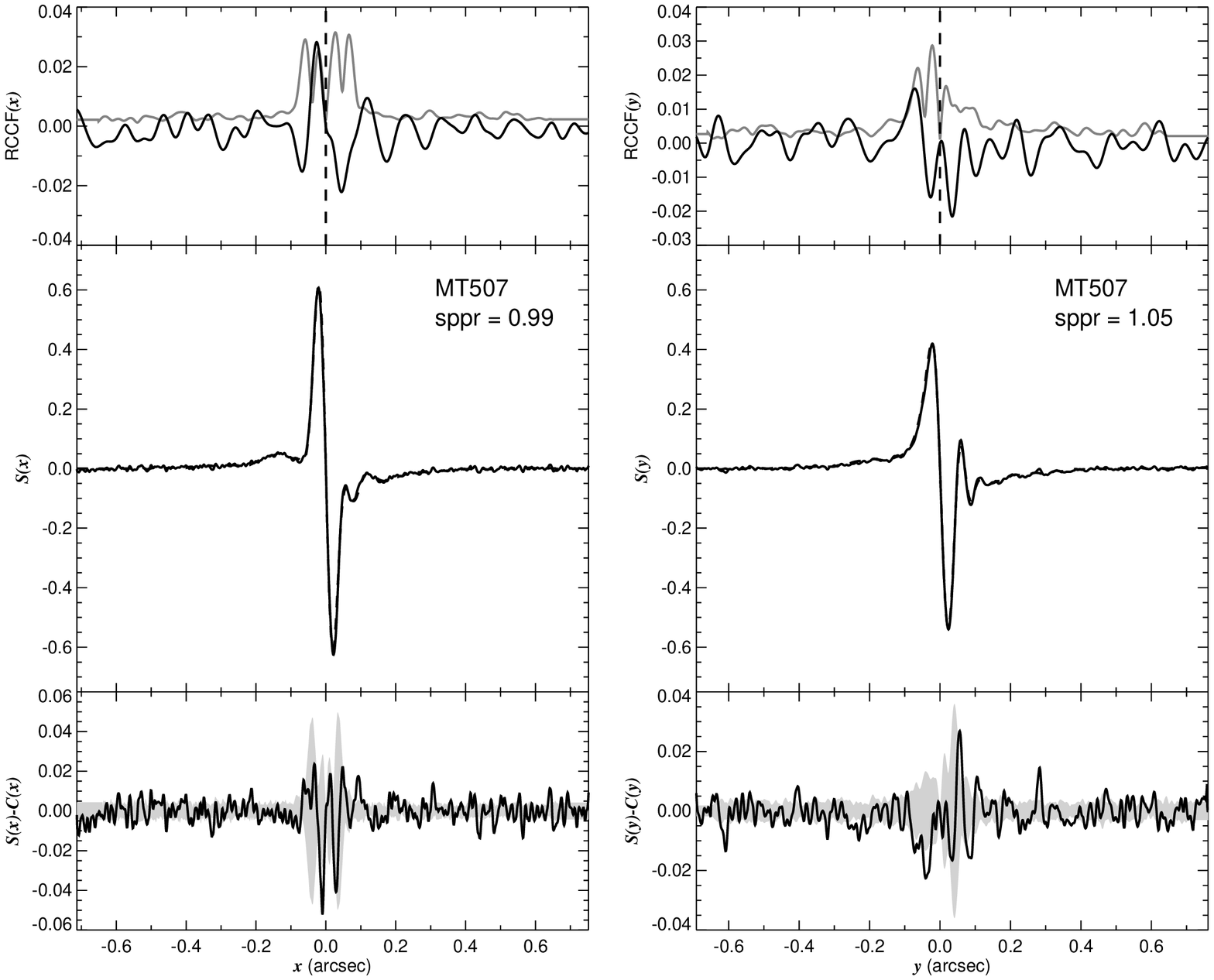}}
\end{center}
\caption{The final $S$-curves for the $x$ and $y$ orthogonal scans for MT 507 
in the same format as Fig.\ 1.1.\label{figmt507}}
\end{figure}

\clearpage
\setcounter{figure}{0}
\renewcommand{\thefigure}{\arabic{figure}.38}
\begin{figure}
\begin{center}
{\includegraphics[angle=90, width=17.5cm]{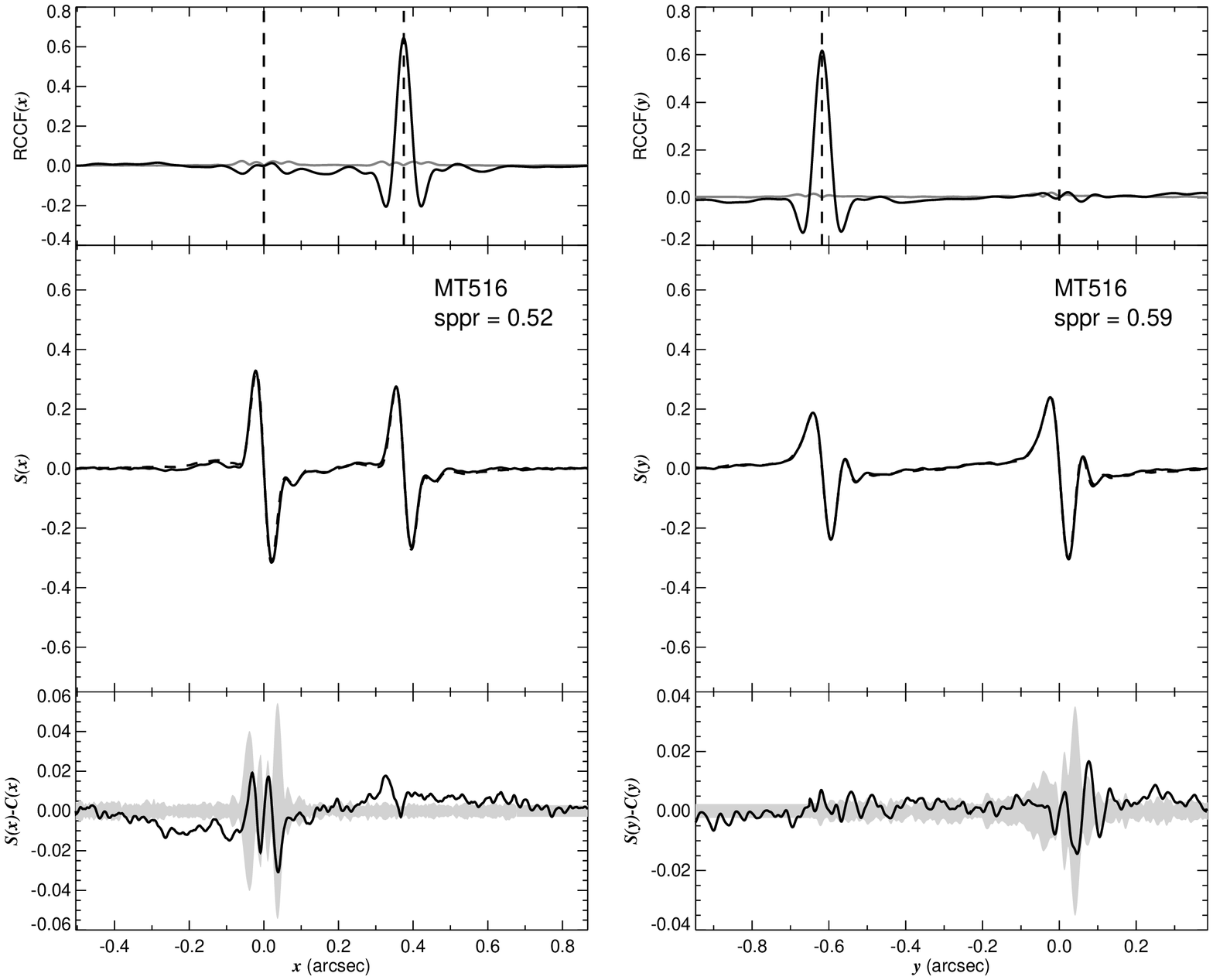}}
\end{center}
\caption{The final $S$-curves for the $x$ and $y$ orthogonal scans for MT516
in the same format as Fig.\ 1.1.\label{figmt516}}
\end{figure}

\clearpage
\setcounter{figure}{0}
\renewcommand{\thefigure}{\arabic{figure}.39}
\begin{figure}
\begin{center}
{\includegraphics[angle=90, width=17.5cm]{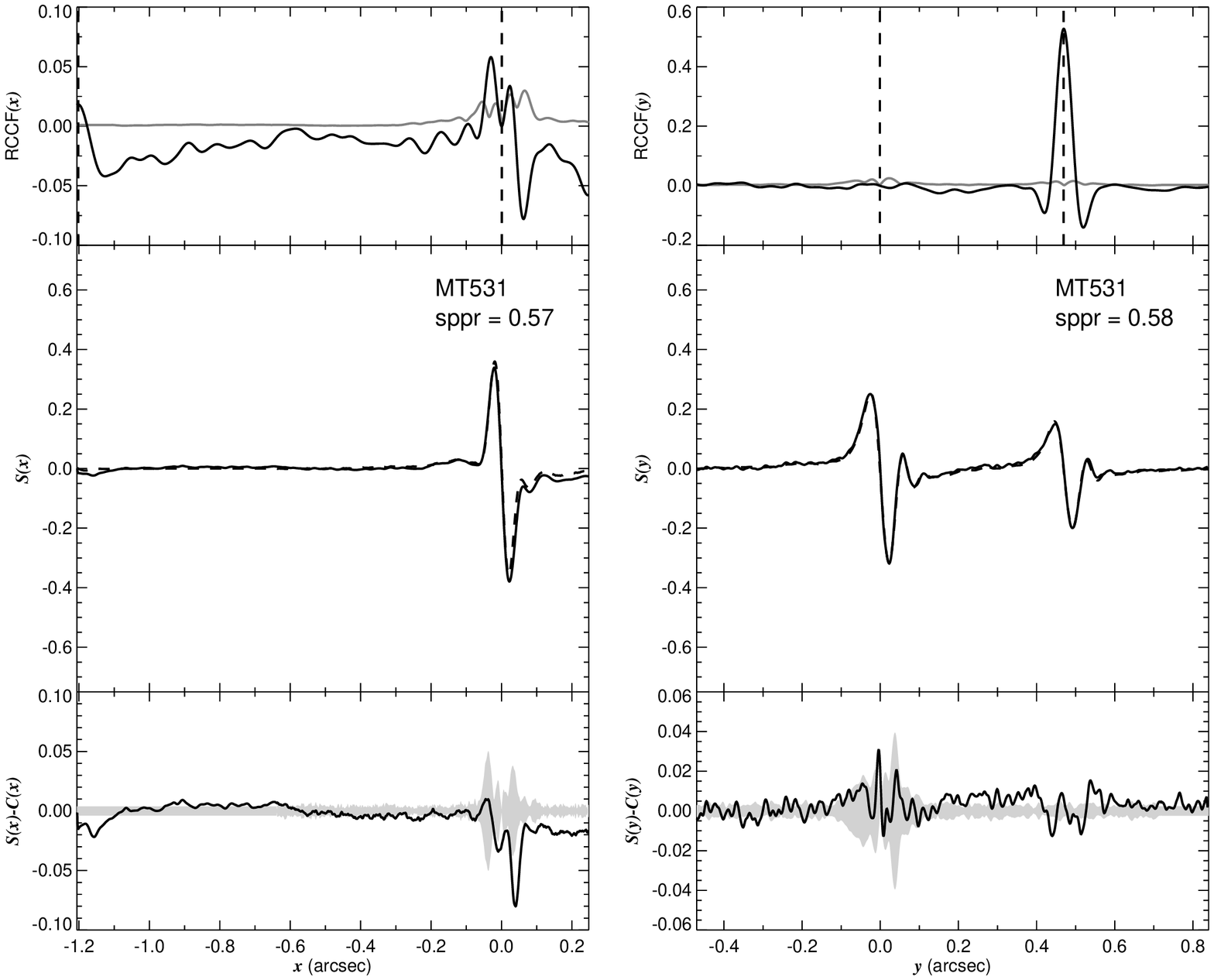}}
\end{center}
\caption{The final $S$-curves for the $x$ and $y$ orthogonal scans for MT 531
in the same format as Fig.\ 1.1.\label{figmt531}}
\end{figure}

\clearpage
\setcounter{figure}{0}
\renewcommand{\thefigure}{\arabic{figure}.40}
\begin{figure}
\begin{center}
{\includegraphics[angle=90, width=17.5cm]{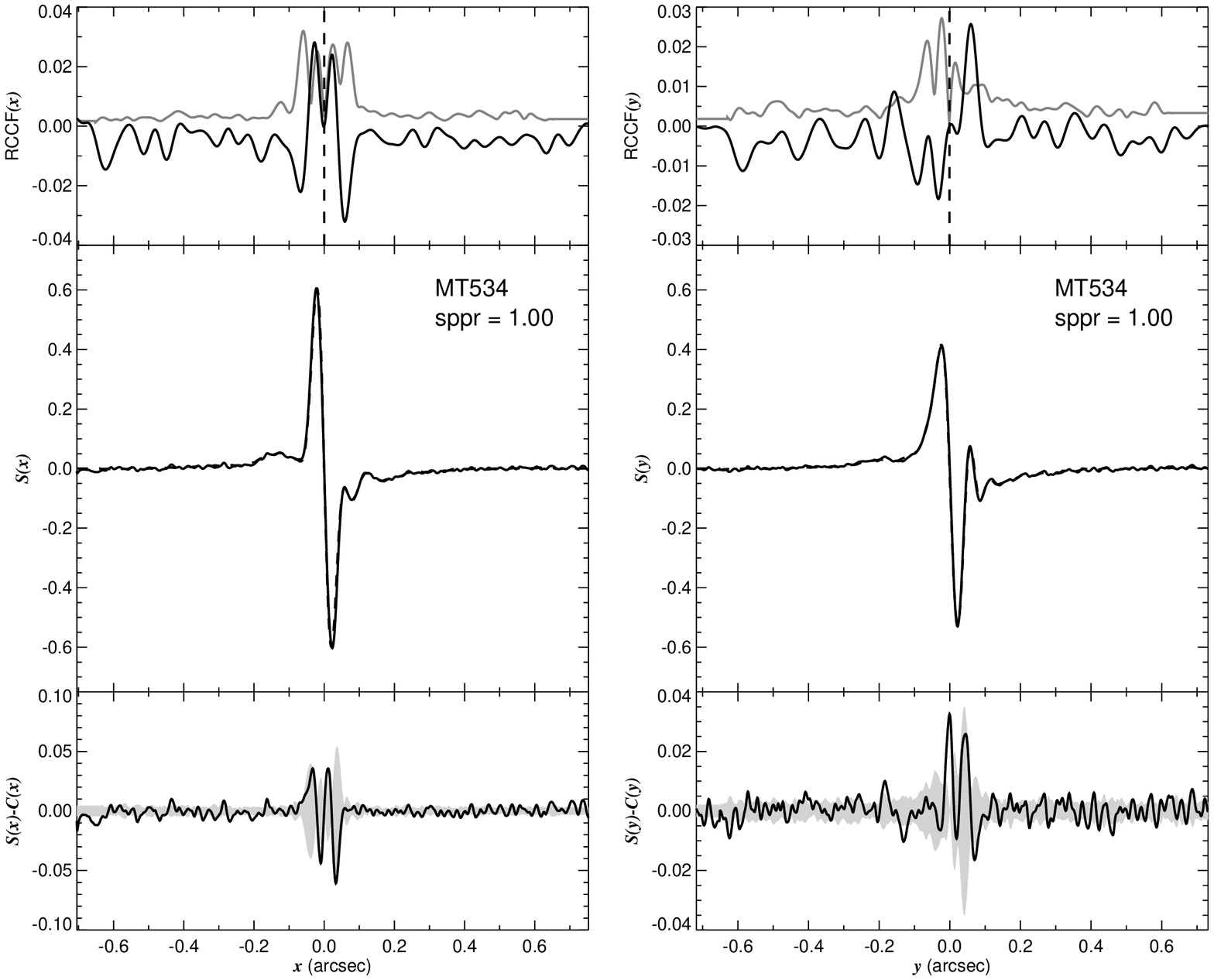}}
\end{center}
\caption{The final $S$-curves for the $x$ and $y$ orthogonal scans for MT534 
in the same format as Fig.\ 1.1.\label{figmt534}}
\end{figure}

\clearpage
\setcounter{figure}{0}
\renewcommand{\thefigure}{\arabic{figure}.41}
\begin{figure}
\begin{center}
{\includegraphics[angle=90, width=17.5cm]{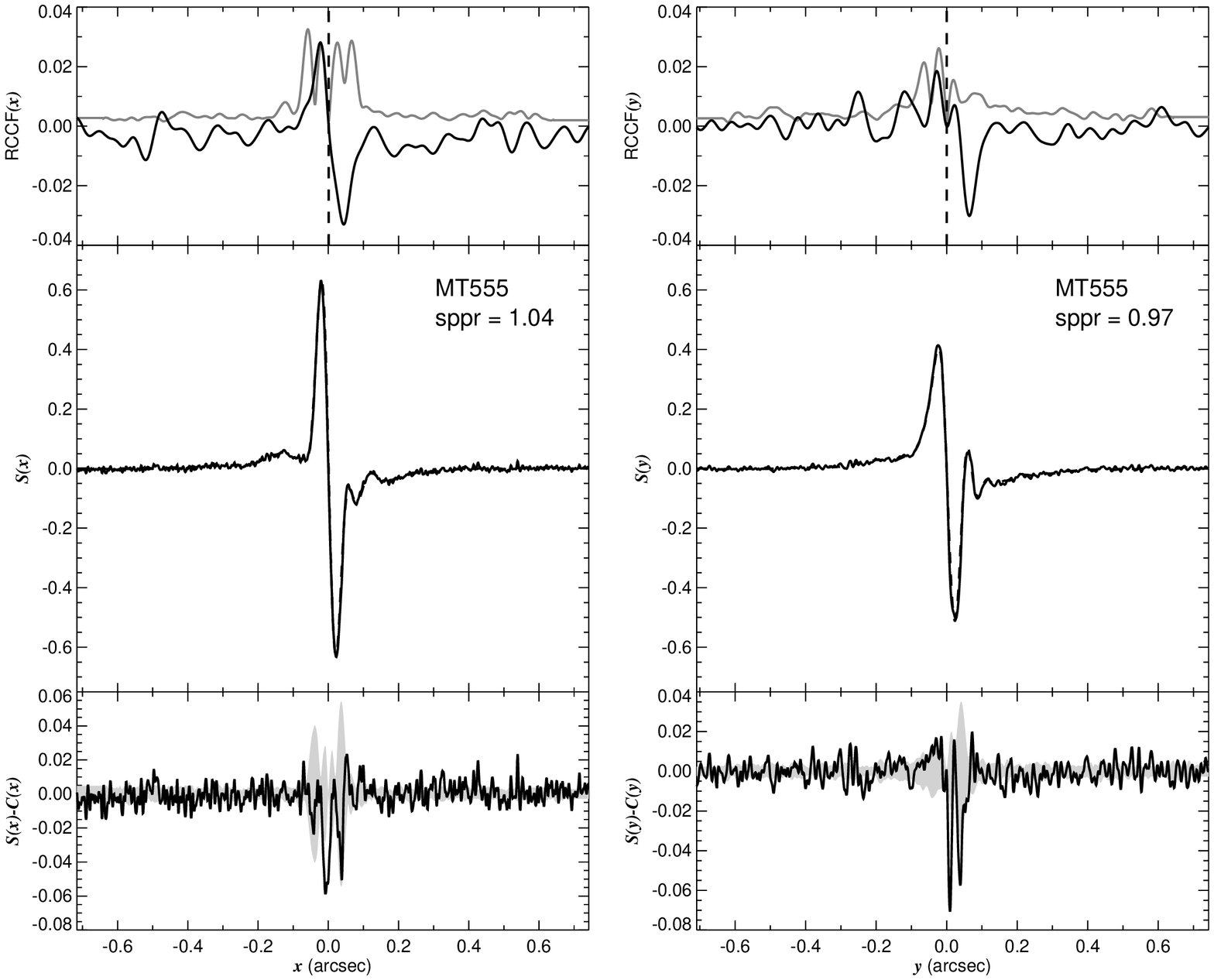}}
\end{center}
\caption{The final $S$-curves for the $x$ and $y$ orthogonal scans for MT 555 
in the same format as Fig.\ 1.1.\label{figmt555}}
\end{figure}

\clearpage
\setcounter{figure}{0}
\renewcommand{\thefigure}{\arabic{figure}.42}
\begin{figure}
\begin{center}
{\includegraphics[angle=90, width=17.5cm]{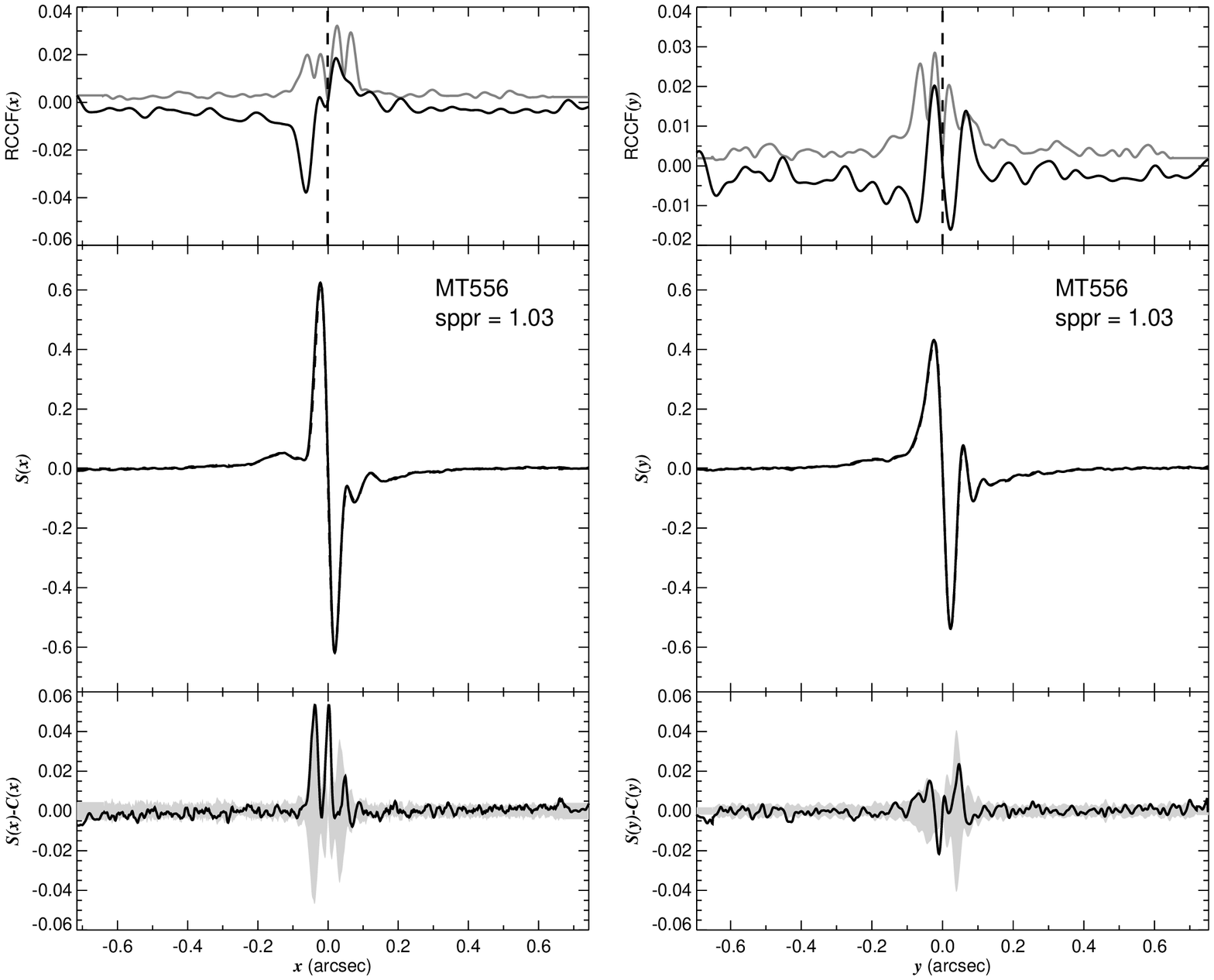}}
\end{center}
\caption{The final $S$-curves for the $x$ and $y$ orthogonal scans for MT 556 
in the same format as Fig.\ 1.1.\label{figmt556}}
\end{figure}

\clearpage
\setcounter{figure}{0}
\renewcommand{\thefigure}{\arabic{figure}.43}
\begin{figure}
\begin{center}
{\includegraphics[angle=90, width=17.5cm]{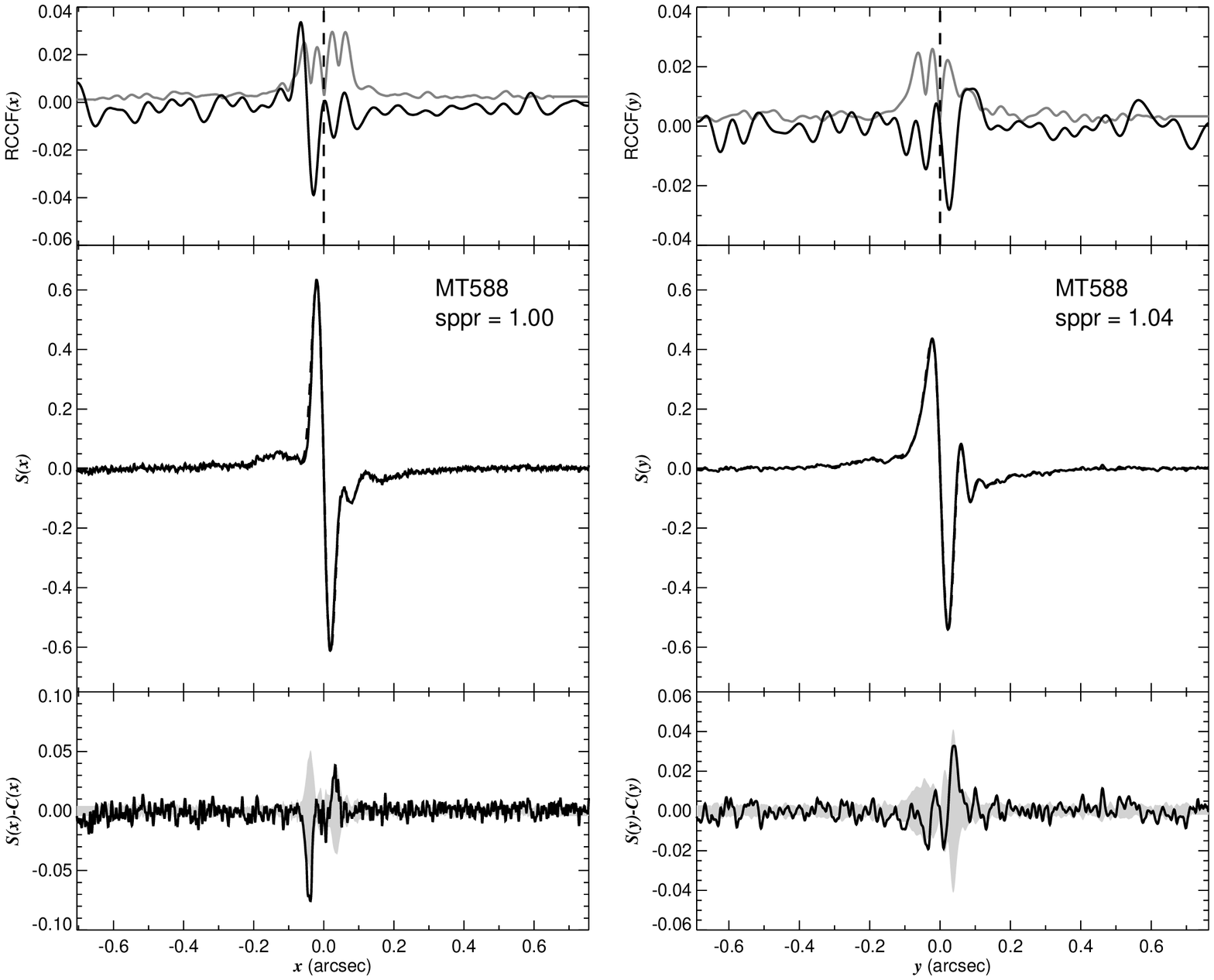}}
\end{center}
\caption{The final $S$-curves for the $x$ and $y$ orthogonal scans for MT 588
in the same format as Fig.\ 1.1.\label{figmt588}}
\end{figure}

\clearpage
\setcounter{figure}{0}
\renewcommand{\thefigure}{\arabic{figure}.44}
\begin{figure}
\begin{center}
{\includegraphics[angle=90, width=17.5cm]{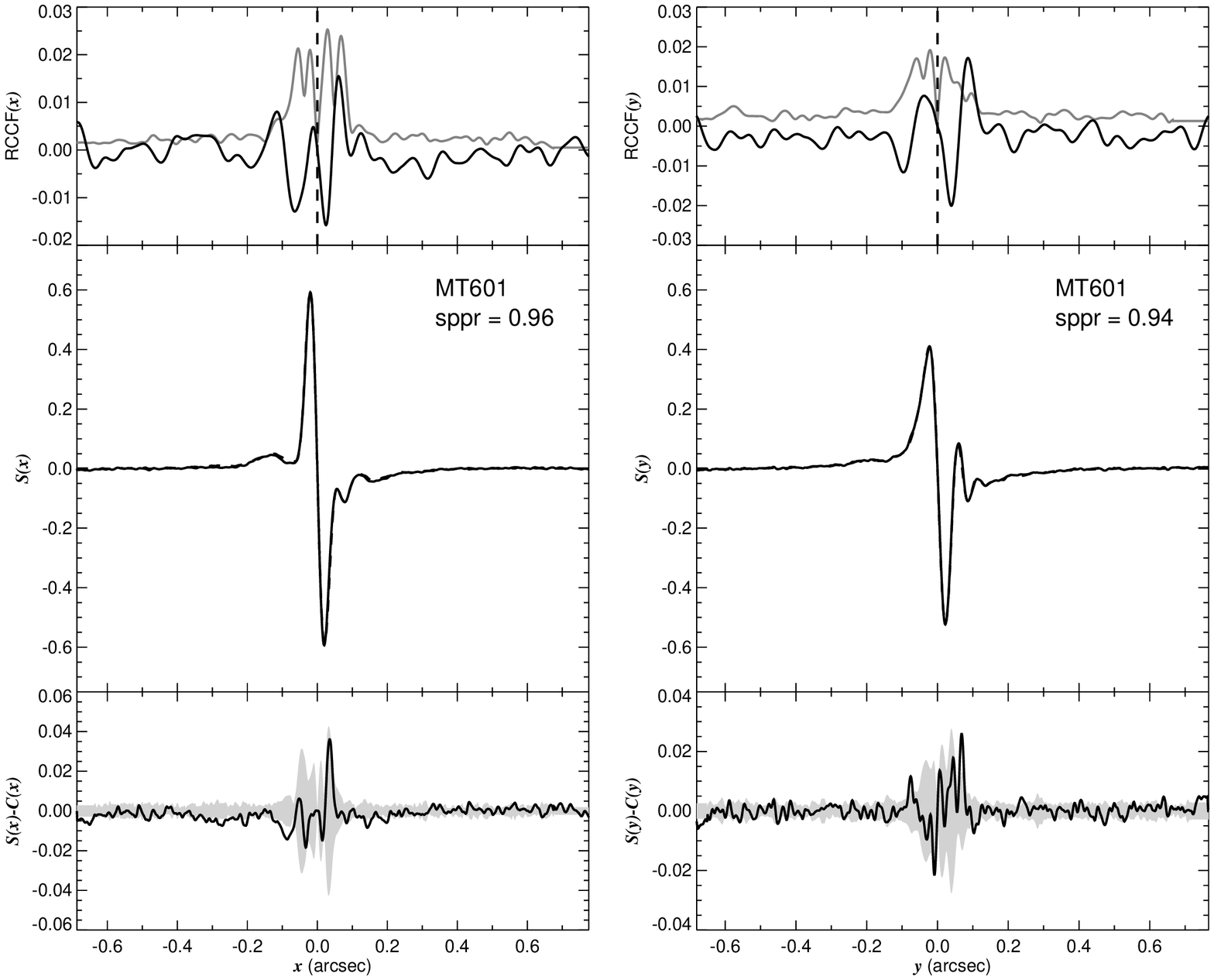}}
\end{center}
\caption{The final $S$-curves for the $x$ and $y$ orthogonal scans for MT 601
in the same format as Fig.\ 1.1.\label{figmt601}}
\end{figure}

\clearpage
\setcounter{figure}{0}
\renewcommand{\thefigure}{\arabic{figure}.45}
\begin{figure}
\begin{center}
{\includegraphics[angle=90, width=17.5cm]{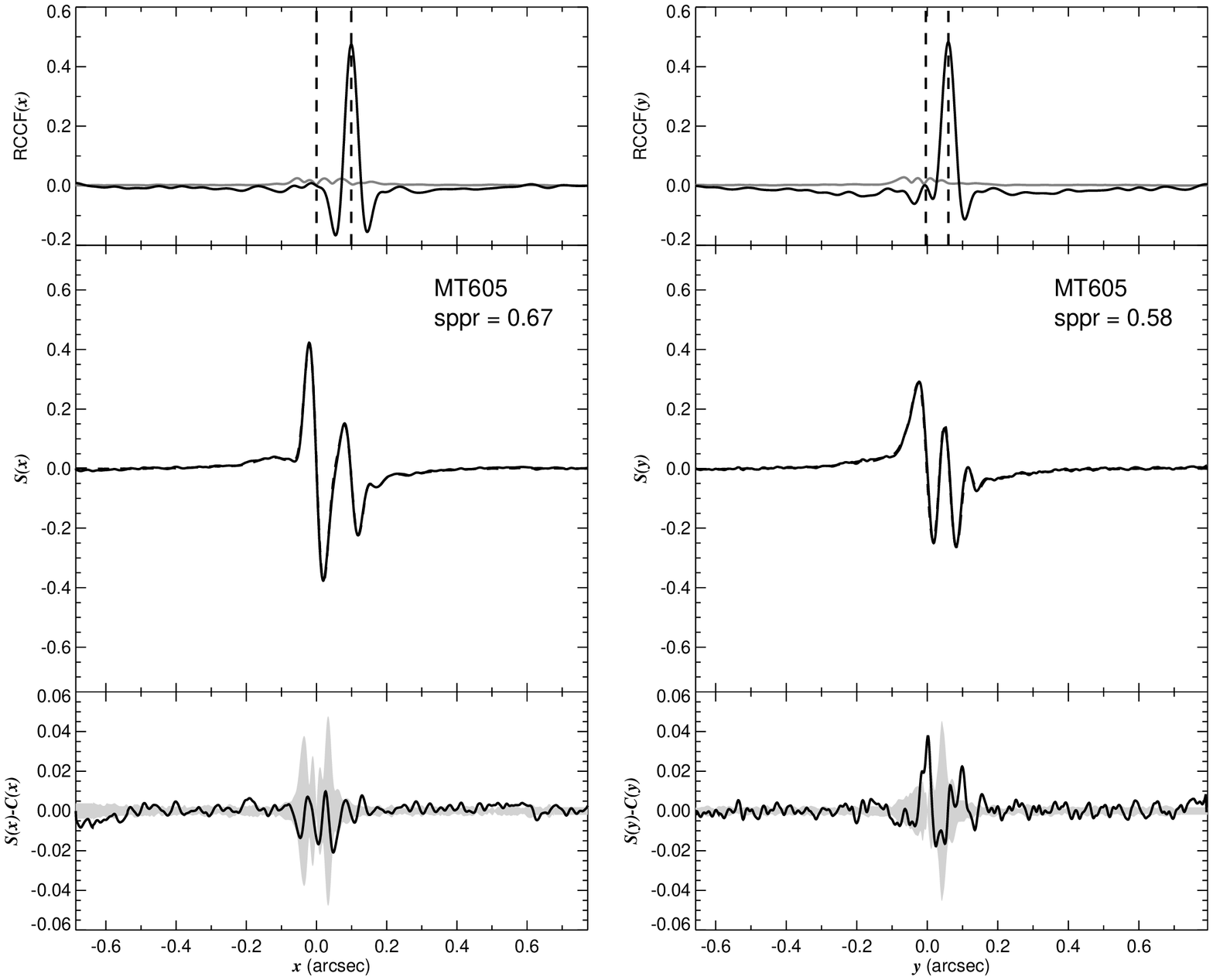}}
\end{center}
\caption{The final $S$-curves for the $x$ and $y$ orthogonal scans for MT 605 
in the same format as Fig.\ 1.1.\label{figmt605}}
\end{figure}

\clearpage
\setcounter{figure}{0}
\renewcommand{\thefigure}{\arabic{figure}.46}
\begin{figure}
\begin{center}
{\includegraphics[angle=90, width=17.5cm]{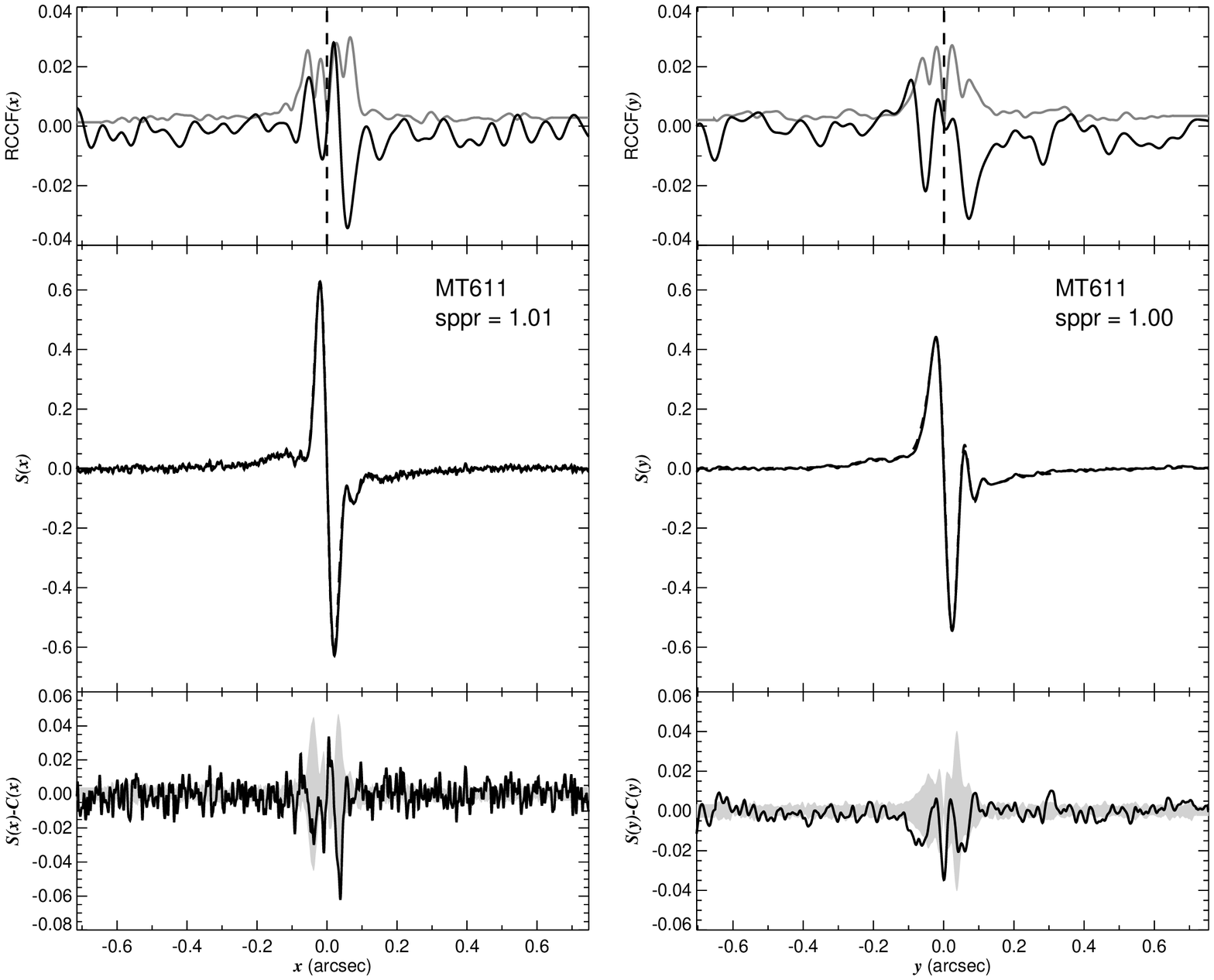}}
\end{center}
\caption{The final $S$-curves for the $x$ and $y$ orthogonal scans for MT 611 
in the same format as Fig.\ 1.1.\label{figmt611}}
\end{figure}

\clearpage
\setcounter{figure}{0}
\renewcommand{\thefigure}{\arabic{figure}.47}
\begin{figure}
\begin{center}
{\includegraphics[angle=90, width=17.5cm]{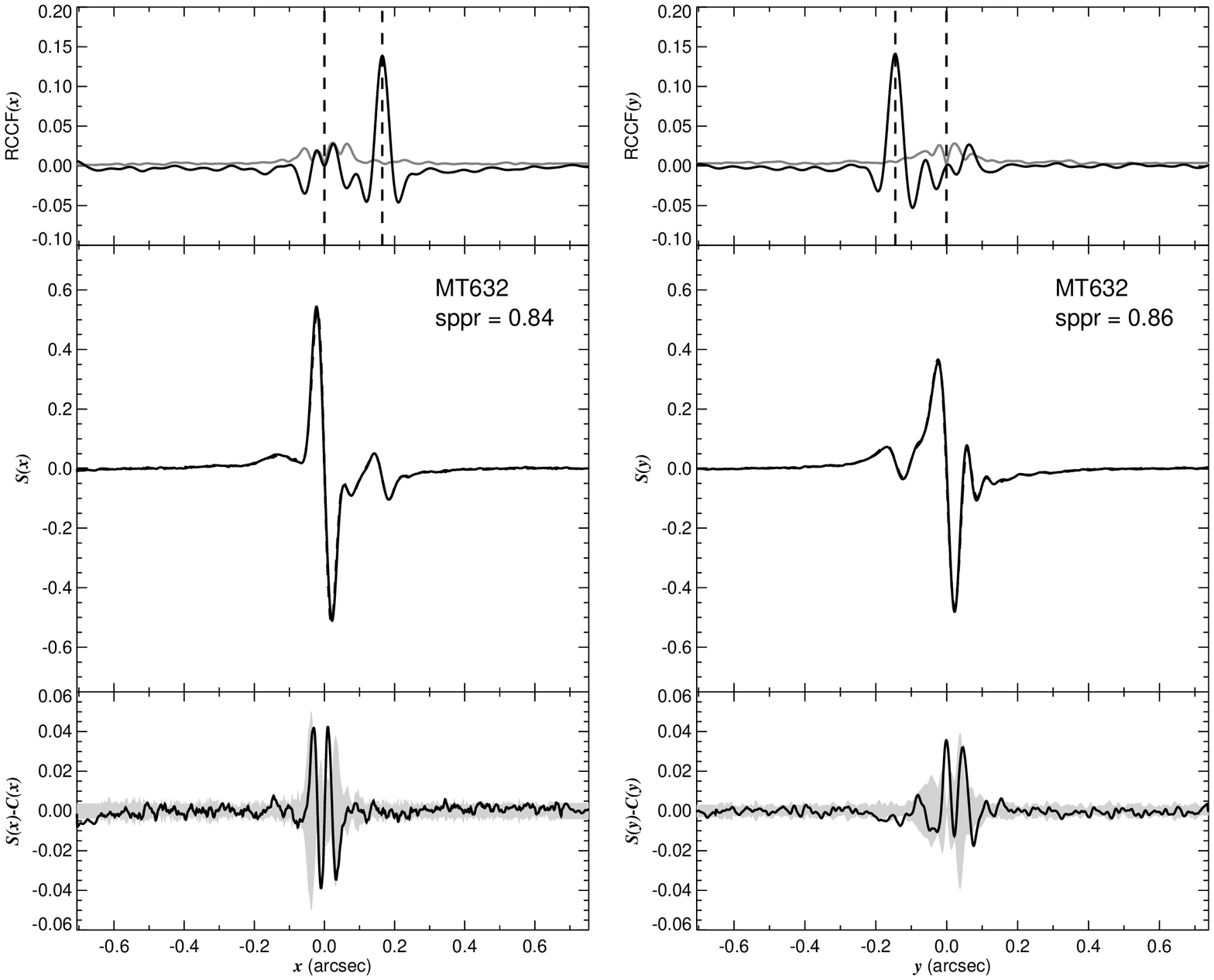}}
\end{center}
\caption{The final $S$-curves for the $x$ and $y$ orthogonal scans for MT 632 
in the same format as Fig.\ 1.1.\label{figmt632}}
\end{figure}

\clearpage
\setcounter{figure}{0}
\renewcommand{\thefigure}{\arabic{figure}.48}
\begin{figure}
\begin{center}
{\includegraphics[angle=90, width=17.5cm]{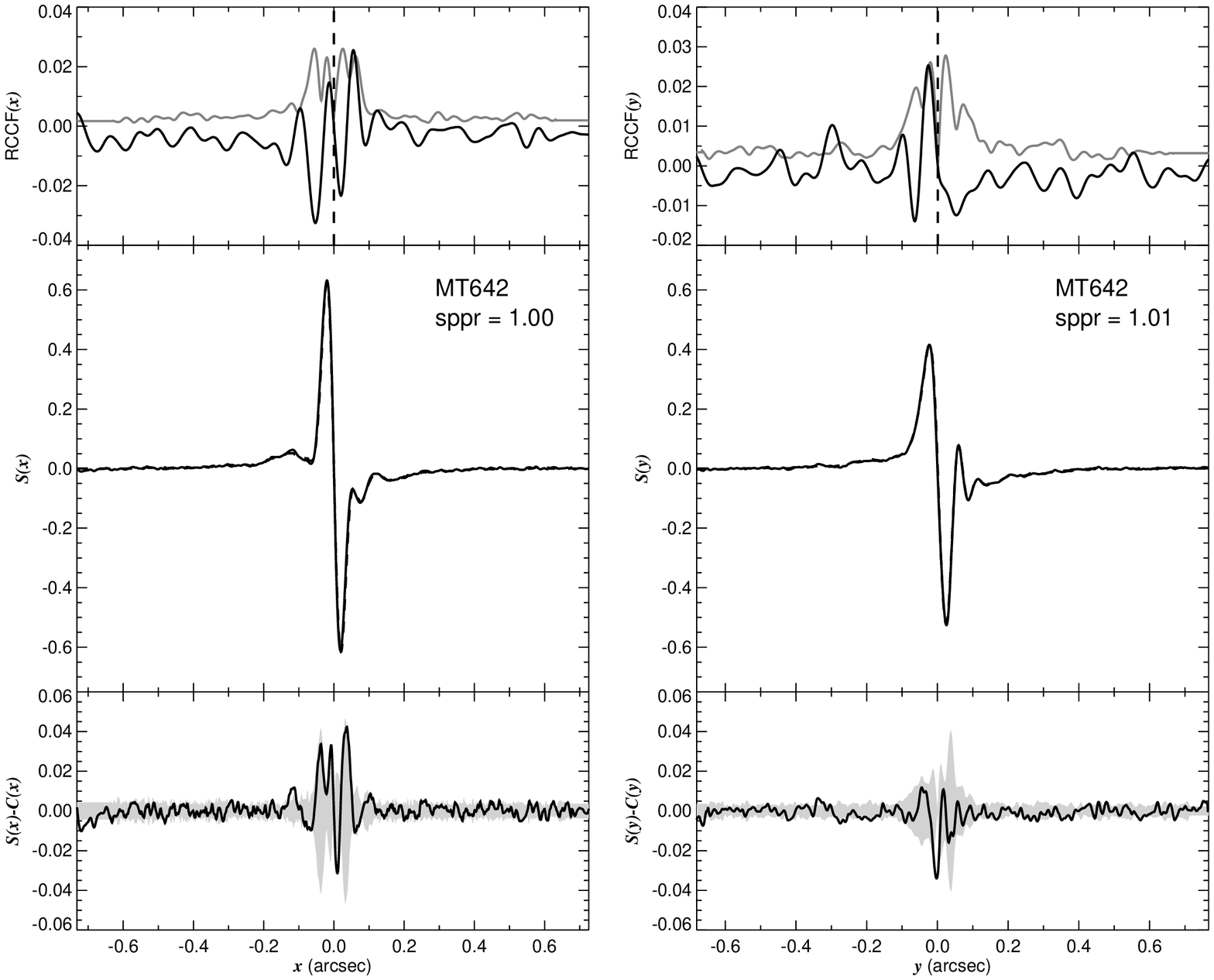}}
\end{center}
\caption{The final $S$-curves for the $x$ and $y$ orthogonal scans for MT 642 
in the same format as Fig.\ 1.1.\label{figmt642}}
\end{figure}

\clearpage
\setcounter{figure}{0}
\renewcommand{\thefigure}{\arabic{figure}.49}
\begin{figure}
\begin{center}
{\includegraphics[angle=90, width=17.5cm]{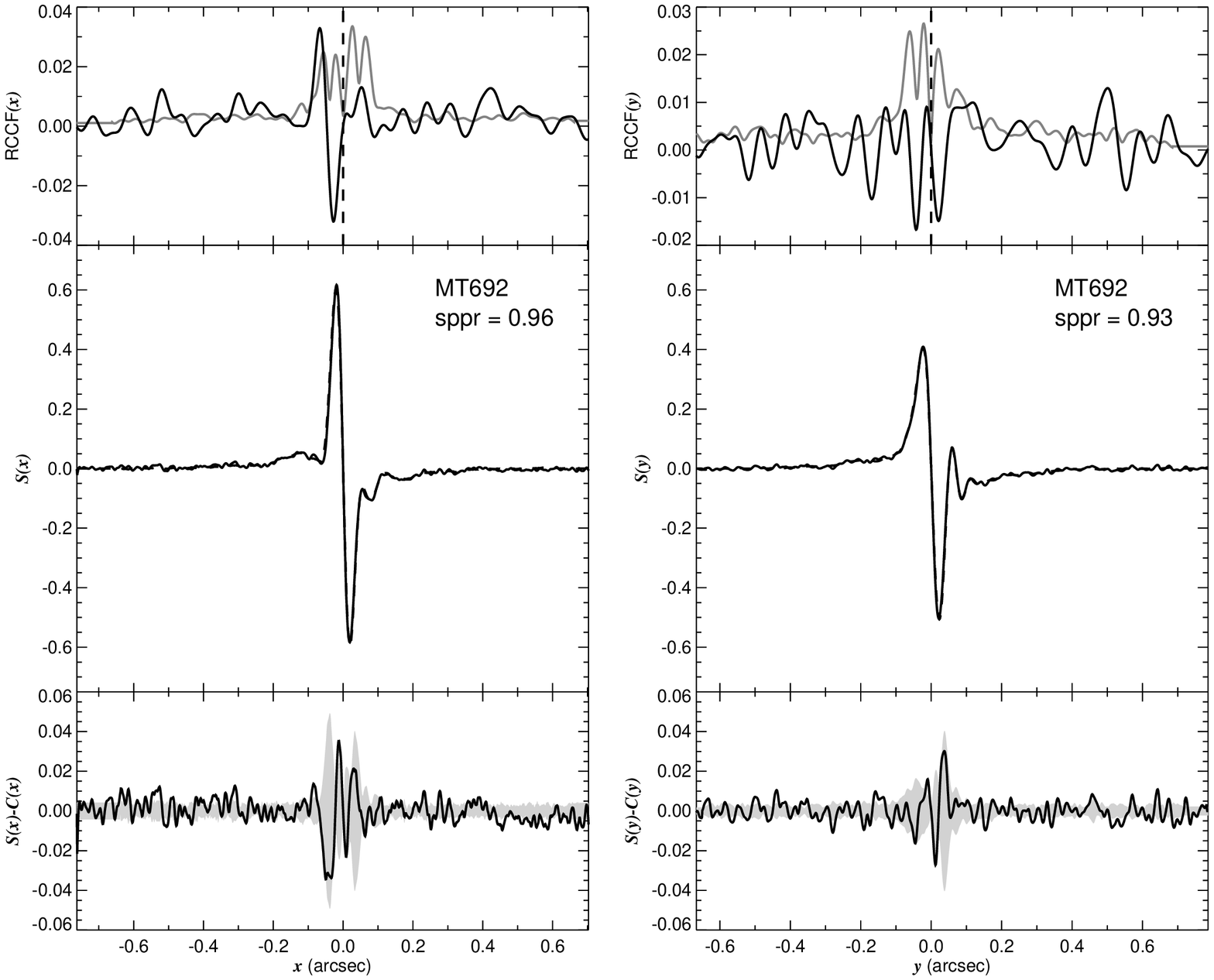}}
\end{center}
\caption{The final $S$-curves for the $x$ and $y$ orthogonal scans for MT 692
in the same format as Fig.\ 1.1.\label{figmt692}}
\end{figure}

\clearpage
\setcounter{figure}{0}
\renewcommand{\thefigure}{\arabic{figure}.50}
\begin{figure}
\begin{center}
{\includegraphics[angle=90, width=17.5cm]{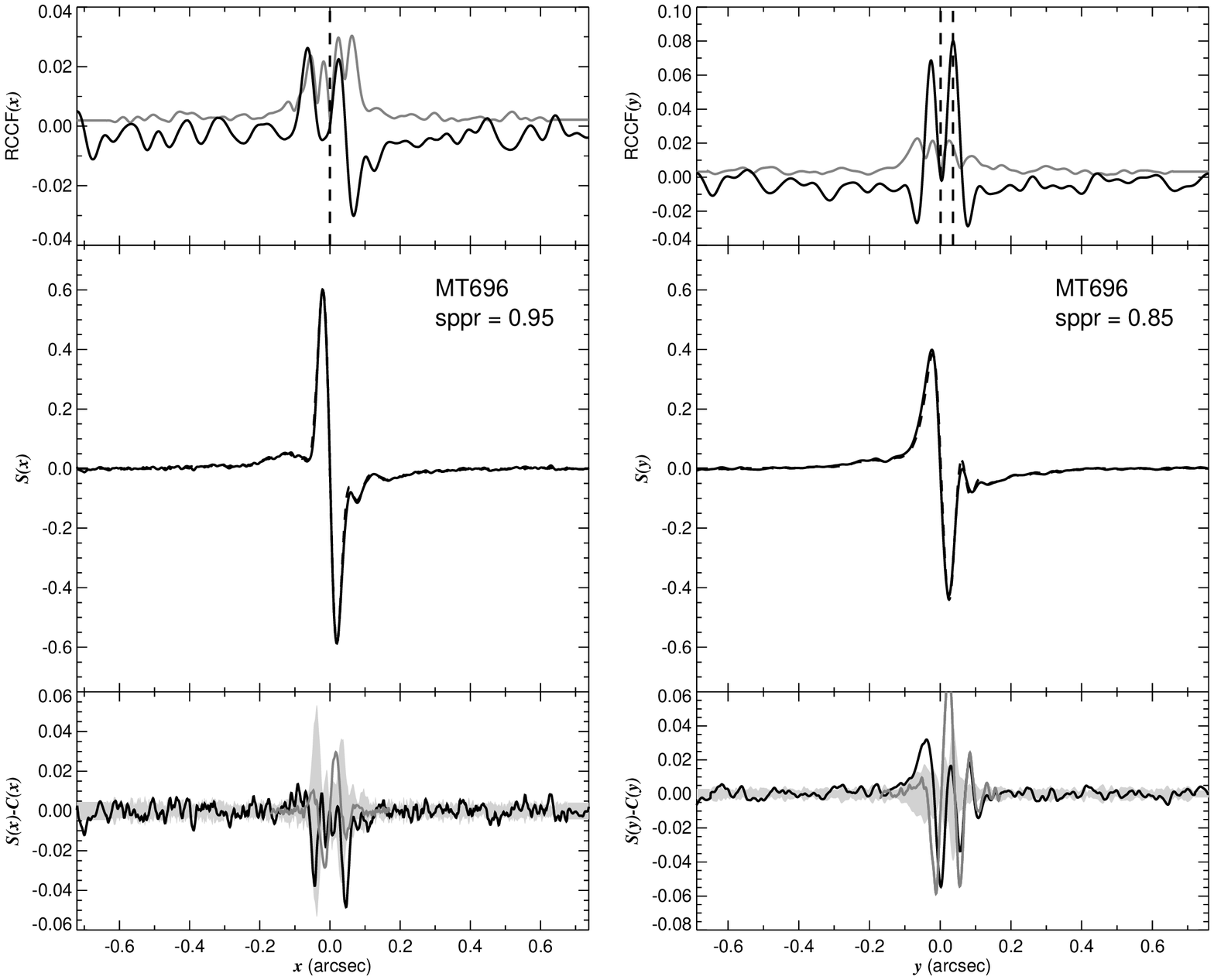}}
\end{center}
\caption{The final $S$-curves for the $x$ and $y$ orthogonal scans for MT 696
in the same format as Fig.\ 1.1.\label{figmt696}}
\end{figure}

\clearpage
\setcounter{figure}{0}
\renewcommand{\thefigure}{\arabic{figure}.51}
\begin{figure}
\begin{center}
{\includegraphics[angle=90, width=17.5cm]{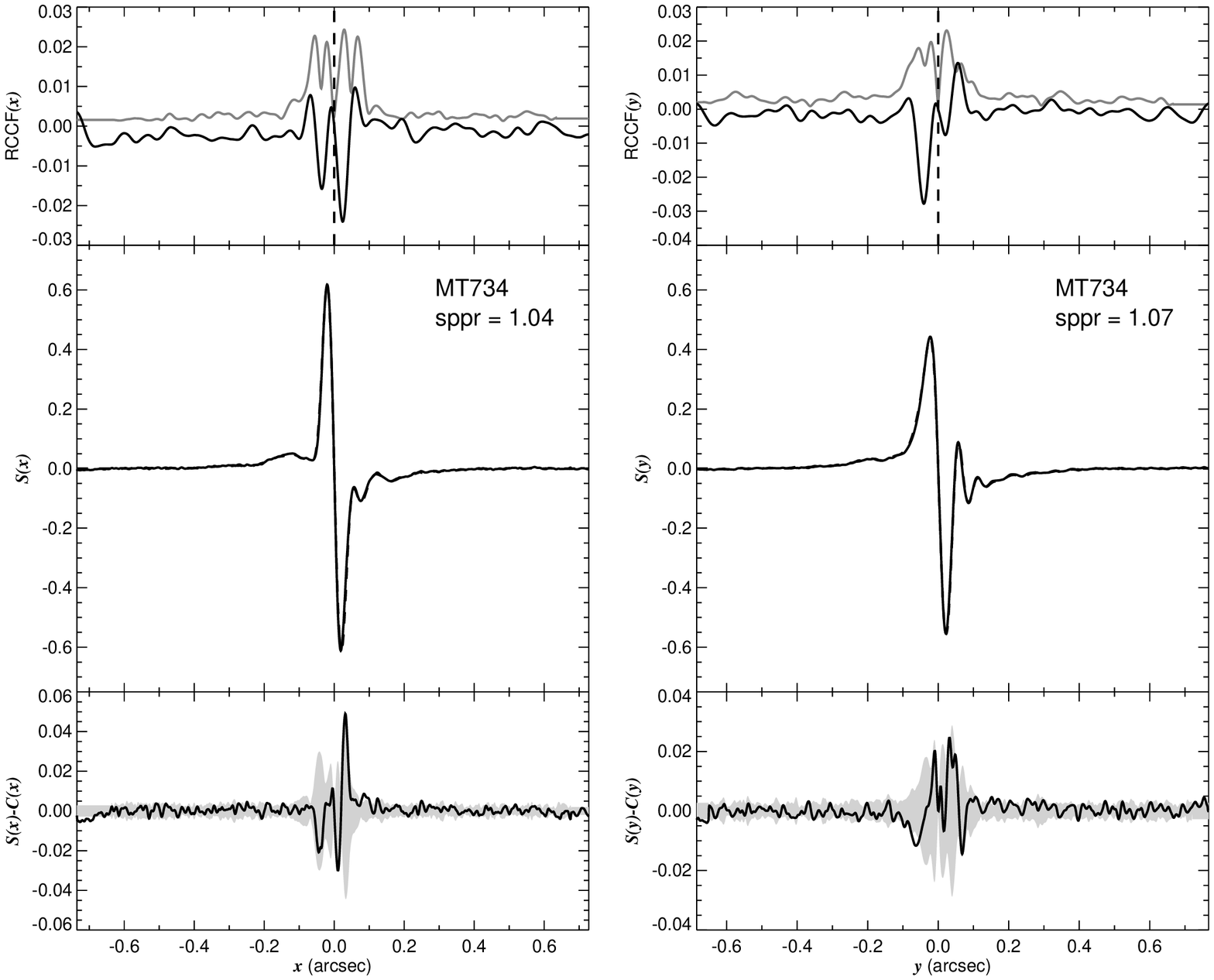}}
\end{center}
\caption{The final $S$-curves for the $x$ and $y$ orthogonal scans for MT 734
in the same format as Fig.\ 1.1.\label{figmt734}}
\end{figure}

\clearpage
\setcounter{figure}{0}
\renewcommand{\thefigure}{\arabic{figure}.52}
\begin{figure}
\begin{center}
{\includegraphics[angle=90, width=17.5cm]{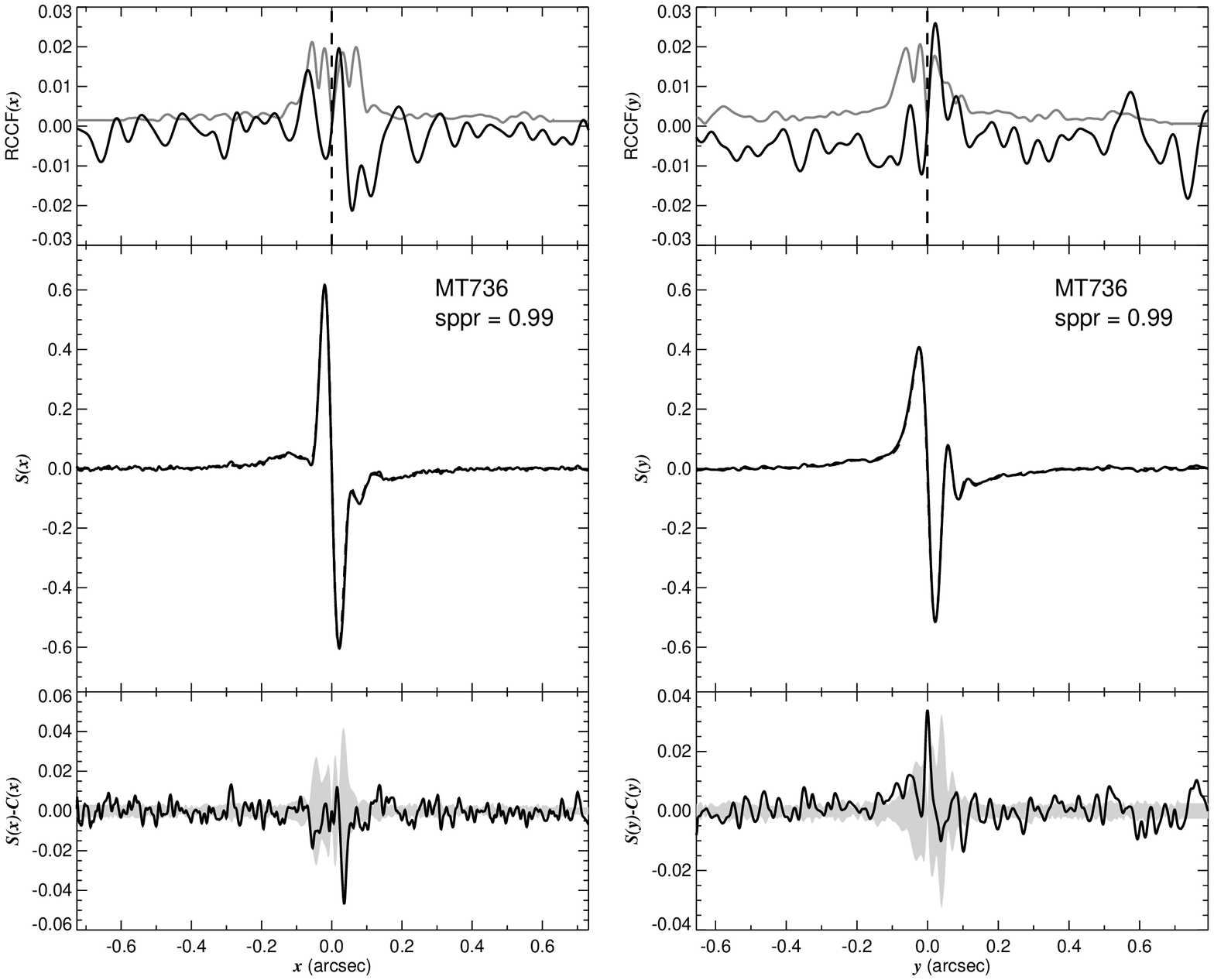}}
\end{center}
\caption{The final $S$-curves for the $x$ and $y$ orthogonal scans for MT 736 
in the same format as Fig.\ 1.1.\label{figmt736}}
\end{figure}

\clearpage
\setcounter{figure}{0}
\renewcommand{\thefigure}{\arabic{figure}.53}
\begin{figure}
\begin{center}
{\includegraphics[angle=90, width=17.5cm]{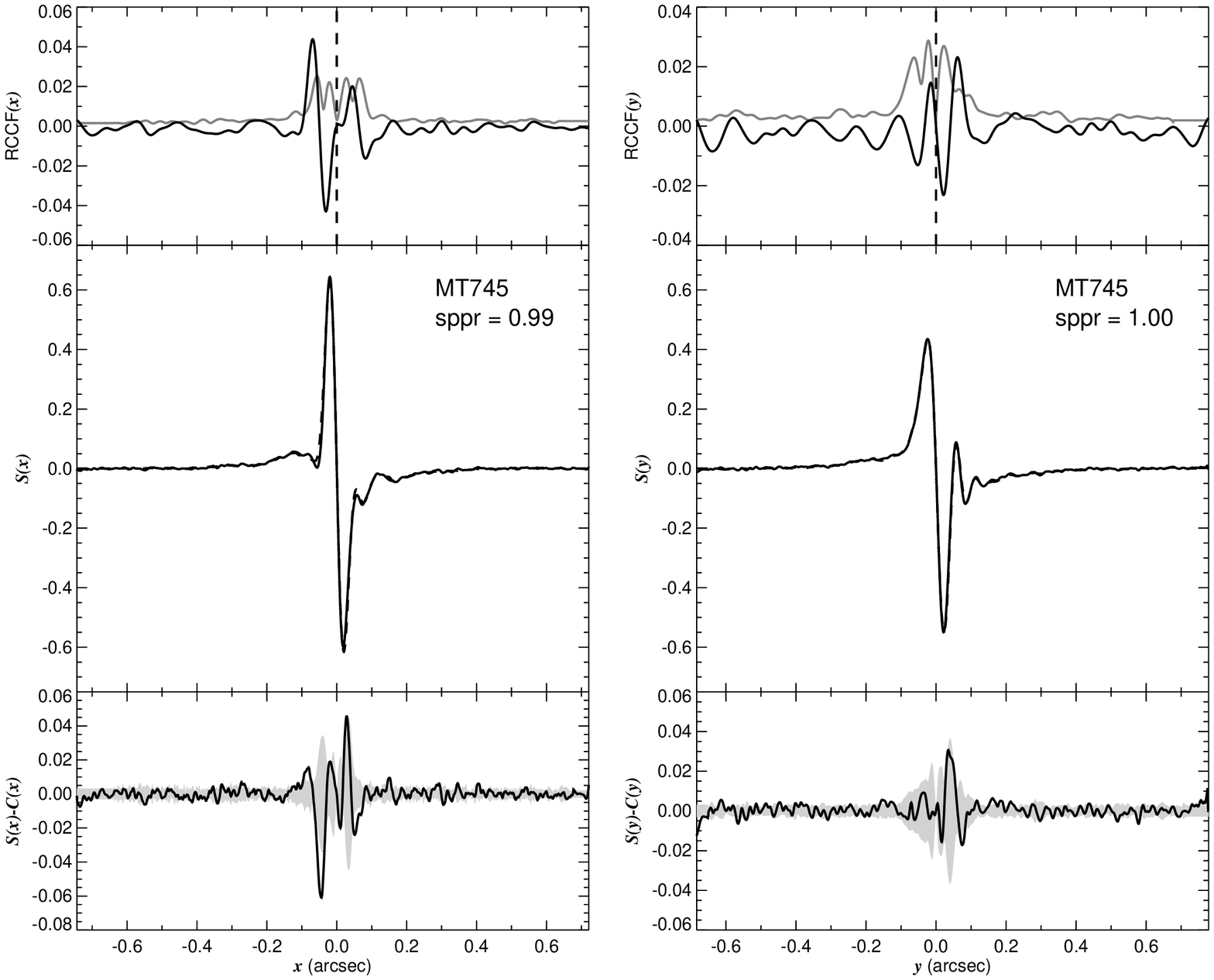}}
\end{center}
\caption{The final $S$-curves for the $x$ and $y$ orthogonal scans for MT 745
in the same format as Fig.\ 1.1.\label{figmt745}}
\end{figure}

\clearpage
\setcounter{figure}{0}
\renewcommand{\thefigure}{\arabic{figure}.54}
\begin{figure}
\begin{center}
{\includegraphics[angle=90, width=17.5cm]{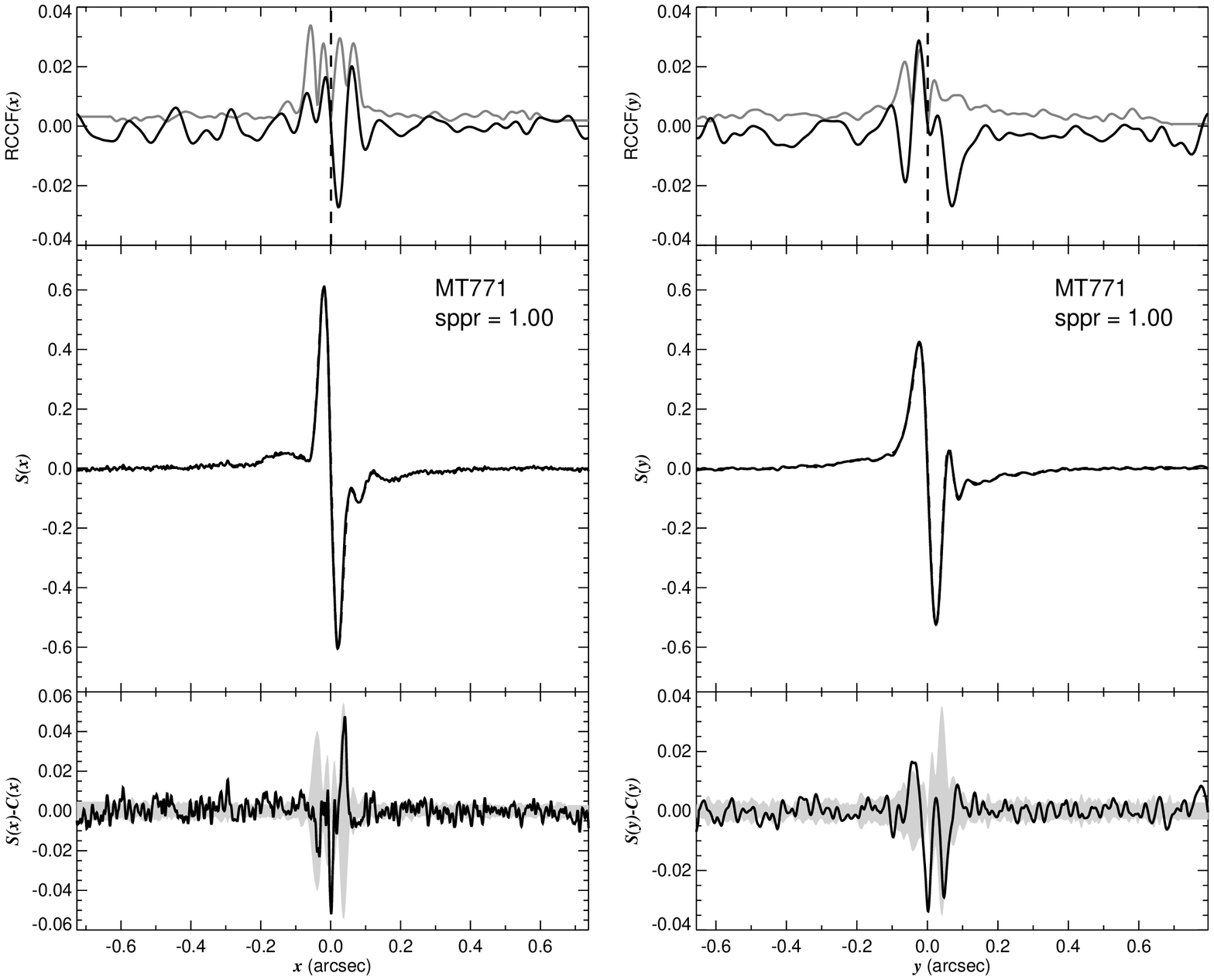}}
\end{center}
\caption{The final $S$-curves for the $x$ and $y$ orthogonal scans for MT 771 
in the same format as Fig.\ 1.1.\label{figmt771}}
\end{figure}

\clearpage
\setcounter{figure}{0}
\renewcommand{\thefigure}{\arabic{figure}.55}
\begin{figure}
\begin{center}
{\includegraphics[angle=90, width=17.5cm]{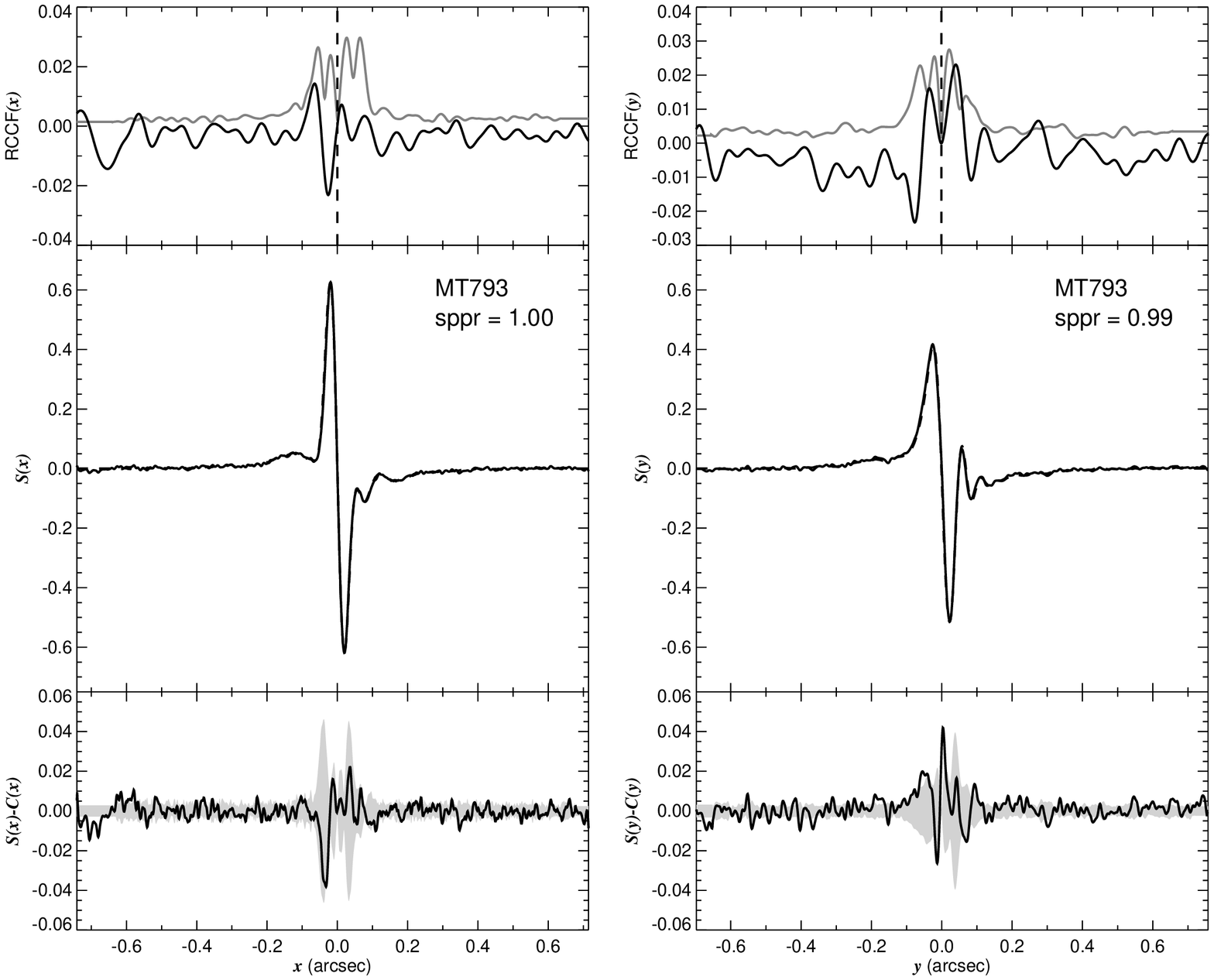}}
\end{center}
\caption{The final $S$-curves for the $x$ and $y$ orthogonal scans for MT 793
in the same format as Fig.\ 1.1.\label{figmt 793}}
\end{figure}

\clearpage
\setcounter{figure}{0}
\renewcommand{\thefigure}{\arabic{figure}.56}
\begin{figure}
\begin{center}
{\includegraphics[angle=90, width=17.5cm]{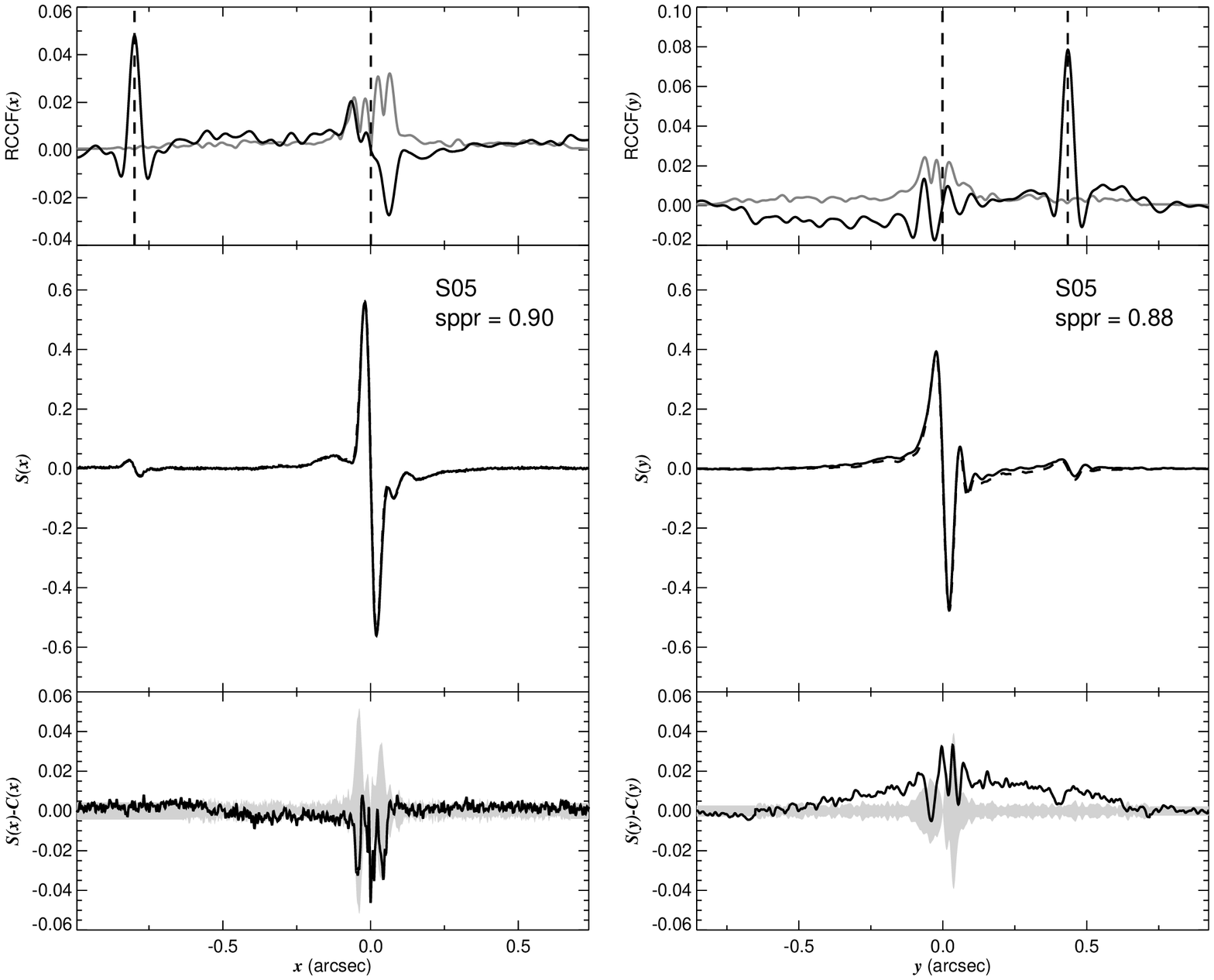}}
\end{center}
\caption{The final $S$-curves for the $x$ and $y$ orthogonal scans for
Schulte~5 in the same format as Fig.\ 1.1.\label{figs05}}
\end{figure}

\clearpage
\setcounter{figure}{0}
\renewcommand{\thefigure}{\arabic{figure}.57}
\begin{figure}
\begin{center}
{\includegraphics[angle=90, width=17.5cm]{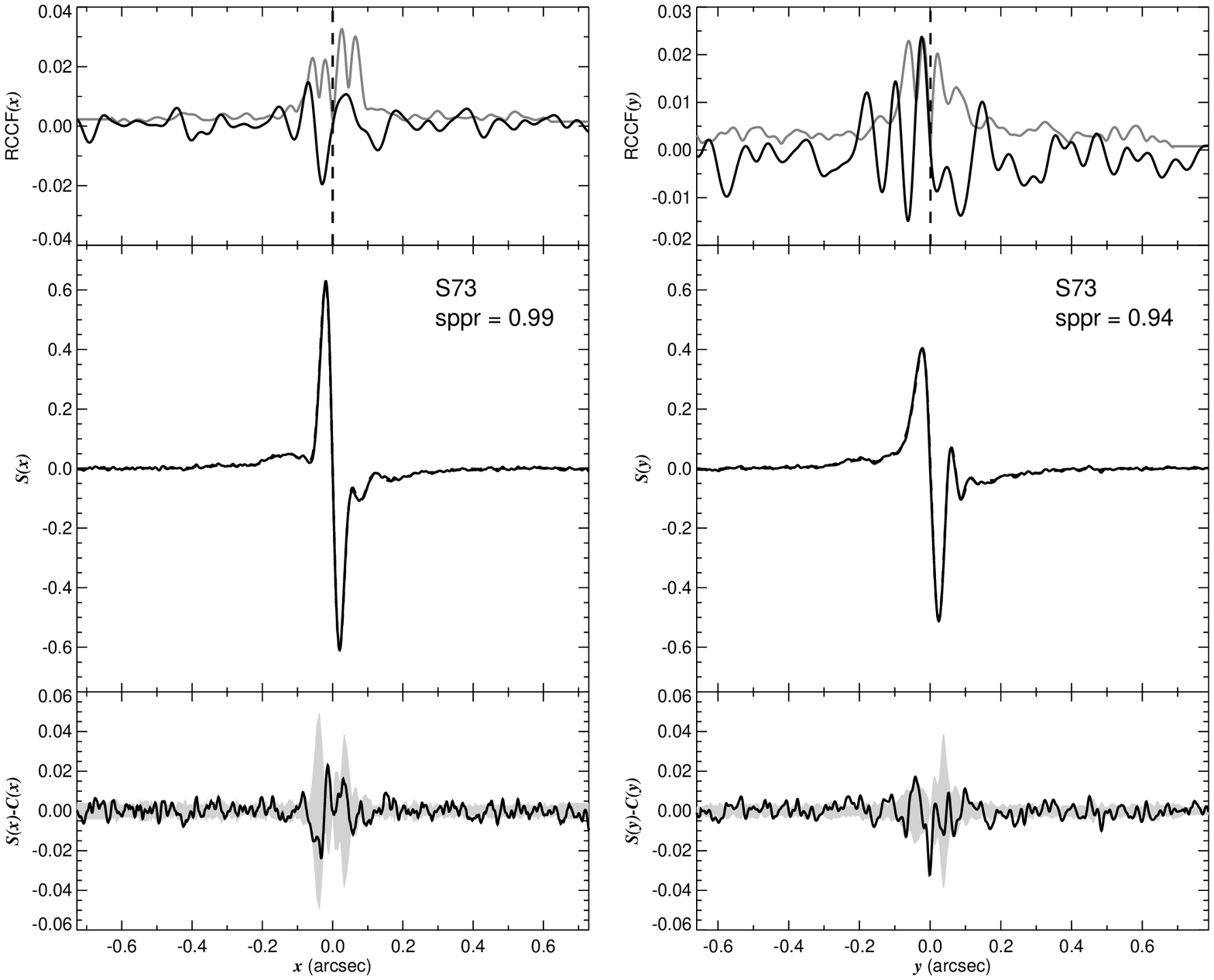}}
\end{center}
\caption{The final $S$-curves for the $x$ and $y$ orthogonal scans for
Schulte~73 in the same format as Fig.\ 1.1.\label{figs73}}
\end{figure}%

\clearpage
\setcounter{figure}{0}
\renewcommand{\thefigure}{\arabic{figure}.58}
\begin{figure}
\begin{center}
{\includegraphics[angle=90, width=17.5cm]{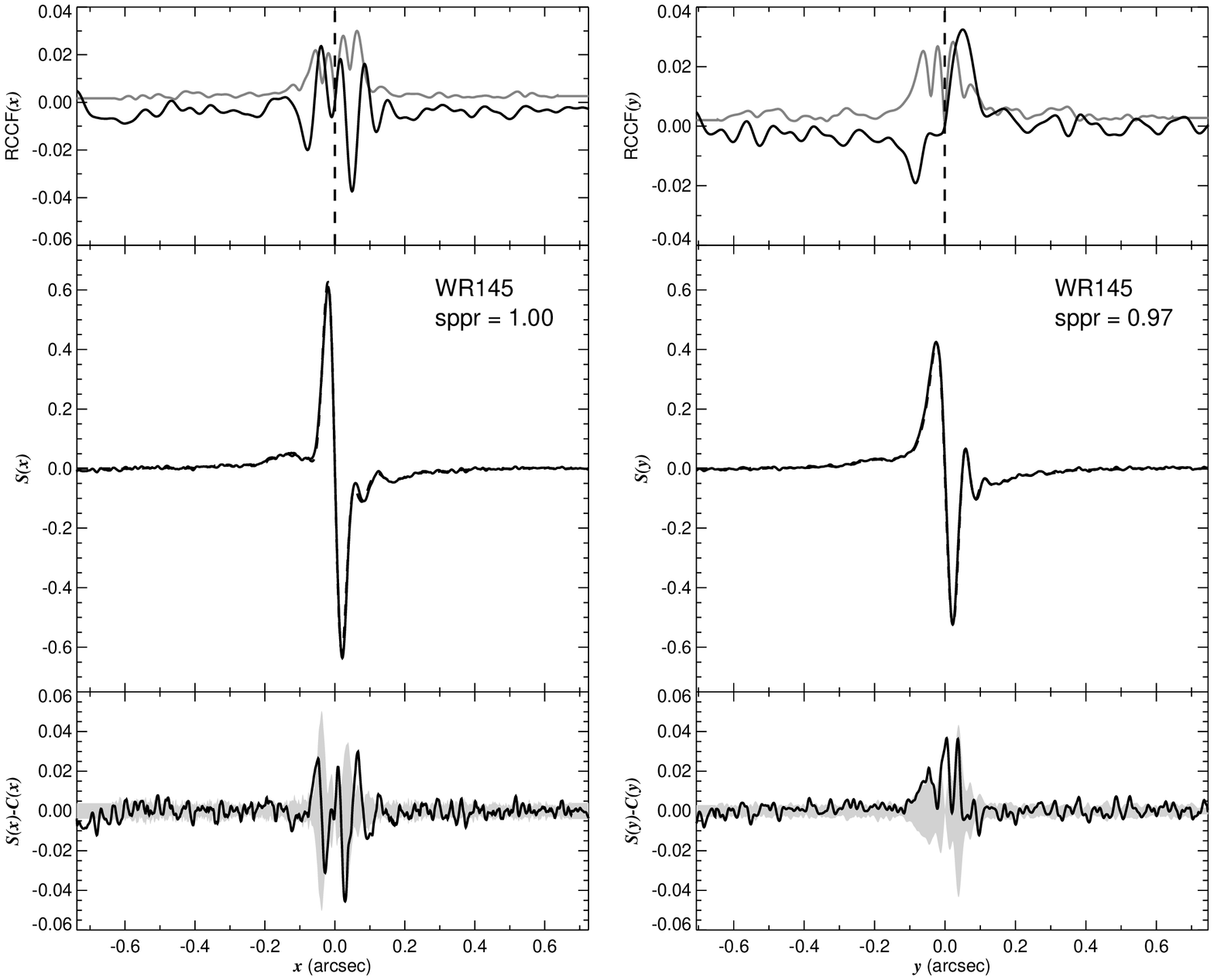}}
\end{center}
\caption{The final $S$-curves for the $x$ and $y$ orthogonal scans for WR~145 
in the same format as Fig.\ 1.1.\label{figwr145}}
\end{figure}


\clearpage

\setcounter{figure}{1}
\renewcommand{\thefigure}{\arabic{figure}}
\begin{figure}
\begin{center}
{\includegraphics[angle=90,width=15cm]{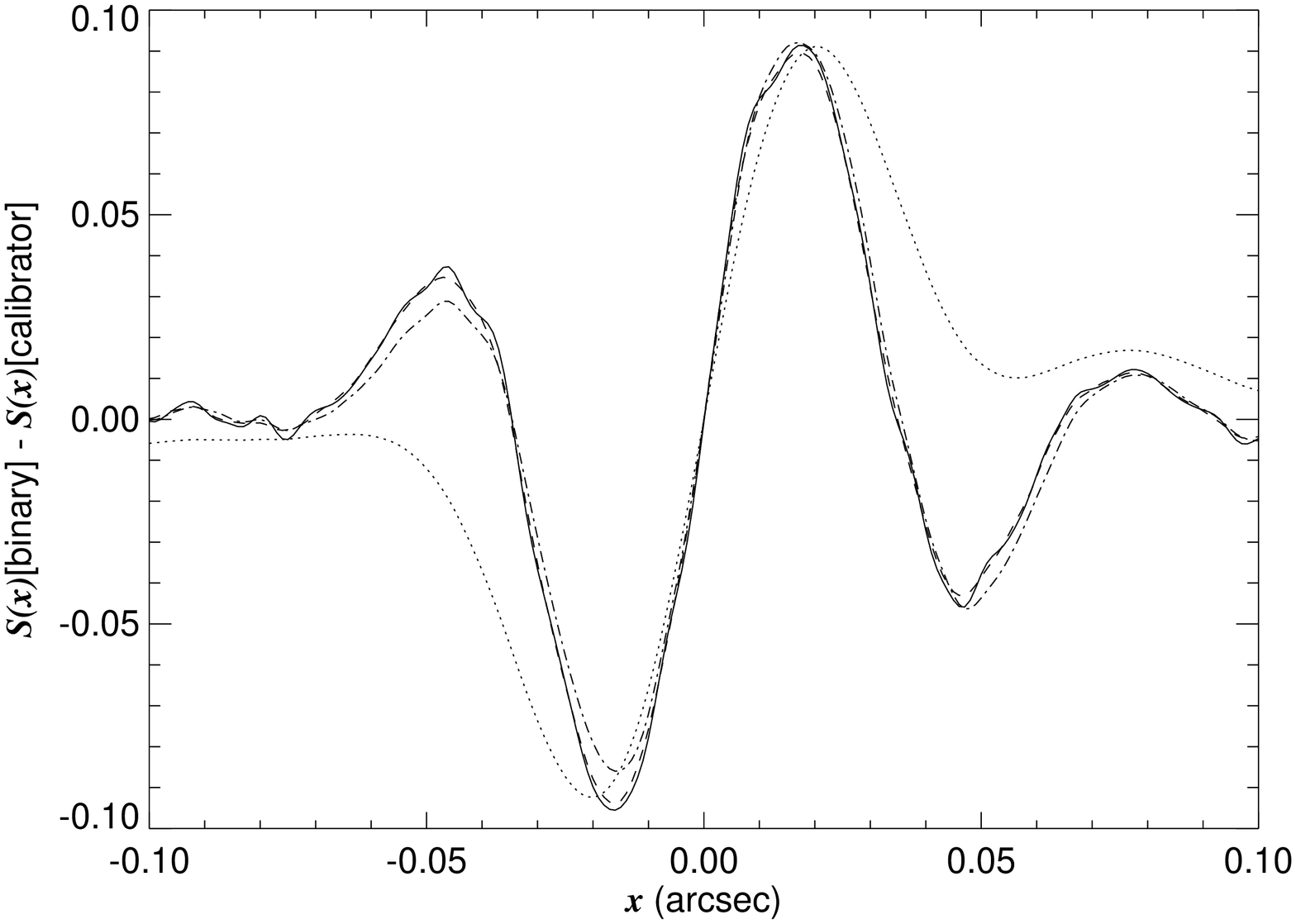}}
\caption{Example plots of the difference between a model binary and
calibrator $S$-curves. The dashed line shows the difference for a
binary with $r = 1$ and $\Delta x=0\farcs015$, and the dotted-dashed
line shows the difference for a binary with $r=0.5$ and $\Delta x
=0\farcs0159$. Both resemble the second derivative of the calibrator
$S$-curve (solid line) for the same amplitude $a$, but differ from
that caused by the dilution of an off-scan companion (dotted
line). \label{f2-fgs}}
\end{center}
\end{figure}

\clearpage

\begin{figure}
\begin{center}
{\includegraphics[width=3.5in]{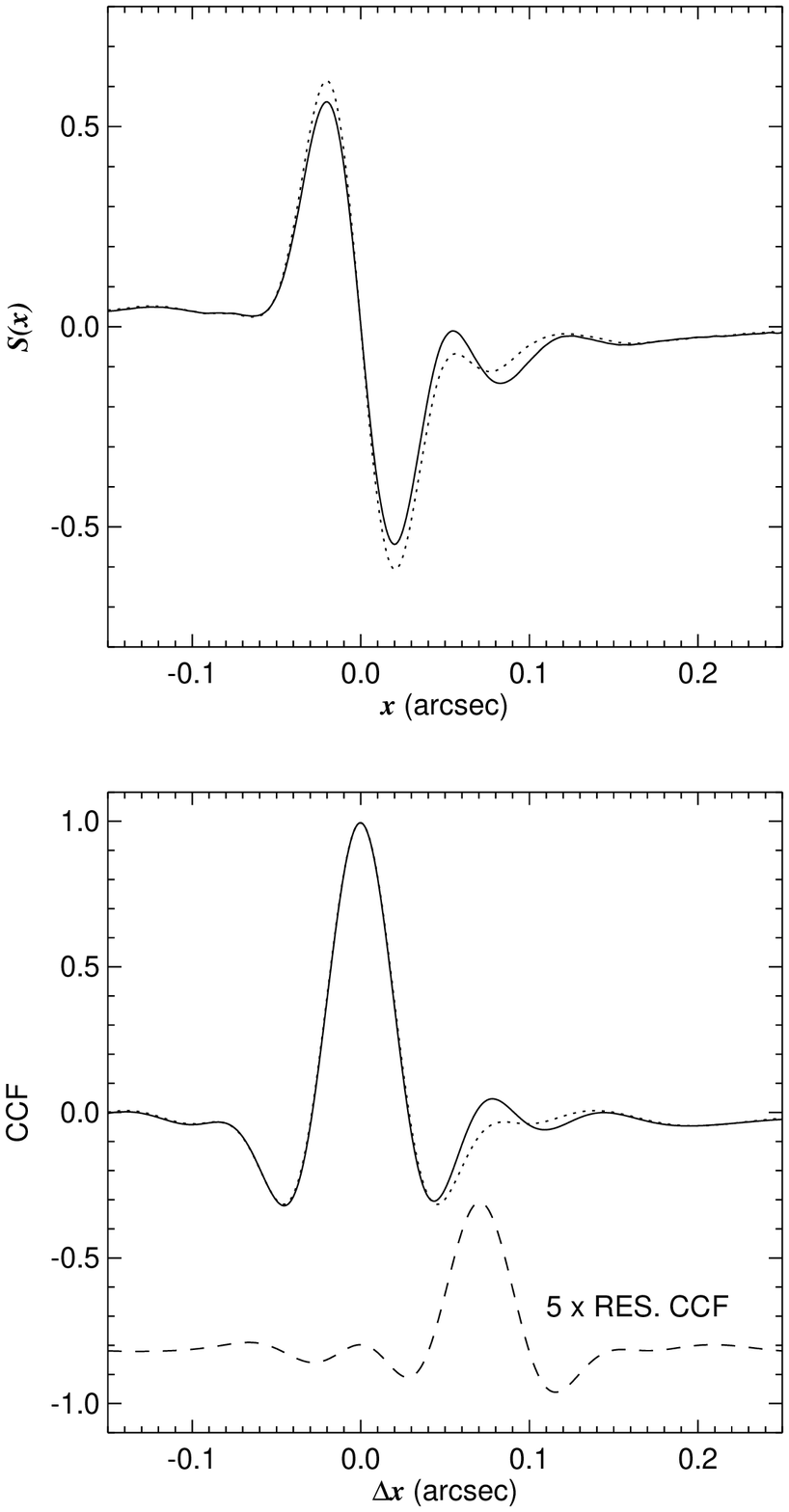}}
\caption{Example of the cross-correlation function. The top panel
  shows an example of a model $S$-curve of a binary star with $\rho =
  +0\farcs07$ and a flux ratio $r=\frac{F_{2}}{F_{1}} = 0.1$ (solid
  line) and that of a single star calibrator (dotted line). The lower
  panel shows the cross-correlation functions of the target with the
  calibrator (solid line) and of the calibrator with itself (dotted
  line). The difference between these two is shown as a dashed line in
  an expanded scale, offset for clarity.\label{f3-fgs}}
\end{center}
\end{figure}

\clearpage

\begin{figure}
\begin{minipage}[b]{0.5\linewidth}
\begin{center}
{\includegraphics[angle=90,width=3.3in]{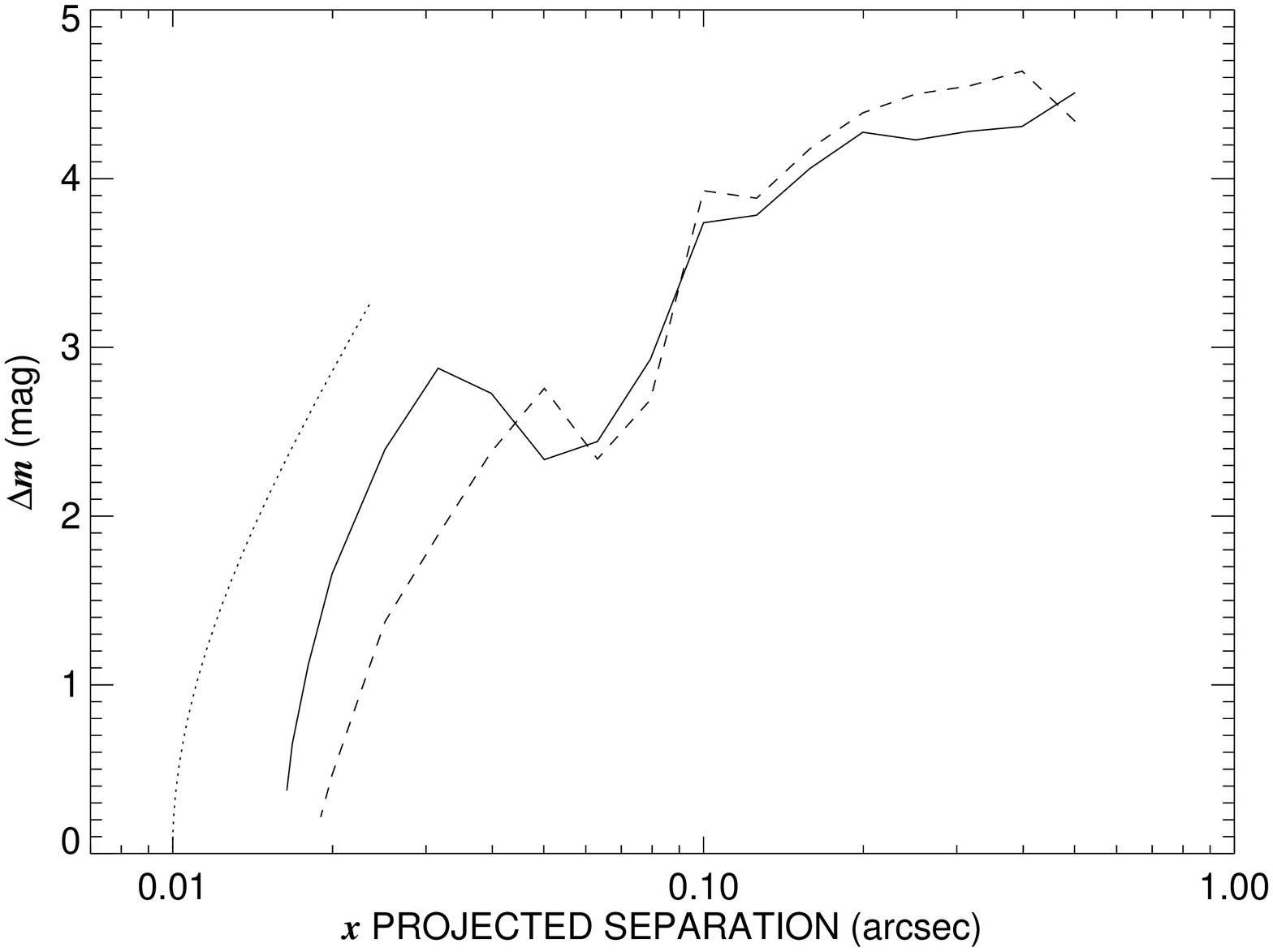}}
\end{center}
\end{minipage}
\hspace{0.5cm}
\begin{minipage}[b]{0.5\linewidth}
\begin{center}
{\includegraphics[angle=90,width=3.3in]{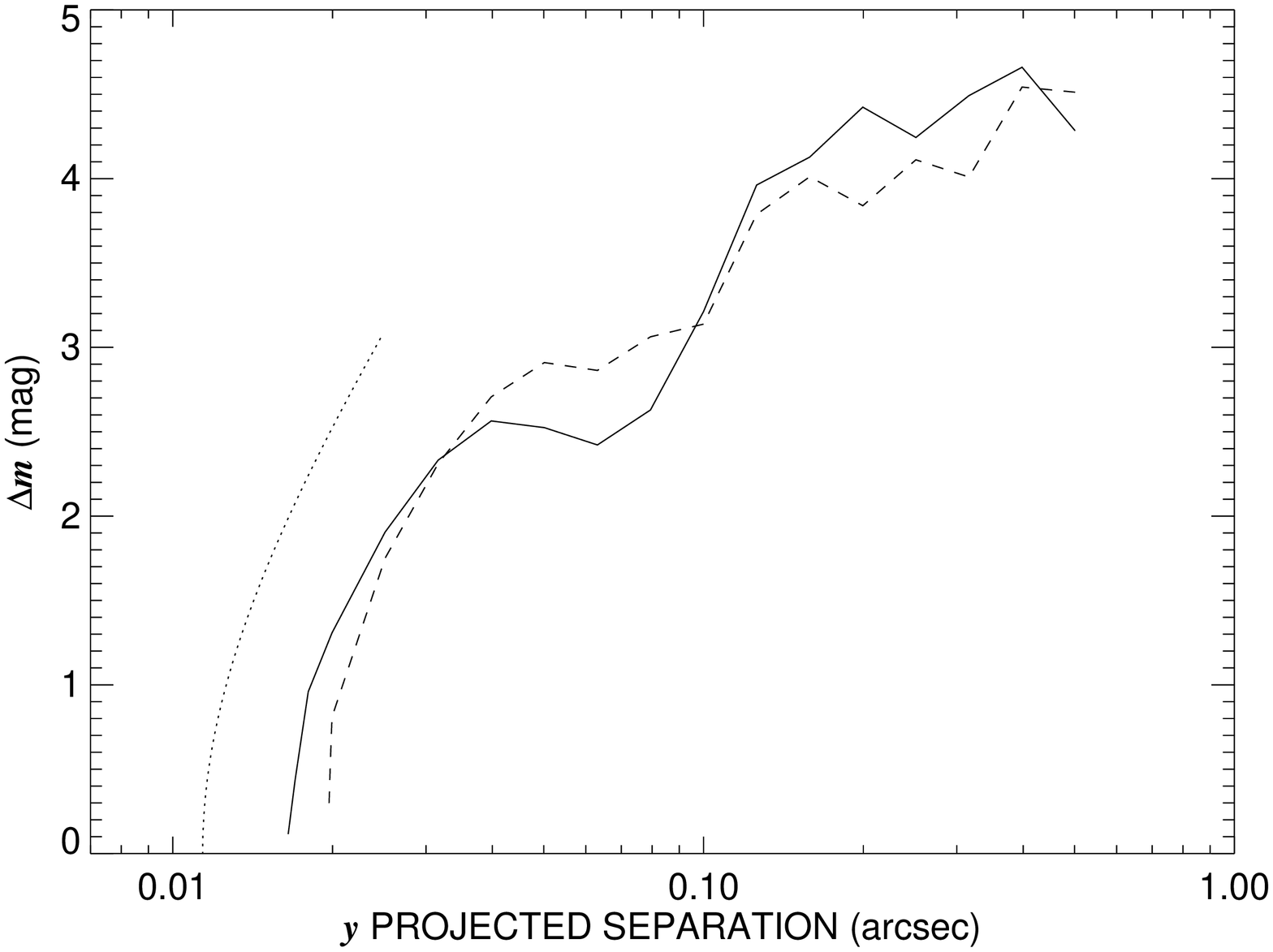}}
\end{center}
\end{minipage}
\caption{Binary detection limits as a function of projected separation
and magnitude difference. The solid (dashed) lines indicate the limits
for positive (negative) offsets in $x$ (left panel) and $y$ (right
panel), respectively. Binaries with separations and magnitude
differences below these limits should be detected. The dotted lines
indicate detection limits for marginally resolved binaries using the second
derivative method. \label{f4-fgs}}
\end{figure}


\end{document}